\documentclass[apj]{emulateapj}
\journalinfo{The Astrophysical Journal, in press}
\slugcomment{Received 2007 November 16; Accepted 2008 February 4}

\newcommand{\lz}{low-$z$}
\newcommand{\hz}{high-$z$}
\newcommand{\ztwo}{$z \sim 2$}

\newcommand{\ha}{H$\alpha$}
\newcommand{\msun}{M$_{\odot}$} 
\newcommand{\msunyr}{\msun\ yr$^{-1}$}

\newcommand{\farcsec}{\hbox{$.\!\!^{\prime\prime}$}}    
\newcommand{\kms}{km~s$^{-1}$}

\newcommand{\vasym}{$v_{asym}$}
\newcommand{\sasym}{$\sigma_{asym}$}
\newcommand{\kasym}{$K_{asym}$}

\newcommand{\nsins}{11}
\newcommand{\perdisks}{$> 50$\%}

\shorttitle{Kinemetry of High-Redshift Galaxies}
\shortauthors{K. L. Shapiro et al.}

\begin{document}


\title{Kinemetry of SINS High-Redshift Star-Forming Galaxies:\altaffilmark{*}

 Distinguishing Rotating Disks from Major Mergers}

\author{Kristen L. Shapiro,\altaffilmark{1} Reinhard Genzel,\altaffilmark{2,3} 
Natascha M. F\"orster Schreiber,\altaffilmark{2} Linda J. Tacconi,\altaffilmark{2} 
Nicolas Bouch\'e,\altaffilmark{2} Giovanni Cresci,\altaffilmark{2} Richard Davies,\altaffilmark{2} Frank Eisenhauer,\altaffilmark{2} 
Peter H. Johansson,\altaffilmark{4} Davor Krajnovi\'c,\altaffilmark{5} Dieter Lutz,\altaffilmark{2} 
Thorsten Naab,\altaffilmark{4} 
Nobuo Arimoto,\altaffilmark{6} Santiago Arribas,\altaffilmark{7} Andrea Cimatti,\altaffilmark{8}
Luis Colina,\altaffilmark{7} Emanuele Daddi,\altaffilmark{9} Olivier Daigle,\altaffilmark{10}
Dawn~Erb,\altaffilmark{11} Olivier Hernandez,\altaffilmark{10} Xu Kong,\altaffilmark{12}
Marco Mignoli,\altaffilmark{13} Masato Onodera,\altaffilmark{14} Alvio~Renzini,\altaffilmark{15} 
Alice~Shapley,\altaffilmark{16} Charles Steidel\altaffilmark{17}}

 \altaffiltext{*}{Based on observations obtained at the Very Large Telescope (VLT) of the European Southern Observatory, Paranal, Chile in the context of guaranteed time programs 073.B-9018, 074.A-9011, 075.A-0466, 076.A-0527, 077.A-0576, 078.A-0600, and 079.A-0341.}

\altaffiltext{1}{Department of Astronomy, Campbell Hall, University of California, Berkeley, California 94720, USA}
\altaffiltext{2}{Max-Planck-Institut f\"ur extraterrestrische Physik (MPE), Giessenbachstr. 1, 85748 Garching, Germany}
\altaffiltext{3}{Department of Physics, Le Conte Hall, University of California, Berkeley, California 94720, USA}
\altaffiltext{4}{Universit\"ats-Sternwarte M\"unchen, Scheinerstr. 1, D-81679 M\"unchen, Germany}
\altaffiltext{5}{Denys Wilkinson Building, University of Oxford, Keble Road, OX1 3RH, UK}
\altaffiltext{6}{National Astronomical Observatory of Japan, Mitaka, Tokyo 181-8588, Japan}
\altaffiltext{7}{Instituto de Estructura de la Materia, Consejo Superior de Investigaciones Cientificas, Serrano 121, 28006 Madrid, Spain}
\altaffiltext{8}{Dipartimento di Astronomia, Alma Mater Studiorum, Universit\`a di Bologna, Via Ranzani 1, I-40127, Italy}
\altaffiltext{9}{Laboratoire AIM, CEA/DSM - CNRS - Universit\'e Paris Diderot, DAPNIA/SAp, Orme des Merisiers, 91191 Gif-sur-Yvette, France}
\altaffiltext{10}{Observatoire du mont M\'egantic, LAE, Universit\'e de Montr\'eal, C.P. 6128 succ. centre ville, Montr\'eal, Qu\'ebec H3C 3J7, Canada}
\altaffiltext{11}{Harvard-Smithsonian Center for Astrophysics, 60 Garden Street, Cambridge, Massachusetts 02138, USA}
\altaffiltext{12}{Center for Astrophysics, University of Science and Technology of China, 230026 Hefei, PR China}
\altaffiltext{13}{Instituto Nazionale di Astrofisica, Osservatorio Astronomico di Bologna, Via Gobetti 101, I-40129 Bologna, Italy}
\altaffiltext{14}{Institute of Earth, Atmosphere and Astronomy, BK21, Yonsei University, Seoul, 120-749 South Korea}
\altaffiltext{15}{Osservatorio Astronomico di Padova, Vicolo dell'Osservatorio 5, I-35122 Padova, Italy}
\altaffiltext{16}{Princeton University Observatory, Peyton Hall, Ivy Lane, Princeton, New Jersey 08544, USA}
\altaffiltext{17}{California Institute of Technology, MS 105-24, Pasadena, California 91125, USA}


\begin{abstract}
	We present a simple set of kinematic criteria that can distinguish between galaxies dominated by ordered rotational motion and those involved in major merger events.  Our criteria are based on the dynamics of the warm ionized gas (as traced by \ha) within galaxies, making this analysis accessible to high-redshift systems, whose kinematics are primarily traceable through emission features.  Using the method of kinemetry (developed by Krajnovi\'c and coworkers), we quantify asymmetries in both the velocity and velocity dispersion maps of the warm gas, and the resulting criteria enable us to empirically differentiate between non-merging and merging systems at high redshift.  We apply these criteria to \nsins\ of our best-studied rest-frame UV/optical-selected \ztwo\ galaxies for which we have near-infrared integral field spectroscopic data from SINFONI on the VLT.  Of these \nsins\ systems, we find that \perdisks\ have kinematics consistent with a single rotating disk interpretation, while the remaining systems are more likely undergoing major mergers.  This result, combined with the short formation timescales of these systems, provides evidence that rapid, smooth accretion of gas plays a significant role in galaxy formation at high redshift.
\end{abstract}




\section{Introduction}

	In recent years, deep observations of galaxies at $z~\sim~2-3$ have provided detailed insight into the growth of structure and galaxy evolution in the early Universe.  High-resolution broad-band imaging and long-slit spectroscopic surveys have shown the population at this redshift to be rapidly evolving and diverse.  This is a reflection of the significant growth of galaxies that occurs at this epoch; at \ztwo, both the cosmic star formation rate and the luminous quasar space density are at their peak \citep[e.g.][]{Fan+01,Cha+05,HopBea06}.  Correspondingly, the stellar mass density in galaxies increases from $\sim 15\%$ its current value at $z \sim 3$ to $50 - 75 \%$ its current value at $z \sim 1$ \citep[e.g.][]{Dic+03,Fon+03,Rud+03,Rud+06}, making this the era when much of the assembly of massive galaxies occurs.  Systems at this redshift consequently show a wide range in properties, with nuclear activities varying from negligible to active to powerful QSOs, star formation rates varying from less than 1 \msunyr\ \citep[e.g. passively evolving BzK-selected objects and quiescent DRGs;][]{Cim+04,Dad+05,Lab+05,Kri+06,Wuy+07} to over $10^3$ \msunyr\ in sub-mm-selected galaxies (SMGs; e.g. \citealt{Bla+02,Sma+02,Tac+06}), and correspondingly large variations in morphology, stellar populations, excitation properties, and dust content \citep[e.g.][]{Red+05,Pap+06,Kri+07}.
	
	Furthermore, the resolving power of 8-10m class telescopes and of millimeter interferometric arrays reveals velocity gradients within many of these systems \citep{Erb+03,Gen+03}.  By coupling such telescopes with high-resolution integral field spectrographs, we are now able, for the first time, to resolve the dynamic structures and internal processes at work within massive galaxies during their critical stages of evolution \citep[][see also e.g. \citealt{Flo+04,Pue+06,Swi+06}]{For+06,Gen+06,Wri+07,Law+07}.  With surveys of the spatially-resolved kinematics of various \hz\  populations \citep{Bou+07}, we can now begin to understand the forces driving such rapid and intense evolution as well as the role of secular evolution and major mergers in these processes.
	
	Differentiating between systems in ordered rotation and those undergoing major merger events has significant ramifications in understanding the evolution of both the baryons and the underlying dark matter distributions.  Kinematic measurements of a system's baryonic component, combined with basic assumptions about the morphology of the system, enables a detailed probe of the mass and angular momentum of dark matter halos at \ztwo\  and of the interaction between these halos and their baryons \citep{For+06,Bou+07}.  Additionally, the baryons themselves can also constrain formation and evolution scenarios, by probing whether the active star formation seen in these objects is triggered by major mergers \citep[e.g.][]{Hop+06} or by smooth accretion \citep{Bir+07}.  In several well-resolved systems at \ztwo, the young age of the stellar population ($\sim 500-1000$ Myr), when combined with the high star formation rate (up to $\sim 200$ \msunyr), suggests extremely rapid ($< 1$ Gyr) formation \citep{For+06,Gen+06}.  This is most surprising since the dynamics of these systems qualitatively appear to be consistent with no recent major merger events, thus indicating a rapid and intense, but still smooth, mass accretion mechanism. 

	However, a quantitative and definitive understanding of the structure of these \hz\ systems is complicated by the limited spatial resolution attainable at this redshift and the lower signal-to-noise (S/N) associated with these faint objects.  Existing prescriptions for measuring galaxy morphology use surface brightness information and rely solely on characteristics of the broad-band emission (CAS: \citealt{Con03}; Gini/M20: \citealt{Lot+04}; S\'ersic fitting: \citealt[e.g.][]{Rav+04,Cre+06}).  At lower redshifts, these techniques have been shown to effectively distinguish disparate populations over a wide range of resolutions and S/N \citep{Lot+04}.  At $z \gtrsim 2$, however, the situation is more complicated, since optical observations probe the rest-frame ultraviolet morphology, which is strongly affected by extinction and by the light from massive stars in star-forming regions.  Near-infrared observations, which correspond to rest-frame optical emission at $z \sim 1-4$, are a significant improvement, but high-resolution near-infrared imaging is still quite observationally expensive.  To reliably probe the dynamical state of a system, a promising alternative comes from integral-field spectroscopic (IFS) observations, yielding spatially-resolved kinematic information.  While this approach is also still observationally expensive, it has the advantage of directly probing the system's dynamics and total enclosed mass.

	Given the detailed kinematic information available with such observations, a technique complementary to morphological methods can be developed to fully exploit the measured two-dimensional velocity structure.  Existing IFS observations of \ha\ emission in a few cases at \ztwo\ reveal dynamics suggestive of either the ``spider diagram" structure found in local disk galaxies or of the complex structures found in mergers \citep{For+06}.  If the observed internal dynamics of the warm gas do in fact reflect those of the system, then it appears possible to distinguish mergers and non-mergers using current data.  Indeed, simulated observations with current technology have predicted that the kinematic differences between non-merging and merging systems should be qualitatively visible in IFS data \citep{Law+06}.
	
	In this paper, we present a scheme for discerning between merging and non-merging galaxies, based on their emission-line kinematic properties and also on the distribution of the stellar continuum intensity, which, with integral-field observations, can be unambiguously separated from the emission lines.  We use kinemetry \citep{Kra+06} to quantify asymmetries in the velocity and velocity dispersion maps, enabling us to differentiate a system in regular, ordered rotation from one disturbed by the complex dynamics of a major merger, even at the $\sim 4$ kpc spatial resolution typical of seeing-limited (FWHM $\sim 0\farcsec5$) observations of \ztwo\ systems.  In Section 2, we describe this method, and in Section 3, we illustrate the power of our technique with a number of template galaxies, drawn from observations of local systems and from simulations (from \citealt{Dai+06,Col+05,Naa+07}).  Section 4 applies this classification scheme to the well-resolved \ztwo\ systems from the Spectroscopic Imaging survey in the Near-infrared with SINFONI (SINS; see \citealt{For+06,Gen+06,Bou+07}; Cresci et al. in prep).  Section 5 discusses these results and addresses the inherent assumptions in our method, including the use of \ha\ kinematics as a tracer of dynamics at high redshift.  Finally, we summarize our conclusions in Section 6.
	
	Throughout this paper, we compare the two classes of systems that we aim to distinguish: those with recent major merger events (mass ratio $\leq 3:1$) and those without.  For simplicity, we refer to these two classes as ``mergers" and ``disks" respectively, but the obvious caveats of such nomenclature are worth mentioning.  Our analyses focus mainly on the gaseous component in galaxies, which relaxes quickly into a flat, disk-like configuration, e.g. in late-stage mergers \citep[e.g.][]{BarHer96,MihHer96,Tac+99,Naa+06}.  The naming scheme adopted here does not reflect an inability to distinguish late-stage mergers from non-merging systems; in fact, Section 3 illustrates that the gas motions in even late-stage mergers are still sufficiently disturbed that they can be clearly identified as such \citep[see also][]{Bar02,ArrCol03}.  In the context of this paper, these systems are never referred to as ``disks," despite their probable geometry; that term is instead reserved for galaxies that show no sign of a recent major merger.
	
	We assume a $\Lambda$-dominated cosmology with $H_{\rm 0} = 70$ km s$^{-1}$ Mpc$^{-1}$, $\Omega_{\rm m} = 0.3$, and $\Omega_{\rm \Lambda} = 0.7$.  For this cosmology, $1\arcsec$ corresponds to $\approx 8.2$ kpc at $z = 2.2$.
	
\begin{figure*}
\centering
\includegraphics[angle=90,width=18cm]{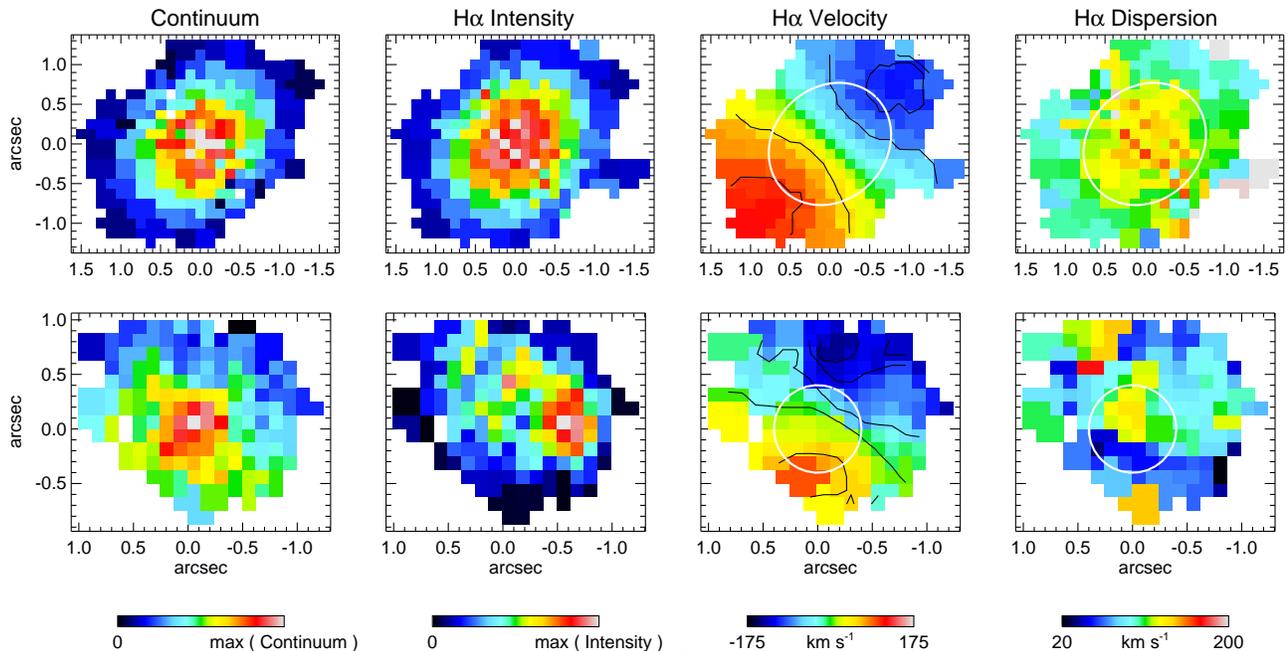}
\caption{{\it From left to right:} Continuum intensity in rest-frame $R$-band, intensity of the \ha\ line emission, and velocity and velocity dispersion of the \ha-emitting component, for a simulated ideal disk ({\it top}) and for the high-redshift galaxy BzK-6004 observed as part of the SINS program ({\it bottom}).   In the ideal disk, the star formation (\ha\ intensity) follows the mass distribution (Continuum), whereas in the observed \ztwo\ system, a significant off-center star-forming region is seen.  Overplotted on the velocity maps are isovelocity contours, whose concave curvature on either side of the rotation axis displays the so-called ``spider diagram" structure characteristic of rotational motion \citep{vdKruAll78}.  Also overplotted on the velocity and dispersion maps are sample ellipses from the expansion with kinemetry.  Along a kinemetry ellipse in the velocity map, the velocity varies as the cosine of the azimuthal angle.  Along an ellipse in the velocity dispersion map, the dispersion is approximately constant with angle.}
\label{figkin}
\end{figure*}

\section{Method}

	To determine whether a particular observed system is a disk or a merger, we use two main criteria: the symmetry of the velocity field of the warm gas, and the symmetry of the velocity dispersion field of the warm gas.  An ideal rotating disk in equilibrium is expected to have an ordered velocity field, described by the so-called ``spider diagram" structure, and a centrally-peaked velocity dispersion field (Figure \ref{figkin}, see also \citealt{vdKruAll78}).
	
	Likewise, such a disk is also expected to have the regular and centrally-peaked continuum distribution characteristic of exponential disks.  Indeed, this feature is the basis for low-redshift morphological classification schemes (e.g. \citealt{Con03,Lot+04,Cre+06}).  However, in our IFS observations, the highest S/N is usually obtained from the \ha\ emission from the warm, star-forming gas, rather than from the underlying stellar continuum whose light is dispersed over many more spectral pixels.  In \citet{For+06}, for example, the emission features have a typical S/N of $\sim 25$, whereas the stellar continuum is often only detected in a small part of a given system, with typical S/N $\leq$ 10 in the brightest region.  We therefore perform only the simplest analysis of the distribution of the stellar component in these systems, using this information to supplement and inform the detailed analysis possible with the high-S/N emission line velocity and velocity dispersion data.

	In our analysis, we do not include constraints based on the intensity distribution of the emission lines.  This tracer of the location and power of star-forming regions is often clumpy and asymmetric in even the most kinematically regular disks (Figure \ref{figkin}; see also \citealt{Dai+06} for local examples) and, consequently, reveals little about the mass distribution and dynamical state of the system.  Rather, it is the kinematics of this gas, and not its spatial distribution, that reflects the dynamical state of the system and therefore forms the basis for our analysis.

\subsection{Quantifying Symmetries with Kinemetry}

\begin{figure*}
  \centering
  \includegraphics[width=0.26\textwidth,angle=90,trim= 20mm 0mm 0mm 0mm]{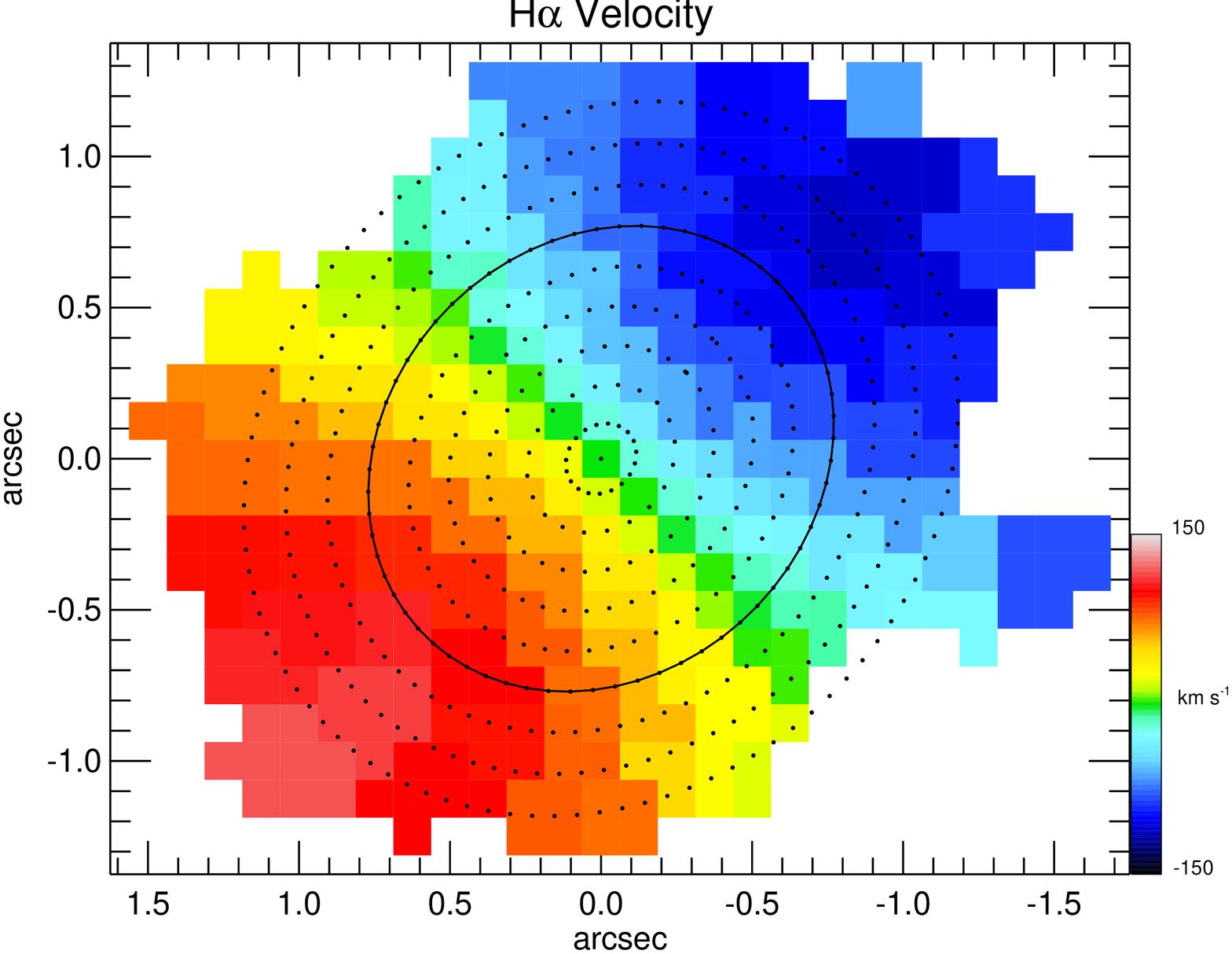}                
  \includegraphics[width=0.3\textwidth]{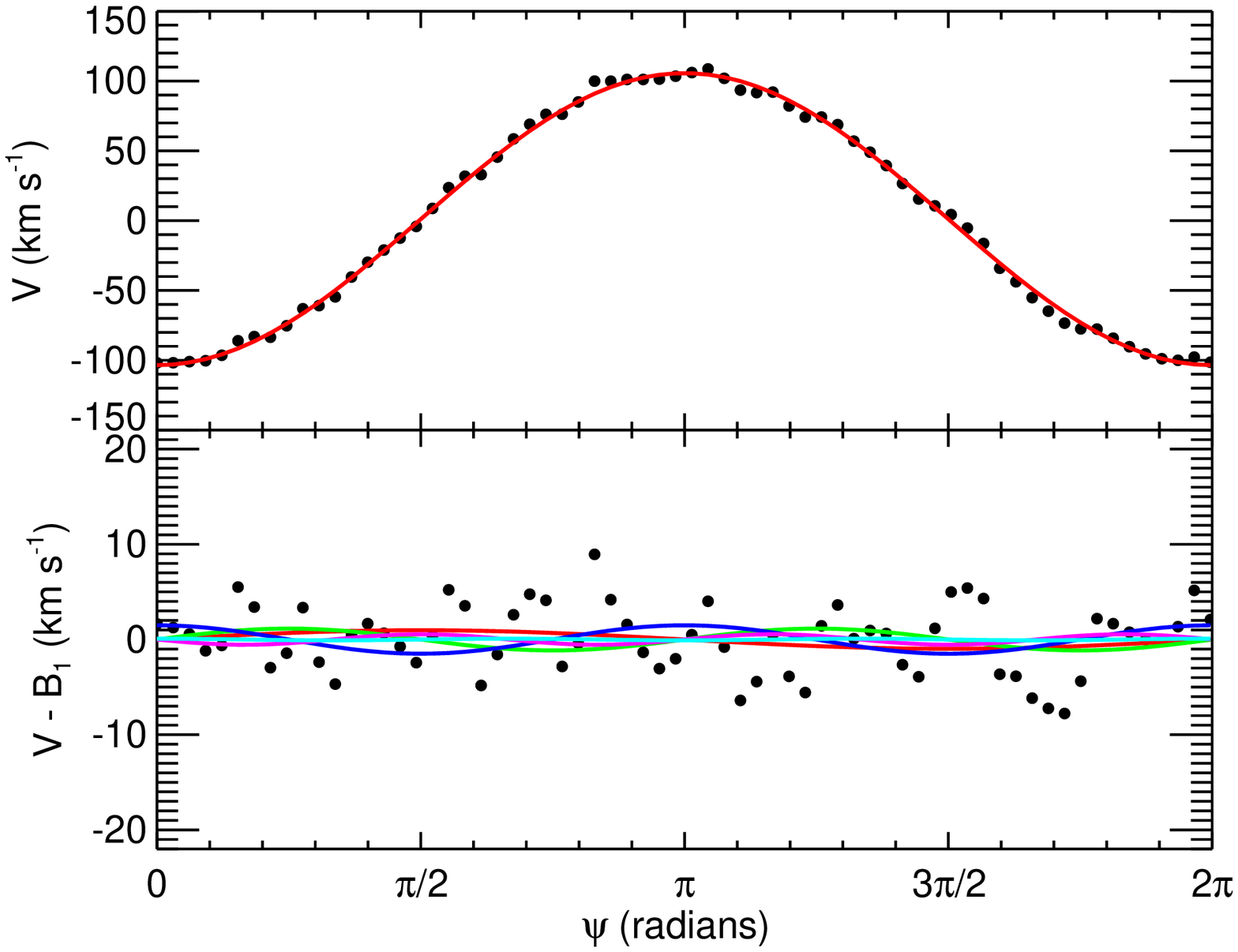}
  \includegraphics[width=0.3\textwidth]{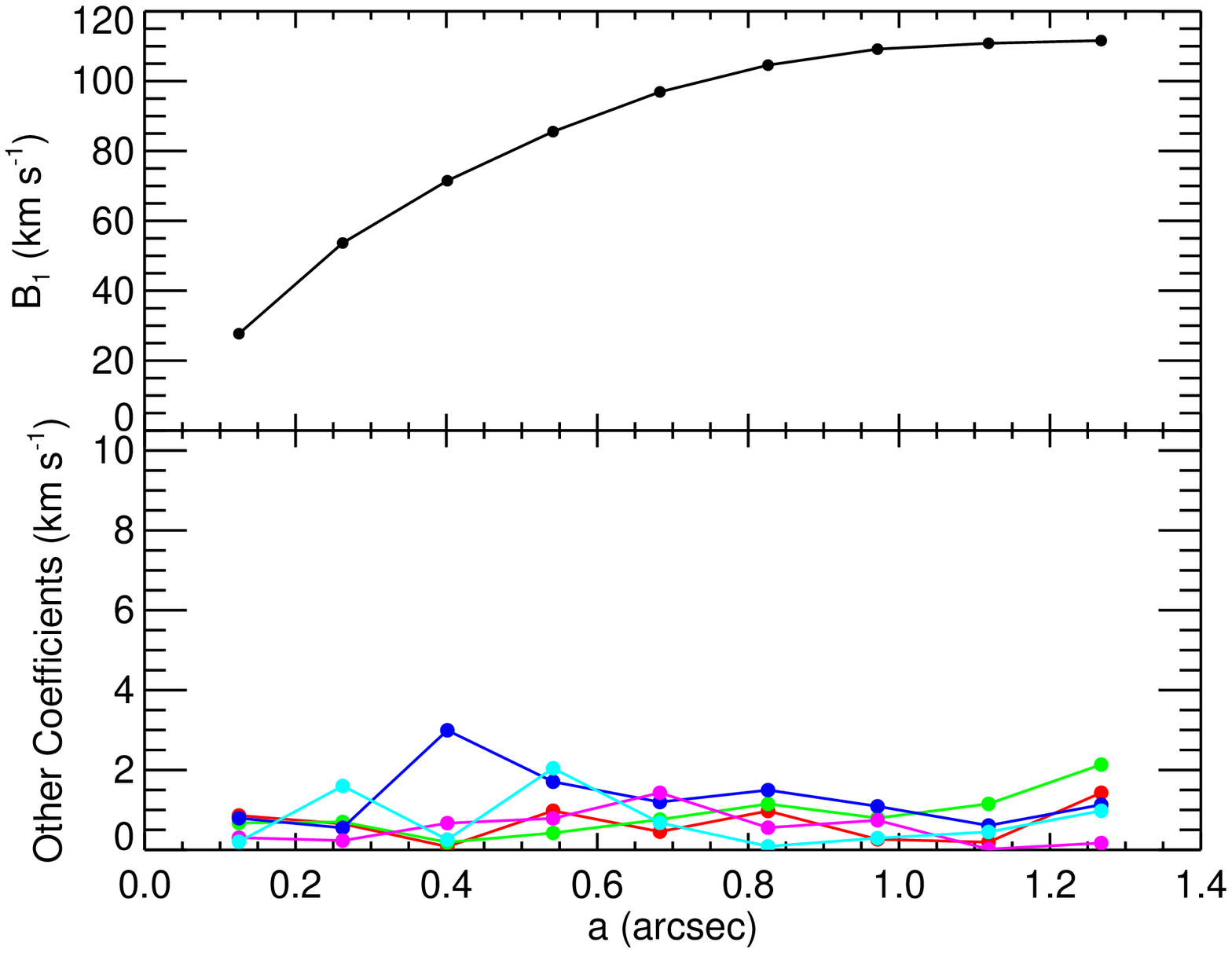}
  \caption{{\it Left:} Velocity field of a toy disk model (see Section \ref{simtemp}), with kinemetry ellipses overlaid.  One ellipse is emphasized with the solid line; the sampling of the rest of the ellipses are shown with the black dots.  {\it Center:} The kinemetry expansion as a function of angle $\psi$ along the solid ellipse.  The top panel shows the measured velocities (black points) and the fit with the $B_1$ coefficient (red); the bottom panel shows the residuals from this fit (black points), and the higher order coefficients measured as a function of $\psi$ ($A_1$ in red, $A_2$ green, $B_2$ blue, $A_3$ magenta, and $B_3$ cyan).  {\it Right:} The kinemetry expansion from the center panel, now shown for all ellipses as a function of semi-major axis length $a$.  The top panel shows $B_1$ as a function of $a$; this is the rotation curve.  The bottom panel shows the strength of the higher order coefficients (same colors as above), all of which are negligible, as would be expected in an ideal disk.  In this toy model, deviations from zero in these coefficients reflect the noise in the velocity field.}
  \label{kintut}
\end{figure*}
	
	The symmetries in kinematic fields can be measured via the kinemetry method developed and described in detail by \citet{Kra+06}.  Briefly, kinemetry is an extension of surface photometry to the higher-order moments of the velocity distribution.  The procedure operates by first describing the data by a series of concentric ellipses of increasing major axis length, as defined by the system center, position angle, and inclination.  The latter two parameters can either be determined a priori and used as inputs or can be measured functions of semi-major axis length as a first step in the kinemetric analysis.  Along each ellipse, the moment as a function of angle is then extracted and decomposed into the Fourier series
	
	\begin{eqnarray}
	K(\psi) = A_0 + A_1 \sin(\psi) + B_1 \cos(\psi)  \nonumber \\
	                   + \ A_2 \sin(2\psi) + B_2 \cos(2\psi) + \ldots  ,
	\end{eqnarray}
	
\noindent	
where the radial dependence of all $A_n$'s and $B_n$'s is implicit, since the above expression is for a single kinemetry ellipse.  Here, $\psi$ is the azimuthal angle in the plane of the galaxy, measured from the major axis; points along the ellipse are sampled uniformly in $\psi$ and are therefore equidistant if the ellipse is projected on to a circle.  The series can be presented as a function of semi-major axis length $a$ and in a more compact way,

	\begin{equation}
	K(a,\psi) = A_0(a) + \sum_{n=1}^N k_n(a) \cos [n(\psi - \phi_n(r))] ,
	\end{equation}

\noindent
with the amplitude and phase coefficients ($k_n, \phi_n$) defined as

	\begin{equation}
	k_n = \sqrt{A_n^2 + B_n^2} \ \ \ \ \ \ \ {\rm and} \ \ \ \ \ \ \ \phi_n = \arctan \left( \frac{A_n}{B_n} \right) .
	\end{equation}

\noindent
The full moment map (i.e. velocity or velocity dispersion) can thus be described by the geometry of the rings and the amplitude of the coefficients $k_n$ (or equivalently, $A_n$ and $B_n$) of the Fourier terms as a function of semi-major axis length $a$ (Figure \ref{kintut}).

	To be more specific, the velocity field in an ideal rotating disk is expected to be dominated by the $\cos \psi$ term, since the velocity peaks at the galaxy major axis ($\psi \equiv 0$) and goes to zero along the minor axis ($\psi = \frac{\pi}{2}$, Figure \ref{kintut}).  The power in the $B_1$ term therefore represents the circular velocity at each ring $a$, while power in the other coefficients (normalized to the rotation curve, $B_1$) represents deviations from circular motion.  In local galaxies with very high S/N observations, for example, various coefficients have been shown to identify bars/radial inflow, lopsidedness/warps, multiple components, and spiral structure \citep{Sch+97,Won+04,Kra+06}.  In lower S/N data with sparser spatial sampling, these higher-order coefficients will also be affected by the rapid variations along each ring induced by the noise.
	
	The velocity dispersion field, on the other hand, is an even moment of the velocity distribution and, as such, its kinemetric analysis is identical to traditional surface photometry.  In an ideal rotating disk, the velocity dispersion will be constant along each ring and will decrease between rings of increasing semi-major axis length (Figure~\ref{figkin}).  For this moment of the velocity distribution, the power in the $A_0$ term as a function of semi-major axis length $a$ will represent the velocity dispersion profile, and all azimuthally varying terms (higher-order kinemetry coefficients) will be zero.  In analogy to the case with the velocity field, non-zero $A_n$'s and $B_n$'s can thus identify expected deviations from symmetry, in the form of lopsidedness and boxy/disky iso-velocity dispersion contours, but are also susceptible to variations caused by noise.

\subsection{Kinemetry of High-Redshift Systems}
\label{KinHighZ}

\begin{figure*}
  \centering
  \includegraphics[width=0.26\textwidth,angle=90,trim= 20mm 0mm 0mm 0mm]{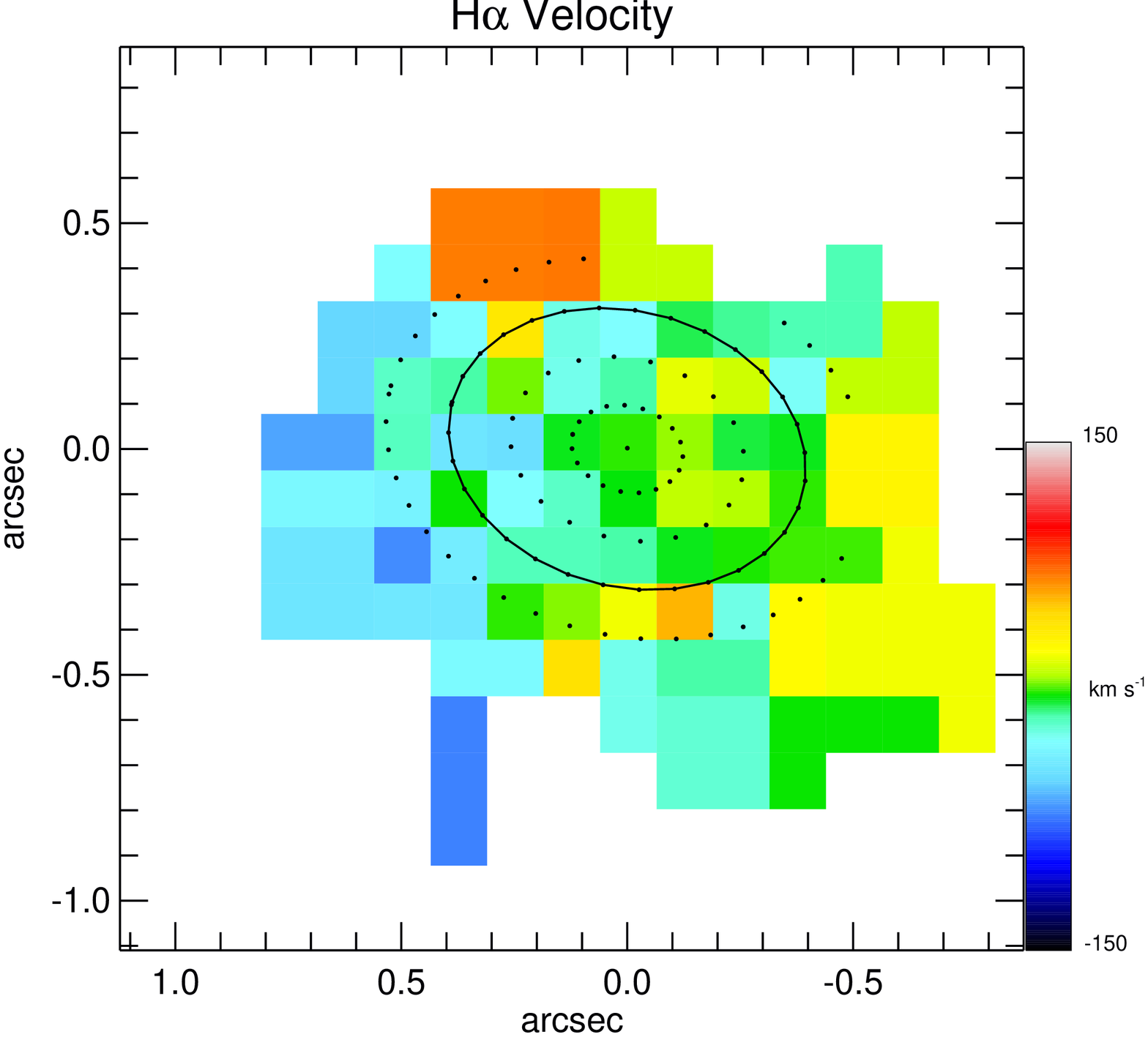}                
  \includegraphics[width=0.3\textwidth]{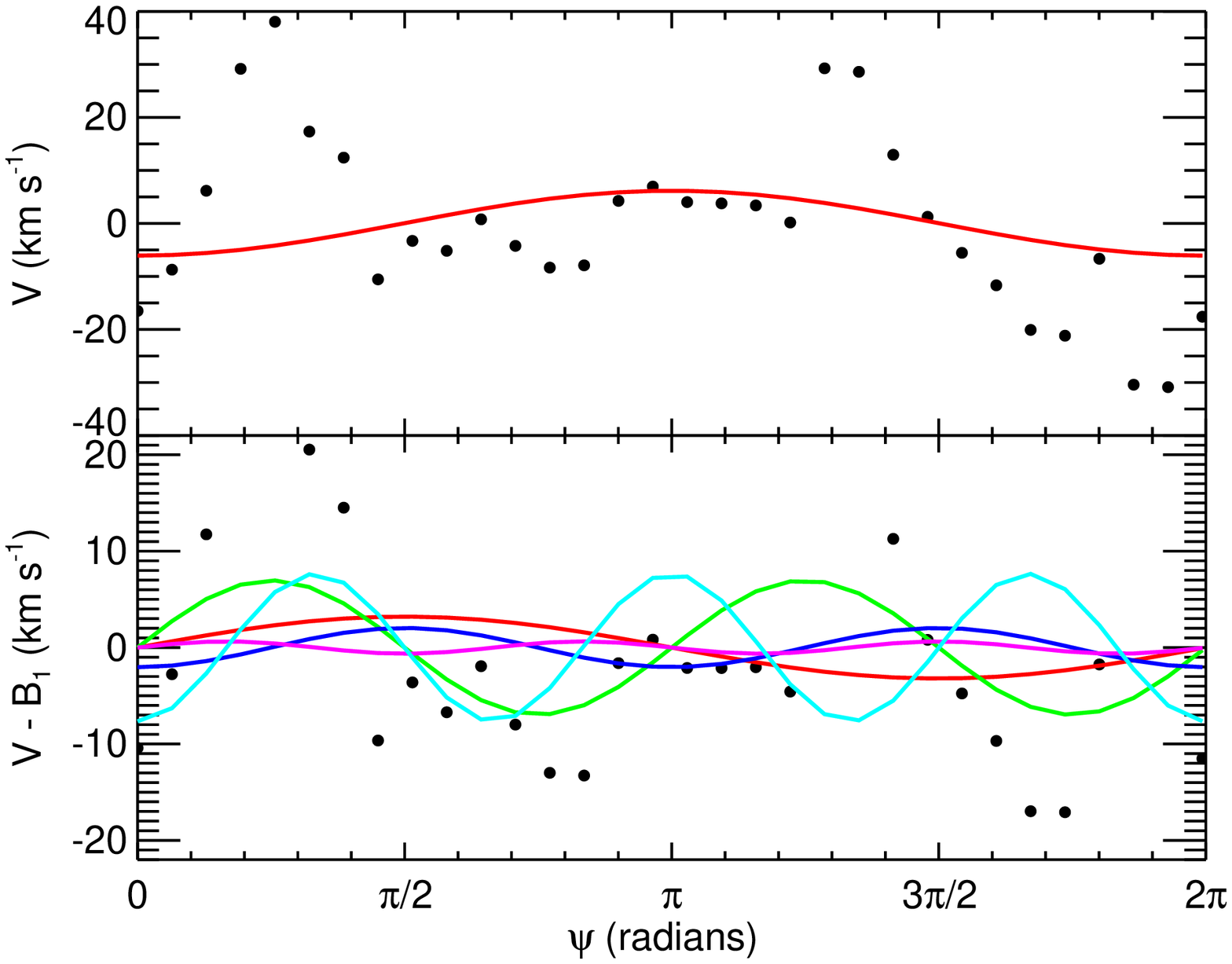}
  \includegraphics[width=0.3\textwidth]{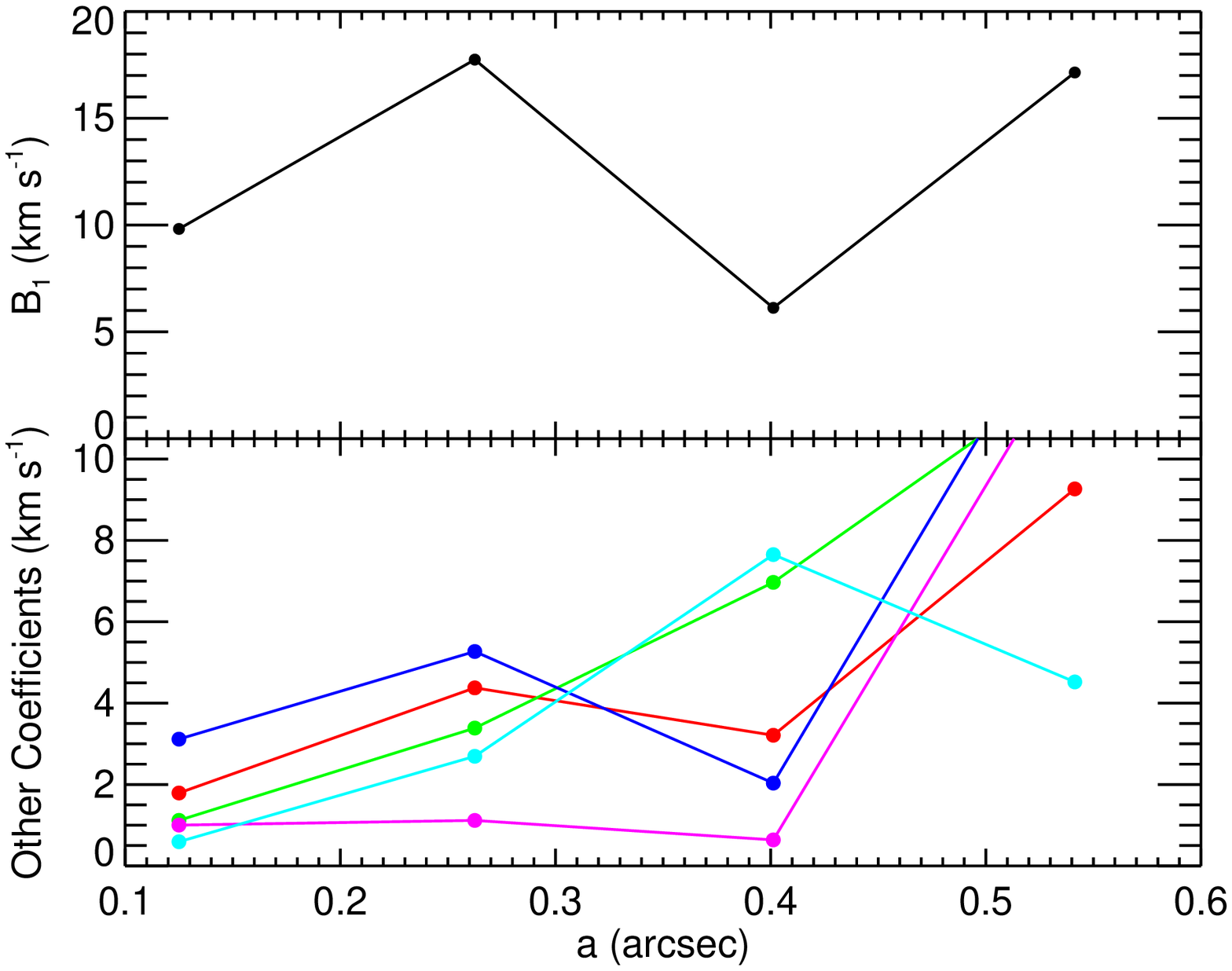}
  \caption{Same as Figure \ref{kintut}, but for an observed merger (IRAS 15206+3342) simulated at \ztwo\ (see Section \ref{obstemp}).  The irregular velocity field ({\it left}) is not well fit by a $\cos \psi$ term and therefore has significant power in all higher order coefficients ({\it center}).  The combined effects of the lack of power in $B_1$ from the poor fit to $\cos \psi$ and the power in all other coefficients ({\it right}) will produce normalized coefficients $k_n / B_1$ that are much higher than those of a disk (Figure \ref{kintut}).}
  \label{kintutmerg}
\end{figure*}

	Kinemetry was originally designed by \citet{Kra+06} for use with very high S/N ($> 100$) stellar kinematic data, as found in observations of bright galaxies in the local Universe.  It has also been used to analyze simulated mergers and merger remnants by \citet{Jes+07} and \citet{Kro+07}.  To apply this method to the much lower S/N emission line data obtained at \ztwo, the breadth of the analysis must be somewhat restricted.  We therefore employ kinemetry in a more limited capacity; rather than using kinemetry to measure and interpret subtle dynamical features of a velocity field, we instead use it to determine the strength of deviations of the observed velocity and dispersion fields from the ideal rotating disk case.  This is identical to assuming that any deviations from the ideal case that might occur in a disk (e.g. lopsidedness, warps, spiral structure) induce less power in the higher Fourier coefficients than those caused by the noise and much less than those that occur in a disturbed, merging system (compare Figures \ref{kintut} and \ref{kintutmerg}).
	
	The first step in the analysis is locating the center of the system, around which the kinemetry ellipses are constructed.  The robustness of this step is critical to the result, since \citet{Kra+06} show that an incorrect assumed center induces artificial power in the derived kinemetry coefficients.  The primary result of a miscentering are elevated $A_2$ and $B_2$, but other coefficients ($A_0$, $A_1$, $A_3$, and $B_3$) are affected as well.  It is therefore important that we use a robust definition of the system center, such that the center of an ideal disk is accurately recovered.
	
	This is a non-trivial task in the clumpy and irregular \ha\ intensity distribution, the spatial distribution of which corresponds only to regions with enhanced star formation rates, is influenced by extinction, and therefore does not necessarily reflect the intrinsic mass distribution in the system (Figure \ref{figkin}).  It is also not straightforward to robustly derive the location of the galaxy center from the velocity and velocity dispersion maps themselves (Krajnovi\'c, private communication).  We therefore take advantage of the continuum distribution, which can be detected in the integral field data at a somewhat lower S/N underneath the emission lines.  In general, the continuum surface brightness distribution is distinct from that of the emission line intensity (Figure \ref{figkin}, see also examples in local spiral galaxies in \citealt{Dai+06}) and has sufficient S/N to differentiate regions of strong continuum emission from those of weaker continuum emission.
		
	At \ztwo, the detected continuum in near-infrared observations corresponds roughly to rest-frame $R$-band.  Observations at this wavelength therefore typically provide an accurate probe of the stellar distribution, although the effects of extinction have been shown to be significant in some local dusty mergers \citep{ArrCol03}.  While it is difficult to know precisely how large a role extinction plays in the observed $R$-band continuum distributions of \hz\ observations, some constraint can be provided from visual analysis of the data themselves.  Several systems observed at \ztwo\ show a compact bright, central region of continuum emission, which remains single-peaked even at $<0\farcsec5$ (4 kpc) resolution (e.g. Figure \ref{figkin}), suggesting that this is an unobscured measure of the stellar (and mass) distribution in these systems.

	As such, the continuum surface brightness distribution can be used to locate the center of a system.  The natural choice for a galaxy's center is the position of the peak in the continuum intensity, which in an ideal disk also corresponds to the centers of the kinematic fields (Figure \ref{figkin}); however, in noisy data, this must be carefully defined.  We identify the center of the continuum distribution by using the data in the brightest 25\% of the pixels and finding the continuum-intensity-weighted average of their positions.  This definition of the system center thus has the added benefit that the center of an early-stage major merging system (with two distinct nuclei) will be directly between the two components and not skewed towards only one of the two mass concentrations.

	In high S/N data, the location of the center is the only parameter that must be determined prior to running a kinemetric analysis; the relevant geometric parameters (position angle and inclination) are derived during the kinemetry analysis.  However, the lower S/N and coarse spatial sampling of the \ztwo\ data (see Figure \ref{figkin}) renders kinemetry's radius-by-radius solution for position angle and inclination rather unstable.  For these systems, a much more robust solution for these parameters is found by considering the entire velocity field at once and solving for a global position angle and inclination.  In nearby spiral galaxies, these quantities are observed to vary slowly throughout the system \citep{Won+04}, thus making global values for position angle and inclination decent approximations.

	To determine the position angle and inclination of a system, we use the known effects of errors in these parameters on the kinemetry coefficients $A_n$ and $B_n$.  \citet{Kra+06} demonstrate that, in the kinemetric expansion of a velocity field, a slightly incorrect assumed position angle generates excess power in the $A_1$, $A_3$, and $B_3$ coefficients, while a slightly incorrect assumed inclination affects primarily the $B_3$ term.  However, large errors in these values can produce significant power in other coefficients as well.  We therefore use the measured power in all coefficients to derive these parameters.
	
	We solve first for the global position angle by stepping through all possible values, in increments of 3$^{\rm o}$, and performing a kinemetric expansion of the velocity field at each assumed value.  Since this procedure focuses on locating the angle of steepest velocity gradient, the axial ratio of the ellipses (i.e. the inclination of the system) does not affect the results and is therefore held constant at unity, forcing the kinemetry ellipses to be circles.  The goodness-of-fit of the assumed position angle is determined as the sum of the squared residuals between the ``circular" velocity field (the 2D image reconstructed from only the $B_1$ term) and the observed velocity field.  These residuals reflect the combined powers in the higher coefficients, which appear as asymmetries in the velocity field.  The curve of the goodness-of-fit as a function of position angle is then smoothed by 3 data points  (=9$^{\rm o}$) to eliminate spurious results induced by the noise in the data, and the best-fit position angle is identified as the position angle that minimizes the smoothed curve.
	
	Assuming this position angle, we then solve for the global inclination of the system, again using kinemetric expansion of the velocity field.  We test $\sim 50$ values for the inclination, evenly spaced between axial ratios of 0.1 and 1.0.  At every assumed ellipticity, kinemetry is performed, with the position angle held constant at the previously-determined best-fit value and the inclination held constant at the assumed value.  The goodness-of-fit is determined as above, using the residuals between the ``circular" velocity field and the actual measured values.  The minimization of this method therefore finds the inclination that produces the smallest deviations from the circular term.  However, since ellipses of similar axial ratios to that measured from the morphology will maximize the kinemetry coverage of the velocity field, and thus minimize the residuals measured near the edges of the system, this process induces a slight bias towards an inclination similar to that of the morphology.  In practice, this bias partially mediates the adverse affects of beam smearing, in which the circular beam increases the opening angle of the isovelocity contours of the inclined velocity field.  As with the position angle determination, we smooth the goodness-of-fit curve by 3 data points to reduced the impact of the noise in the velocity field, and we find the inclination that minimizes the smoothed residuals curve.
	
	With the position angle and inclination held constant at these best-fit values, we can then perform a robust kinemetric analysis of the low-S/N \hz\ velocity and velocity dispersion fields.  As in \citet{Kra+06}, kinemetry is carried out to the fifth order terms, $A_5$ and $B_5$, which include most of the physical deviations from symmetry in the field but are not overly affected by rapid point-to-point variations induced by noise.  Since the kinemetry coefficients are by definition orthogonal, this somewhat arbitrary choice of where to stop the expansion will not alter the measured values of the kinemetry coefficients.

\subsection{Criteria for Differentiating Disks and Mergers}
\label{Criteria}

	The main component of the analysis thus consists of running kinemetry, with pre-defined system center, global position angle, and global inclination, on both the velocity and velocity dispersion field of a galaxy.  Since the quantity of interest is the deviations of these fields from those of ideal disks, we find the average deviation, defined as the average of the kinemetry coefficients that would be identically zero in a noiseless ideal disk.

\begin{figure*}
\centering
\includegraphics[width=11cm,angle=90]{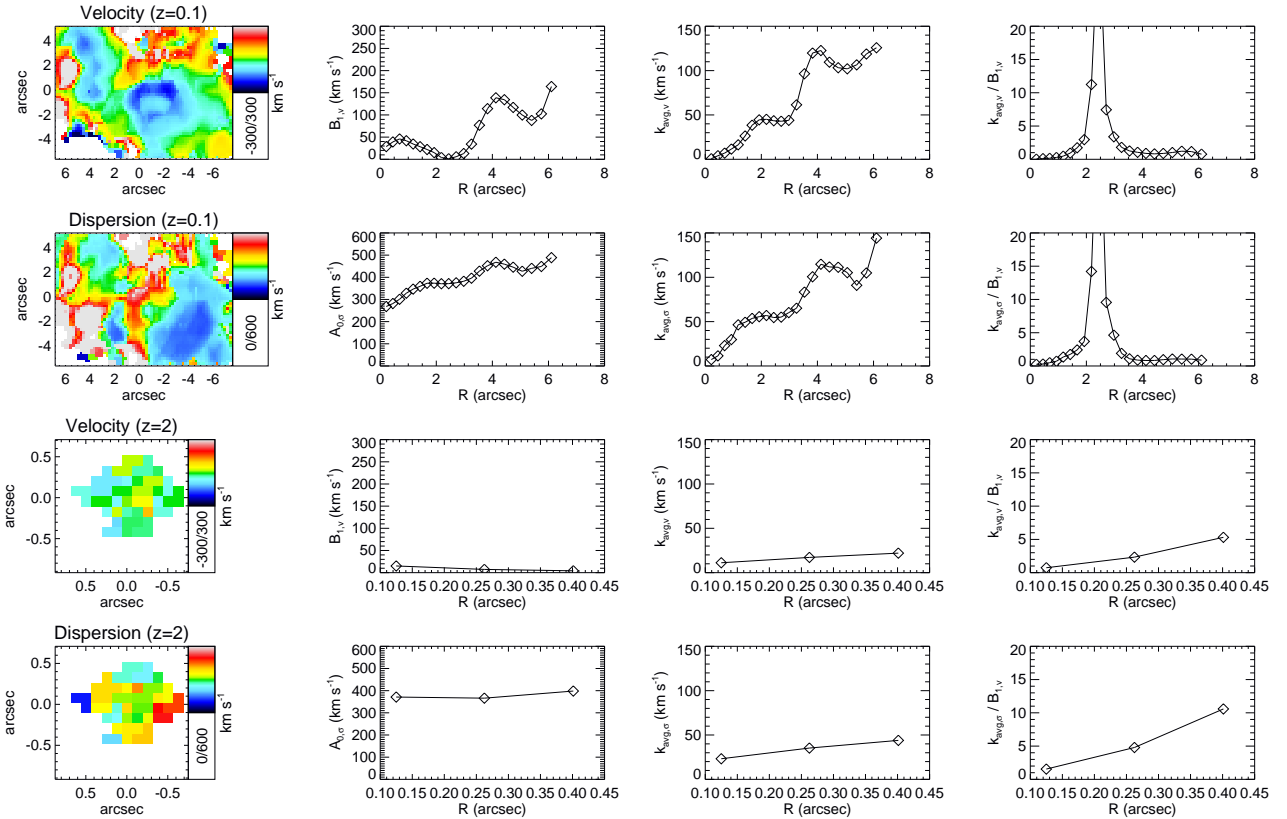}
\caption{{\it From left to right:} The map of a given velocity moment, followed by the circular term of its kinemetry expansion, the higher order coefficients of the expansion $k_{avg}$, and the derived asymmetry parameter, all as functions of radius from the system center.  This is shown, from top to bottom, for the velocity and velocity dispersion of the $z \sim 0.1$ ULIRG Mrk 273, and for the velocity and velocity dispersion of the simulated version of this system at \ztwo\ (see text for details).  In the velocity maps, the diminished spatial extent of the high-redshift data corresponds to less resolution of both the rotation curve and the higher order coefficients.  On the other hand, the velocity dispersion maps show negligible change in the dispersion profile with changing spatial coverage, while the higher order terms, as with those from the velocity field, lose dynamic range.}
\label{figdynrange}
\end{figure*}

	For a velocity field in an ideal, rotating disk, the only non-zero kinemetry coefficient ought to be $B_1$, which we denote here as $B_{1,v}$ to indicate that this is the kinemetry result from the velocity field (Figure \ref{kintut}).  Information about additional asymmetries are therefore contained in the higher-order terms, $k_{2,v}$ -- $k_{5,v}$.  Although \citet{Kra+06} use only odd kinemetry terms (e.g. $k_5$) to describe odd moments of the velocity distribution (velocity field), we include both the even and the odd terms in the kinemetry expansion here, since mergers produce extremely disturbed velocity fields, with power in all kinemetry coefficients (Figure \ref{kintutmerg}).  In principle, one could also include the $A_{1,v}$ term as a measure of asymmetry, since this term represents any velocity gradients (inflows/outflows) found along the minor axis.  In practice, however, we choose not to use the presence of a non-zero $A_{1,v}$ coefficient in the definition of non-ideal disks, since significant radial flows may be the result of such phenomena as outflows related to AGN/starburst winds or inflows induced by bar instabilities (see discussion in Section \ref{discussha}), and so do not provide information as to whether a system is undergoing a merger.  We therefore compute the average $k_{avg,v}$ = $(k_{2,v} + k_{3,v} + k_{4,v} + k_{5,v}) / 4$ as a measure of the non-ideal components of a system.
	
	This average deviation $k_{avg,v}$ is normalized to the rotation curve $B_{1,v}$ in order to assess the relative level of deviation, following the prescription of \citet{Kra+06}.  This normalization has the added benefit of accounting for the loss of dynamic range in $k_{avg,v}$ at high redshifts, where the dim outer regions of a galaxy are too faint to be detected.  In merging systems, greater radial coverage will correspond to greater detectable deviations from the ideal disk geometry, and it will also correspond to a greater dynamic range in the velocity gradient $B_1$ (Figure \ref{figdynrange}).  By normalizing $k_{avg,v}$ to $B_{1,v}$, then, a system will roughly retain the same amount of detectable asymmetry regardless of the radial extent of the data.  We therefore define the asymmetry, or level of deviation from an ideal disk, in the velocity field to be
	
	\begin{equation}
	v_{asym} = \left< \frac{k_{avg,v}}{B_{1,v}} \right>_r ,
	\end{equation}
	
\noindent
where the average is over all radii (relative to the continuum peak) of the kinemetry ellipses.

	For the velocity dispersion field, the only non-zero kinemetry coefficient in an ideal, rotating disk is $A_{0,\sigma}$, which quantifies the velocity dispersion profile.  All higher-order coefficients $k_{1,\sigma}$ -- $k_{5,\sigma}$ therefore measure any asymmetries in the field.  For this moment of the velocity distribution, then, $k_{avg,\sigma}$ = $(k_{1,\sigma} + k_{2,\sigma} + k_{3,\sigma} + k_{4,\sigma} + k_{5,\sigma})/5$ will contain information about the deviations from the ideal case.
	
	\citet{Kra+06} note that, when looking at the velocity dispersion of the stellar component, the appropriate normalization for this even moment is the $A_{0,\sigma}$ coefficient.  In the stellar case, the $A_{0,\sigma}$ term traces the mass of the system; however, this is not generally true of the velocity dispersion of a gas component, which can also be affected by shocks, especially in the violent environment of a major merger.  In the kinemetry of gas kinematics, if there is significant rotation, the potential is often more reliably probed by the rotation curve, $B_{1,v}$.  Furthermore, in Figure \ref{figdynrange}, the weakness of using a normalization to $A_{0,\sigma}$ in the presence of varying radial coverage is clear.  As with the velocity field deviations $k_{avg,v}$, the velocity dispersion deviations $k_{avg,\sigma}$ become stronger with radius; a loss in radial coverage therefore directly corresponds to a loss in dynamic range in the asymmetries.  In contrast, the value of the $A_{0,\sigma}$ circular term remains roughly constant, even when observed with the broader PSF of \ztwo\ observations.  The value of $k_{avg,\sigma}/A_{0,\sigma}$ consequently decreases significantly for a given system when observations are less sensitive to the outer regions.  A more appropriate normalization is the rotation curve of the velocity field, $B_{1,v}$, which is both a more reliable measure of the system's mass and responsive to the loss of dynamic range with decreased sensitivity.  We therefore define the asymmetry in the velocity dispersion field as
		
	\begin{equation}
	\sigma_{asym} = \left< \frac{k_{avg,\sigma}}{B_{1,v}} \right>_r ,
	\end{equation}
	
\noindent
where the average over all radii is unaffected by the combination of velocity and velocity dispersion kinemetric coefficients, since the kinemetry ellipses for the two maps are identical by the construction described in Section \ref{KinHighZ}.

\section{Application to Template Galaxies}
\label{LowZ}

	To assess the capabilities of these criteria, we draw on a sample of observed low-redshift disks and mergers, which are then ``observed" as if they were at \ztwo, as well as on a sample of synthetic \hz\ systems, created from simulations of varying complexity.  This set of template galaxies spans a large range in morphology, merging history, nuclear activity, and star-formation rate.

\subsection{Observed Systems}
\label{obstemp}

	Our template local galaxies are drawn from two samples: the SINGS spiral galaxy survey ($z \sim 0$; \citealt{Ken+03}), as observed in \ha\ by \citet{Her+05,Dai+06,Che+06}, and the \ha\ observations of \lz\ ultraluminous infrared galaxies (ULIRGs, $z \sim 0.1$) of \citet{Col+05}.  Together, these data sets provide high-quality local observations of disks and mergers, respectively.
	
	The SINGS galaxies were observed in \ha\ emission with the FANTOMM Fabry-Perot scanning interferometer \citep{Her+03}.  The template galaxies used here were all observed at the 1.6m telescope of the Observatoire du mont M\'egantic in Qu\'ebec, Canada \citep{Her+05,Dai+06,Che+06}.  The pixel scale of the velocity maps is 1\farcsec61, and the spectral resolution $R$=12200-20000.  From this sample, we use six spiral galaxies (NGC~925, NGC~3198, NGC~4321, NGC~4579, NGC~4725, and NGC~7331), which together span a range of Hubble types, nuclear activity, star formation rates, and distribution of star-forming regions throughout the system.

	The ULIRG merger systems were observed in \ha\ emission and optical continuum with the INTEGRAL fiber-fed integral field system \citep{Arr+98}, mounted on the 4.2m William Hershel Telescope of the Roque de Los Muchachos Observatory of La Palma, Spain.  These data have resolution elements (fibers) of 0\farcsec9 diameter and spectral resolution $R$=1500 \citep{Col+05}.  From this sample, we use eight mergers (Mrk~273, Arp~220, IRAS~08572+3915, IRAS~12112+0305, IRAS~14348-1447, IRAS~15206+3342, IRAS~15250+3609, and IRAS~17208-0014), which, in analogy with the spiral sample, span a range of merger stage, nuclear activity, star formation rates, and distribution of star-forming regions throughout the system.

	For both the disk and merger template galaxies, we simulate observations of these systems as if they were at \ztwo\ and were observed with the SINFONI integral field unit \citep{Eis+03,Bon+04}, on the Very Large Telescope (VLT) at Cerro Paranal, Chile.  To artificially ``redshift" the template galaxies, we convolve the datacubes to the mean seeing of such observations (FWHM of 0\farcsec5 $\simeq 4$ kpc; \citealt{For+06} achieved this beam size as typical seeing, while \citealt{Gen+06} used adaptive optics to reach a seeing FWHM of 0\farcsec15) and to the spectral resolution of the instrument (75 \kms\ at $2.2 \mu$m), interpolate the data onto the larger pixel scale (spatial scale of 0\farcsec125 $\simeq 1$ kpc; velocity scale of 33 \kms), account for the cosmological surface brightness dimming, and add Gaussian noise such that the S/N of the resulting datacubes is comparable to that of VLT/SINFONI observations.  The \ha\ kinematics are extracted from these datacubes using the same technique as that of \citet{For+06} and \citet{Bou+07} on actual SINFONI observations and are shown in Appendix Figures \ref{fig:spirz2} and \ref{fig:mergz2}.

	The resulting sample of observed galaxies thus totals 14; six of local disks and and eight of local mergers, all ``redshifted" to \ztwo.

\subsection{Simulated Systems}
\label{simtemp}

	Because this naive ``redshifting" of local systems cannot truly emulate the proto-galaxy population at \ztwo, in which such phenomena as massive ($10^{8-9}$ \msun) clumps of star formation affect the structure and kinematics of the systems \citep{BouElmElm07}, we supplement our template galaxies with synthetic systems of various complexity.  For this, we use both toy disk models, in which we have a complete understanding of all aspects of the data, as well as the detailed hydrodynamic cosmological simulations of \citet{Naa+07}, from which we can ``observe" synthetic systems at \ztwo.

	For the toy disk models, we use the modeling routines described in \citet{For+06}, which generate simple models of azimuthally symmetric rotating disks, parametrized by mass, inclination, scale length, scale height, and isotropic velocity dispersion.  Appropriate pixel sizes, beam smearing, spectral resolution, and noise levels are also included in the construction of the toy datacube.  We improve on this model by allowing both an azimuthally-symmetric coupled mass and light distribution, and an overlaid light-only distribution (not azimuthally symmetric).  The former therefore represents the stellar component of the system, while the latter produces very bright regions corresponding to areas of increased star formation.  This addition allows us to generate models that more realistically simulate the clumpy and irregular \ha\ intensity distribution of systems at \ztwo.  We create five models with this technique, all of which have a centrally-peaked and azimuthally-symmetric mass distribution.  We vary their light distribution (star-forming regions) as follows: one galaxy is nearly edge-on and is azimuthally symmetric around a centrally-peaked light distribution, one is nearly edge-on with a varying light distribution, one is of intermediate inclination and is azimuthally symmetric around a centrally-peaked light distribution (shown in Figure \ref{figkin}), one is of intermediate inclination with a light distribution much more extended on one side of the galaxy, and the last is of intermediate inclination with the light distribution illuminating only a single side of the galaxy.  From these models, we extract the emission line kinematics and continuum intensity by fitting a Gaussian with a constant offset to each spectrum (see Appendix Figure \ref{fig:toy} for the models and their derived kinematics).

\begin{figure*}
\centering
  \includegraphics[width=0.45\textwidth]{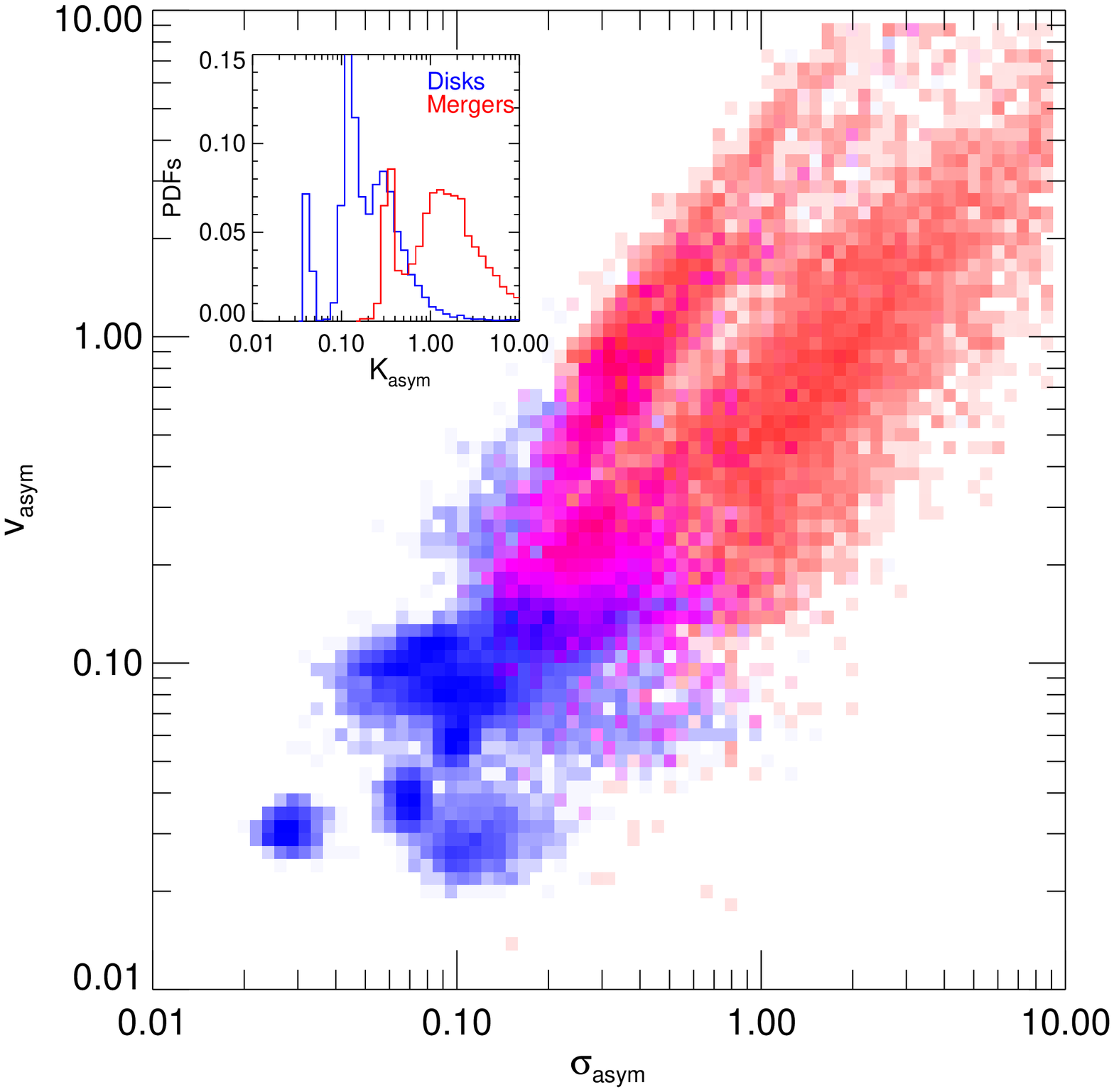}       
  \includegraphics[width=0.45\textwidth]{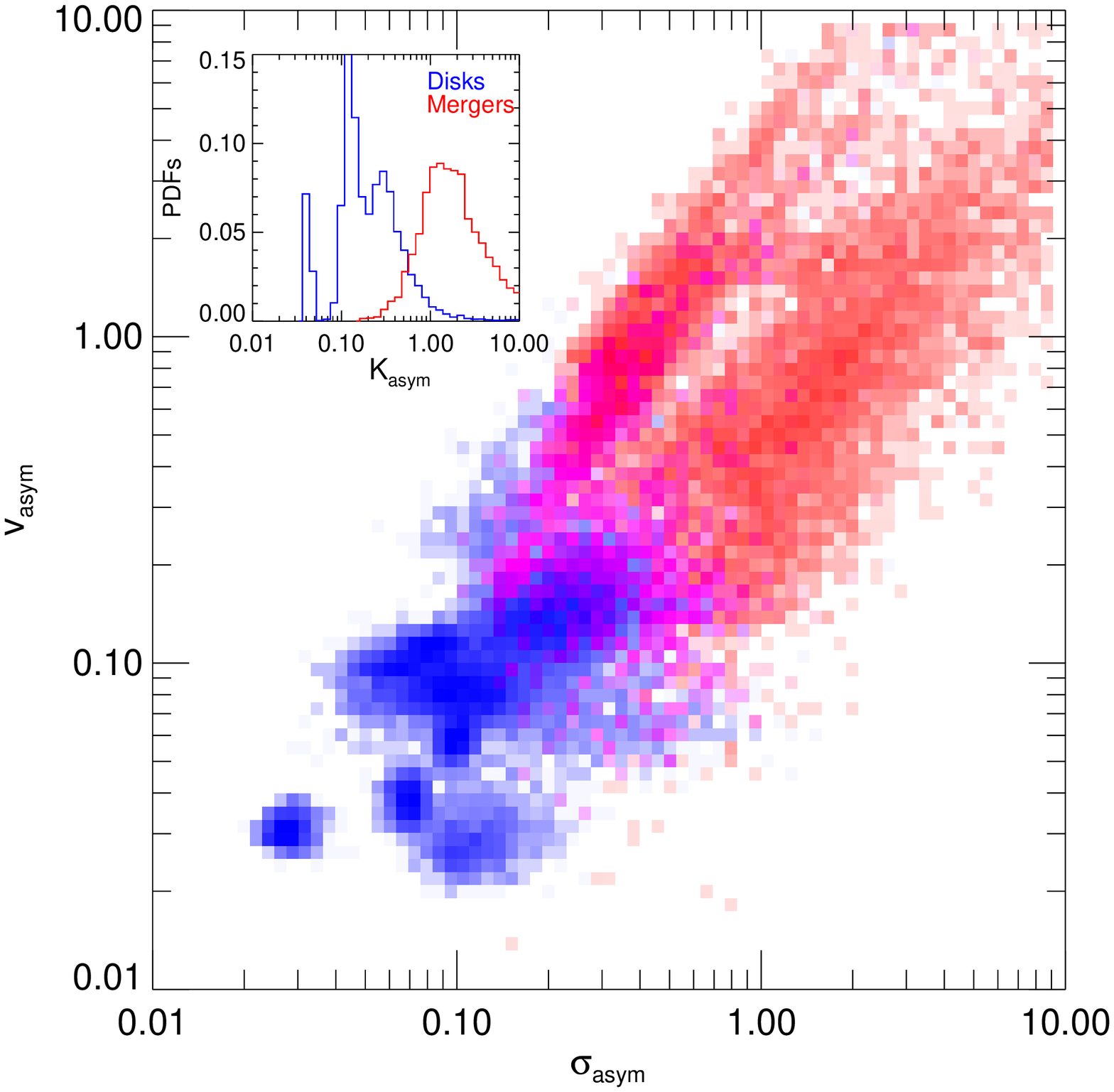}         
  \caption{Asymmetry measure of the velocity and velocity dispersion fields for ({\it left}) all of the template galaxies and ({\it right}) all of the template galaxies except the obviously misclassified ULIRG, IRAS 12112+0305.  The probability distributions in this space are shown with shading for the template disks (blue) and mergers (red), as derived from the Monte Carlo realizations.  Inset are the PDFs for the total kinematic asymmetry (\kasym) for disk and mergers.  The empirical delineation of \kasym = 0.5 cleanly separates the two classes, as is especially visible in the right panel.}
\label{figtemp}
\end{figure*}

	Our cosmologically-simulated models are drawn from \citet{Naa+07}.  The synthetic \ztwo\ systems come from high resolution ($10^6$ gas particles and $10^6$ dark matter particles per halo) SPH simulations of 8 halos with a variety of mass accretion histories.  Based on these mass accretion histories, we select time snapshots in which the halos have been evolving without a major merger in recent history or in the near future as well as snapshots in which the halos are visibly in the process of a major merger.  For the latter group, this selection criteria essentially require the system to have a double nucleus, separated by $\lesssim 5$ kpc.  From the 8 halos, we find 6 snapshots between $z = 1.8$ and $z = 2.8$ in which the systems are unambiguous disks and 4 snapshots between $z = 1.6$ and $z = 3.0$ in which the systems are unambiguous mergers.  These 10 snapshots of the model galaxies are ``observed" by converting the star-formation rate to \ha\ emission using the conversion factor from \citet{Ken98}, accounting for cosmological surface brightness dimming, binning to SINFONI-size pixels, convolving the data to the appropriate spatial and spectral resolutions, and adjusting the (Gaussian) noise level such that the S/N in \ha\ is comparable to that of the SINS observations.  The continuum intensity for each system is ``observed" through a similar process, using the stellar mass and converting to $R$-band luminosity with an assumed $M/L_R = 1$, typical for a Kroupa IMF in star-forming systems at this redshift \citep{Fon+04}.  We then rotate the halo to a random inclination, as given by a $\sin(i)$ probability distribution function, and extract the \ha\ kinematics from the resulting datacube (see Appendix Figures \ref{fig:cd} and \ref{fig:cm}).

	The resulting sample of model galaxies totals 15; five toy disks, six cosmologically simulated disks at $z = 1.8-2.8$, and four cosmologically simulated mergers at $z = 1.6-3.0$.

\subsection{Classification}
\label{ClassLowZ}

	We test the criteria described in Section \ref{Criteria} on the 29 template galaxies to determine how well we can differentiate disks and mergers based on warm gas kinematics.  For each system, we perform the analysis of Sections \ref{KinHighZ} and \ref{Criteria} to measure \vasym\ and \sasym\ for these templates.  (The results of the kinemetric analysis are summarized in Appendix Table \ref{tabkin} and shown in Appendix Figures \ref{fig:spirz2} through \ref{fig:cm}.)  Since this method does not lend itself to straightforward error propagation, we use Monte Carlo realizations of the noise in the data to measure the probability distribution functions (PDFs) of the asymmetries in these systems.
	
	For each template system, the Monte Carlo realizations consist of creating 1000 different realizations of the moment maps - the continuum intensity, and the emission line intensity, velocity, and velocity dispersion - based on their corresponding error maps.  These error maps correspond to the measurement errors of the velocity moments, as derived when fitting the kinematics from the datacubes.  For each moment map, we perturb the observed data points by randomizing them, using Gaussian noise parametrized by the measured (1-$\sigma$) errors.  The new set of maps is then rerun through the entire analysis described in Sections \ref{KinHighZ} and \ref{Criteria}.

	Figure \ref{figtemp} illustrates the resulting \vasym\ and \sasym\ measurements for the template systems.  In this figure, all of the results from the Monte Carlo realizations are plotted, with red shading indicating the resulting PDF of the merger templates and blue shading indicating that of the disk templates.  These two classes can be cleanly separated by the empirical delineation of total kinematic asymmetry \kasym\ = $\sqrt{v_{asym}^2 + \sigma_{asym}^2} = 0.5$, as visible in the inset.

	The majority of the disks (89\%) are located in the lower left (low \vasym, low \sasym) of the diagram, with the small deviations from the ideal case (\vasym\ $\equiv 0$, \sasym\ $\equiv 0$) coming from noise, thickness of the disk, and other kinematic features such as warps and multiple components.  The mergers, for the most part, show strong deviations from zero in both \vasym\ and \sasym\ and are located towards the upper right of the plot.  However, 20\% of merging systems remain indistinguishable from disks.  This is largely due to a single ULIRG, IRAS 12112+0305, whose velocity and velocity dispersion fields appear regular at \ztwo\ (see Appendix Figure \ref{fig:mergz2}), although other systems contribute to a lesser extent as well (Figure \ref{figtemp}).  Based on these results, we can roughly estimate the errors in these criteria and can expect to correctly classify  $\sim 89 \%$ of disks and $\sim 80 \%$ of mergers.

	Because these conclusions are based on a detailed and complicated analysis (Section \ref{KinHighZ}), we tested on several systems how changes in the assumed center, position angle, and inclination would affect the classification of the system as a disk or a merger.  We first examined variations in the assigned center and found that, for disks, the classification of a system as such is virtually independent of the center, except in extreme miscenterings when the center is assigned to the very edge of the system.  However, the location of the center is more important in the case of mergers with a double nucleus.  In these systems, if the center is skewed too far towards one of the two mass concentrations (both of which have some ordered rotational motion), the system can be misclassified as a disk.  This reinforces the necessity of choosing the center of mass of the system, via the continuum-intensity-weighted center as described in Section \ref{KinHighZ}.

	The test of variations in position angle indicated that the classification of a system as a disk or merger is even more robust against changes in this parameter.  In the case of disks, these systems maintain their low \kasym\ through extreme variations in position angle, until the position angle is aligned within $\sim 10^{\rm o}$ of the minor axis.  In mergers, the position angle has no physical meaning but, by the process described in Section \ref{KinHighZ}, has been defined such that it minimizes \kasym.  As a results, in this systems, all variations in position angle only increase their measured asymmetries, making their classification as mergers even stronger.  Finally, variations in inclination were also examined and were shown in nearly all cases to have no effects on the classification of either disks or mergers.  These tests of the effects of assumed inclination, position angle, and center on the classification of a system therefore suggest that, even in systems where there remains some uncertainty about the values of these parameters, our method will provide a robust classification with \kasym.

\begin{figure*}
\centering
\includegraphics[width=15cm]{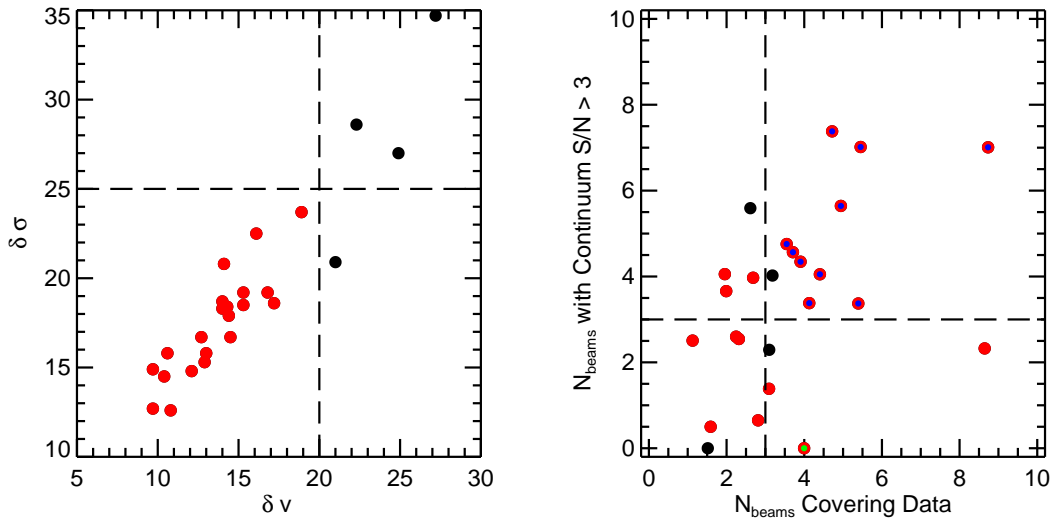}
\caption{{\it Left:} Comparison of the ``redshifted" template galaxies' typical errors in velocity and velocity dispersion (dashed lines) with the measured median errors of each of the SINS program galaxies.  SINS systems with smaller errors than those of the templates are shown in red.  {\it Right:}  The horizontal axis compares the median number of beams covering each system in the template galaxies (vertical line) with those of the SINS galaxies.  The vertical axis compares the median number of beams covering data with continuum S/N$ > 3$, as used to determine the galaxy centers, in the template galaxies (horizontal line) with those of the SINS systems.  Red points are the same galaxies shown in red in the left panel; red and blue points are those systems with data of better quality than that of the templates, based on both panels.  Only these systems are included in our analysis, with one additional inclusion.  We also analyze BzK-15504, shown with the red and green point, which fulfills three of the four data quality requirements but has insufficient S/N in its continuum emission.  The lower S/N per pixel is in part due to the finer spatial sampling of this system, which \citet{Gen+06} observed with the 0\farcsec1 pixel scale.  Fortunately, the continuum emission of this system has been observed in deep VLT/NACO imaging with laser guide star adaptive optics, which revealed a distribution consistent with that measured from the SINFONI data.  We are therefore confident that the continuum center measured from the SINFONI data does in fact reflect the center of the mass distribution in this system.}
\label{figlimits}
\end{figure*}

\section{Application to High-Redshift Galaxies}
\label{HighZ}

	Having tested these criteria on our template galaxies, we now apply them to the `unknowns,' the \hz\ systems observed in the SINS program (\citealt{For+06,Gen+06,Bou+07}; Cresci et al. in prep).  Here we examine those galaxies with sufficient spatial resolution and data quality to perform the kinemetric analysis described above.  Galaxies that are unresolved, have large errors in the kinematics, or have significantly lower S/N than the majority of the sample are omitted from the present discussion; in future work, we will expand our criteria to include variations in these quantities.

\subsection{Data}
\label{DataHighZ}

	The SINS \ztwo\ sample is taken from large photometric samples, in which \hz\ objects are identified through either their rest-frame ultraviolet color/magnitude \citep[BM/BX criterion:][]{Ade+04,Ste+04,Erb+06b,Erb+06a}, or their rest-frame optical properties (s-BzK: \citealt{Dad+04b,Dad+04a,Kon+06}, GDDS: \citealt{Abr+04}).  These selection criteria sample luminous (L $\sim 10^{11-12} L_\odot$) galaxies with an range of star formation rates (SFR $\sim 10 - 200$ \msun/yr) and ages (50 Myr - 2Gyr; \citealt{Erb+06b,Dad+04b,Dad+04a}).  From these photometric samples, our selection criteria emphasize somewhat brighter ($<$F(\ha)$>$ of $10^{-16}$ compared to $6 \times 10^{-17}$ erg/s/cm$^2$) systems, with broader line widths ($<$$v_c$$>$ of $175 \pm 68$ compared to $140$ km/s) than the average galaxy in the \citet{Erb+06a} sample.  Both samples have similar mean dynamical masses.

	The SINS galaxies were observed in \ha\ emission, which at \ztwo\ is redshifted to the $K$-band, with VLT/SINFONI.  Approximately $R$-band continuum emission is visible beneath the strong emission lines.  Most of the data have 0\farcsec125 x 0\farcsec250 pixels, sampling a typical PSF FWHM of 0\farcsec5, and have a spectral resolution of $R \sim 4000$.  Additionally, a few systems have been observed with adaptive optics, enabling the use of the finer pixel scale of 0\farcsec05 x 0\farcsec10 to sample the typical PSF FWHM of 0\farcsec15.  The data analysis is described in \citet{For+06}; here we add the additional step of spatially binning the datacube to a minimum amplitude-to-noise ratio of 5 with the Voronoi binning technique of \citet{CapCop03}.  This reduces the spatial resolution at the fainter edges of a system, where several spatial elements must be summed, but the amplitude-to-noise requirement ensures a more robust measurement of the kinematics in each bin.

\begin{table*}
\caption{Properties of the High-$z$ Galaxies}
\label{sinsprops}
\begin{center}
\begin{minipage}{13cm}
\begin{center}
\begin{tabular*}{13cm}{ccccccc}
\tableline
Galaxy & $z$ & SFR (\msun/yr)\footnotemark[a] & M$_{\rm dyn}$ ($10^{10}$ \msun)\footnotemark[b] & $R_{1/2}$ (kpc) & Selection\footnotemark[c] \\
\tableline
\tableline
SSA22a-MD41 	& 2.17	& 34	 	& 4.0 	& 5.8 	& UV \\
Q1623-BX528 		& 2.27 	& 28	 	& 2.3 	& 6.2 	& UV  \\
Q2343-BX389  		& 2.17 	& 93	 	& 11.0 	& 6.2 	& UV  \\
Q2343-BX610 		& 2.21 	& 115 	& 10.2 	& 5.4 	& UV  \\
Q2346-BX482 		& 2.26 	& 69	 	& 7.8 	& 6.4 	& UV  \\
BzK-6004 		& 2.39 	& 157 	& 16.2 	& 6.6 	& optical \\
BzK-12556 		& 1.59 	& 38	 	& 1.8 	& 5.1 	& optical  \\
BzK-15504\footnotemark[d] 		& 2.38 	& 101 	& 10.0 	& 5.3 	& optical  \\
D3a-6397 		& 1.51 	& 65	 	& 6.3 	& 7.6 	& optical  \\
K20-ID7 			& 2.22 	& 84	 	& 3.5 	& 4.5 	& optical \\
K20-ID8			& 2.22	& 42		& 1.8		& 5.4		& optical  \vspace{1.5mm} \\
Average			& 2.12 	& 81	 	& 6.8 	& 5.9 	& -- 	\vspace{1.5mm} \\
SINS UV Sample\footnotemark[e]  & 2.26	& 60$\pm$36 	& 4.1$\pm$3.7 	& 4.3$\pm$2.1 & UV  \\
SINS Optical Sample\footnotemark[f]  & 2.02 & 46$\pm$49 & 4.2$\pm$4.5 & 4.7$\pm$1.6 & optical  \\
\tableline
\end{tabular*}
\end{center}
\footnotetext[a]{As derived from the total \ha\ flux measured from the SINS data.  These values are converted to SFR via the \citet{Ken98} conversion, with an additional conversion to a Chabrier IMF.  Extinction $A_V$ is measured from SED modeling (see F\"orster Schreiber et al. in prep), and the Calzetti extinction law is used to estimate extinction $A_{H\alpha}$, from which the extinction correction is calculated.}
\footnotetext[b]{From the measured circular velocity, $R_{1/2}$, and axial ratio of the intensity distribution.}
\footnotetext[c]{Rest-frame waveband in which the object was identified.}
\footnotetext[d]{Observed with LGS, yielding a PSF FWHM of 0\farcsec15; all other systems were obtained in seeing-limited mode with a mean PSF FWHM of 0\farcsec5.}
\footnotetext[e]{Systems observed in the SINS program that were originally identified with rest-frame UV photometry.  Values given are mean and standard deviation of the sample.}
\footnotetext[f]{Systems observed in the SINS program that were originally identified with rest-frame optical photometry.  Half of this sample has $z \sim 2.3$, and half $z \sim 1.5$.  Values given are the mean and standard deviation of the sample.}
\end{minipage}
\end{center}
\end{table*}

	The SINS systems were observed with integration times ranging from 1.5 hours to over 8 hours (\citealt{For+06,Gen+06,Bou+07}; Cresci et al. in prep), which, when coupled with differences between systems, results in a wide range of errors on the kinematic measurements, S/N levels, and number of beams covering the systems.  In Figure \ref{figlimits}, we compare the assumed values for our template galaxies (typical values for the entire sample) to the actual values for individual SINS galaxies (see figure caption for details).  For this sample, the most critical and most limiting requirement is that the systems are well resolved, which we define as covered by $\gtrsim 3$ beams, although we also require  that the continuum be detected with a significant S/N level and that the kinematic measurements be sufficiently precise.  Analyzing data that do not meet these standards will adversely affect the spatial sampling of the kinemetry ellipses, their centering, and the precision of the derived asymmetry measures $v_{asym}$ and $\sigma_{asym}$; the calibration of our analysis to such lower quality data is outside the scope of this paper.  We find that \nsins\ of the \ztwo\ systems have equal or higher quality data than the templates and thus their kinematic asymmetries can be reliably determined.  We therefore perform our analysis on only these systems.
	
	The properties of this subsample are listed in Table \ref{sinsprops} and compared to the average properties of the UV-selected and optically-selected samples observed in the SINS program.  In most cases, the star formation rates, dynamical masses, and half-light radii of the objects analyzed here are slightly higher than the mean values for these parameters from the full SINS sample but still within the standard deviations.  Since the SINS sample contains a number of spatially unresolved compact objects \citep{For+06}, these statistics are consistent with the data requirements outlined above, in which only the well resolved (and thus large half-light radius and likely high mass) objects are selected for analysis.  It is these resolved objects in which the most progress can be made; for this reason, these systems were also the focus of \citet{For+06} and \citet{Gen+06}.
	
	In their analyses, \citet{For+06}, \citet{Gen+06}, and \citet{Bou+07} study the SINS data using the framework of rotating disks.  These authors have argued from qualitative examination of the data, along with analysis of the rotation curves and comparison to model disks in several cases, that many of these UV/optical-selected systems are candidate disk galaxies, although they note that a minority do in fact have the disturbed velocity fields expected of mergers.  We are now in a position to quantitatively test these conclusions.

\subsection{Classification}
\label{ClassHighZ}

\begin{figure*}
\centering
\includegraphics[width=15cm]{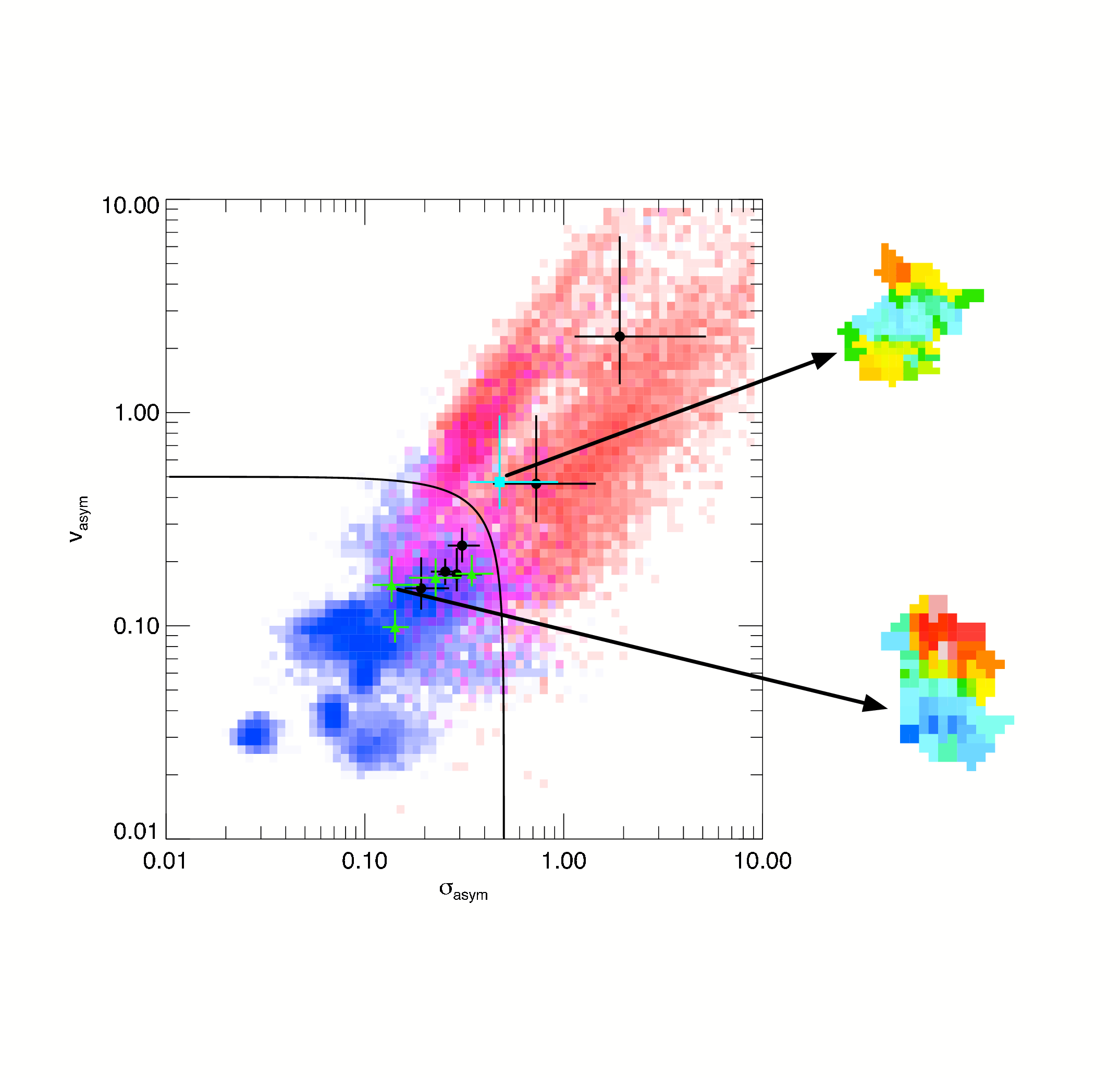}
\caption{Asymmetry measure of the velocity and velocity dispersion fields for the SINS program galaxies that have high enough quality data for such analysis, overplotted on the disk and merger template PDFs from Figure \ref{figtemp}.  The line indicates the division between disks and mergers at \kasym\ = 0.5.  The probable disks identified by \citet{For+06} and \citet{Gen+06} are indicated here as green triangles.  The merger identified by \citet{For+06} is shown with the cyan square.  Sample velocity fields of SINS disk-like and merger-like systems are shown at right; the full analysis of the SINS sample is recorded in Appendix Table \ref{tabkin} and shown in Appendix Figures \ref{fig:zd} and \ref{fig:zm}.}
\label{figsins}
\end{figure*}

	We perform the analysis of Sections \ref{KinHighZ} and \ref{Criteria} on the SINS systems, with the same 1000 Monte Carlo realizations, as done on the template \ztwo\ systems (Section \ref{ClassLowZ}).  In Figure \ref{figsins} we plot the resulting \vasym\ and \sasym\ measurements for these galaxies.  These results confirm the analyses of \citet{For+06} and \citet{Gen+06} in that \perdisks\ (8/11) of galaxies in the SINS subsample studied here are consistent with a rotating disk interpretation (see Appendix Table \ref{tabkin} for a list of results for each system).  Given the error rate of these criteria, as found with the template systems, this suggests that these results sample a parent population composed of 8.4 disks and 2.6 mergers, such that $89\% \times 8.4 + 20\% \times 2.6 = 8$ systems are observed as disks and $80\% \times 2.6 + 11\% \times 8.4 = 3$ systems are observed as mergers, making the fraction of disks in the subsample of Table \ref{sinsprops} as high as $8.4/11 = 75\%$.  Additionally, we expect that $20\% \times 2.6 \sim 1$ merger will be misclassified as a disk and that $11\% \times 8.4 \sim 1$ disk will be misclassified as a merger.

	It is of special interest to specifically investigate the most qualitatively convincing disks in the sample identified by \citet[][SSA22a--MD41, Q2343--BX389, Q2343--BX610]{For+06} and \citet[][BzK--15504]{Gen+06}.  We highlight these systems in Figure \ref{figsins} and indeed find that all four are consistent with a rotating disk interpretation.  \citet{For+06} also point out that two of their systems that may in fact be mergers; the one such system with sufficient data quality, Q1623--BX528, is identified in Figure \ref{figsins} and is in fact a merger.  Our criteria thus make intuitive sense based on visual analysis of the observed velocity and velocity dispersion fields and therefore demonstrates quantitatively the validity of the rotating disk interpretation as applied to the SINS sample.

\section{Discussion}	

	Sections \ref{ClassLowZ} and \ref{ClassHighZ} illustrate the efficacy of our dynamical criteria in identifying systems undergoing major mergers and their applicability to the SINS \hz\ survey; however, there are a number of caveats in this method that merit closer investigation.  Here we expand the discussion of Section \ref{ClassHighZ} to highlight the impact of this technique on \hz\ studies of galaxy formation, as well as address the most critical issues with this technique: the ability of the \ha\ kinematics to reflect the underlying dynamics of a system, and the observational constraints on probing the structure of a distant system in detail.

\subsection{Implications for Galaxy Formation at High-Redshift}

	Of the \ztwo\ systems analyzed here, we find that $8/11$ (\perdisks) are consistent with a rotating disk interpretation.  These results, coupled with the large stellar and dynamical masses of these systems ($M_{dyn} \sim 10^{10-11}$~\msun; Table \ref{sinsprops}), quantitatively confirm that massive disks were already in place at this redshift.  It furthermore strengthens the conclusions of \citet{For+06} and \citet{Gen+06} that the high (up to $100$ \msunyr) star formation rates of these systems are not the consequence of a recent violent merger but rather are happening within disks dominated by ordered rotation.  
	
	In the case of a typical massive ($\sim 10^{11}$ \msun) star-forming galaxy, BzK-15504, which \citet{Gen+06} studied in detail, modeling of the spectral energy distribution (SED) from broad band photometry was used to measure the stellar mass and age of the system.  With the stellar mass M$_*$ $8 \times 10^{10}$ \msun\ and the star formation rate of $\sim 100$ \msunyr\ (as measured from \ha), together with the assumption of a constant star formation rate, all the stars in this system could have been formed in $\sim 500$ Myr \citep{Gen+06}.  This number agrees well with the stellar age measured from the SED fitting ($300-800$ Myr), making it likely that this system formed rapidly with continuous star formation (and therefore mass inflow) at its current rate \citep{Gen+06}.  Given that the above analysis (Figure \ref{figsins}) indicates that this system has not undergone any recent major merger activity, and given that this system is typical for its population, our results provide new and direct empirical evidence that the smooth accretion mechanism can play an important role in the early stages of the evolution of massive galaxies.
	
	The diagnostic tool described here is critical in expanding our understanding of structure formation and evolution in the early Universe.  With current extensive data sets of \hz\ systems, including broad-band photometry and integral-field kinematic observations, much can be learned about the stellar populations, star formation processes, and nuclear activity of the galaxies evolving in a critical epoch of the Universe's history.  We now add another crucial piece to the study of \hz\ systems, by introducing a method to quantitatively evaluate the dynamical state of a system and therefore to link that system's observed properties with a major merger event or with a more quiescent evolutionary history.

\subsection{\ha\ as a Probe of a System's Dynamics}
\label{discussha}

	The reliability of using \ha\ emission to study the structure of a galaxy is, at first glance, rather unclear.  The motions of the warm gas are not guaranteed to reflect those of the underlying stellar distribution, since the former component is much more easily disturbed - with gravity or pressure fluctuations - than the more massive, collisionless stars.  On the other hand, the dissipative gas component also more efficiently relaxes into a thin disk and could conceivably demonstrate ordered rotational motion while disturbances in the stellar (and mass) distribution persist \citep[e.g.][]{MihHer96}.  These competing effects, together with the ambiguity caused by the limitations of spatial resolution, could conceivably render it difficult to interpret the kinematics of the warm gas as uniquely representing a disk or a merger.

	Furthermore, the kinematics of the warm gas are also expected to reflect such phenomena as the large-scale gas flows that feed active nuclei and the powerful galactic winds from AGN/starburst activity, both of which at \ztwo\ are thought to play an important role in regulating the star formation history of the Universe.  For this reason, we designed our criteria with input from templates that were likely to include as many of these phenomena as possible.  Our sample of disks includes a kinematically perturbed system, several barred systems, and two Seyfert galaxies.  One of these active systems (NGC~4579) is a barred galaxy in which the radial motion of the gas is clearly visible as a strong velocity gradient along the minor axis \citep{GonPer96,Dai+06}.  We also observe such an inflow in a SINS system, BzK-15504, as described by \citet{Gen+06}.  In this system, a strong velocity gradient is also seen along the minor axis, in the form of a high $A_{1,v}$ term, presumably corresponding to the driving of fuel towards a growing bulge with an embedded active nucleus \citep{Gen+06}.  We can nevertheless robustly identify both the ``redshifted" NGC~4579 and BzK-15504 as disks, since the kinemetric signature of the inflows is restricted to the $A_{1,v}$ coefficient, which is excluded from our analysis for precisely this reason (and could potentially be used in future work as a tracer of AGN feedback at high redshift).  As shown in Section \ref{ClassLowZ}, this omission does not diminish our ability to detect mergers, since the disturbed velocity fields of such systems also produce power in all the higher-order kinemetry coefficients.
	
	To discount the possibility of contamination by large-scale outflows in the \ztwo\ kinematics, we compare with observations of local systems with substantial galactic winds.  In general, line emission from these ``superwinds" in local starbursting galaxies dominates neither the total \ha\ surface brightness nor the emission line kinematics along the galaxy major axis \citep{LehHec95,LehHec96a,LehHec96b}.  The gas flowing outward along the minor axis has been shown to contribute only a small fraction of the total \ha\ luminosity; in even the extreme cases of local infrared-luminous and highly-extincted starburst galaxies, \citet{Arm+90} show that the extended emission-line gas outside of a few kpc from the nucleus accounts for $<25\%$ of the total \ha\ emission.  In a few cases in our SINS sample, our data reveal a high-velocity outflow component, but this wind contributes $<10\%$ to the total \ha\ luminosity of the system.  Similar results have been found with integral field data of a \ztwo\ SMG by \citet{Nes+07}.
	
	For the majority of the SINS \ztwo\ systems, then, it is likely that we can rule out possible contamination from superwinds and that we have sufficiently accounted for the affects of large-scale gas inflows.  The \ztwo\ data are thus more consistent with the scenario in which the emission line kinematics trace dynamical features similar to those of the template galaxies.

	This argument suggests that the kinemetric analysis of a system is largely independent of SFR-driven phenomena and therefore of the overall SFR, a fact that is born out by the data themselves.  In local spiral galaxies, where the relatively low SFRs ($\sim 1-10$ \msunyr) reflect low gas fractions, the dynamics of the warm gas are driven by those of the dominant, rotating stellar component.  In contrast, the SINS galaxies, with SFRs characteristic of ULIRGs (up to and exceeding $100$ \msunyr), have significantly higher gas fractions ($f_{gas} \sim 0.4$; \citealt{Tac+06,Bou+07}) and consequently a dynamically important dissipative component.  The gas in these systems quickly concentrates into massive, powerful star-forming regions, which drive turbulent motion and thus high velocity dispersions in the warm gas.  However, the clumps detected in SINS systems have not fatally disrupted the dynamics of their host galaxy; \perdisks\ of SINS systems display velocity fields consistent with ordered rotation and regular (though elevated) velocity dispersions.  Despite orders of magnitude difference in SFR between local spiral galaxies and SINS systems, it appears that the processes governing the \ha\ kinematics in the two regimes are remarkably similar.  This is consistent with results from \citet{Bou+07}, who demonstrated that star formation at high SFR and high redshift is governed by the same physics as in the local Universe.  For the purpose of the analysis presented here, then, \ha\ emission can be effectively employed as a tracer of a system's dynamics for a large range of SFR and redshift.

 \subsection{Continuum Surface Brightness Distribution}
 
	One of the unique capabilities of the integral-field data used in our analysis, beside providing spatially resolved  kinematics of a system, is separating the emission coming from star-forming regions, as visible in \ha\ emission, from that of the underlying stellar background, visible in approximately $R$-band continuum emission.  Morphological criteria using the continuum distribution to detect mergers have been implemented at lower redshift by e.g. \citet{Con03} and \citet{Lot+04} using HST data.  However, at \ztwo\ the stellar distribution is difficult to probe at optical (rest-frame UV and $B$) wavebands, since this emission can be severely affected by extinction in gas-rich systems \citep[e.g.][]{Col+05}.  It is therefore of interest to examine the rest-frame $R$-band continuum emission measured directly in our integral field data.
	
	Unfortunately, these methods require both high S/N and resolution elements smaller than $1$ kpc/pixel, which with SINFONI at \ztwo\ is attainable only in AO-assisted observations.  Since our observations were optimized for the analysis of the line emission properties of the sources and were mostly obtained in seeing-limited mode, we tested several simpler criteria, based on kinemetry of the continuum distribution (which by definition is identical to surface photometry).  Any asymmetries in the continuum distribution would likely be due to the presence of multiple mass concentrations, as in an early-stage merger.  We find that the systems identified as potential early-stage mergers are, as expected, a subset of those identified as mergers by the kinematic criteria.  This result highlights the unique capabilities of directly probing the dynamical properties of \hz\ systems with integral field observations.

	We do not develop further any analysis of the continuum emission in our SINS IFS data, as the low S/N of this emission limits the strength of the conclusions that can be drawn from such analyses.  Forthcoming sensitive high-resolution (HST and ground-based AO-assisted) imaging of the SINS systems in the near-infrared will provide the necessary resolution and S/N required for the quantitative morphological analyses of broad-band emission (F\"orster Schreiber et al. in prep).  The comparison of such results with those from the kinemetric analysis developed here will provide a valuable further probe into the nature of \hz\ systems.

\subsection{Limitations of the Method}

	For the analysis described here, we designed our template systems to have S/N, resolutions, and spatial extent comparable to typical values of the SINS survey at \ztwo.  In Section \ref{DataHighZ}, we illustrated that a subsample of \nsins\ SINS observations have higher quality data than the templates and thus have morphologies that can be reliably measured.
	
	For the remainder of the systems, and for systems at other redshifts or measured under different observing conditions, the current criteria cannot be blindly applied.  We expect that there will be fundamental data limitations beyond which these criteria cannot distinguish disks from mergers.  In future work, we will therefore generalize this methodology to a wide range of S/N, spatial resolutions, and spatial sampling in order to investigate the effectiveness of our criteria for a more complete range of redshifts and observing conditions.

\section{Conclusions}

	We present a simple set of kinematic criteria that can be used to distinguish mergers and disks in the SINS survey and in similar observations (i.e. with Keck/OSIRIS).  The reliance of our criteria on dynamical information, rather than on surface brightness distributions, takes full advantage of the wealth of information provided with integral field data.
	
	We show, via a large set of template galaxies, that our criteria can reliably distinguish the majority of major mergers (80\%) from disks.  When applied to the SINS galaxies, this tool provides quantitative support for the rotating disk / smooth accretion scenario that the interpretation of recent results has suggested \citep{For+06,Gen+06,Bir+07}.  In the subset of the SINS systems studied here, we quantitatively show that \perdisks\ likely have not had a major merger in their recent history to fuel their rapid star formation, providing direct evidence for this scenario.
	
	Looking forward, the differentiating of disks and mergers on a galaxy-by-galaxy basis that is now possible is useful both immediately and in the future with next-generation \hz\ surveys on 30m class telescopes.  As an increasing number of galaxies in the high-redshift Universe are probed with spatially resolved kinematics, the tool presented here can be used to observationally constrain merger fractions, as well as to understand the effect of mergers on star formation rates, nuclear activity, and growth of structure within proto-galaxies.

\acknowledgements
	We are very grateful to the entire SINS team, whose input greatly improved the quality of this work.  We also thank Carl Heiles for many useful discussions and the anonymous referee for helpful comments.  NMFS acknowledges support by the Schwerpunkt Programm SPP1177 of the Deutsche Forschungsgemeinschaft.  NA acknowledges support from a Grant-in-Aid for the Scientific Research (No. 19540245) by the Japanese Ministry of Education, Culture, Sports, and Science.  SA and LC have been supported by the Spanish Ministry for Education and Science under grants PNE2005-01480 and ESP2007-65475-C02-01.


	


\section*{Appendix}

\begin{figure*}
  \centering
  \includegraphics[width=0.3\textwidth]{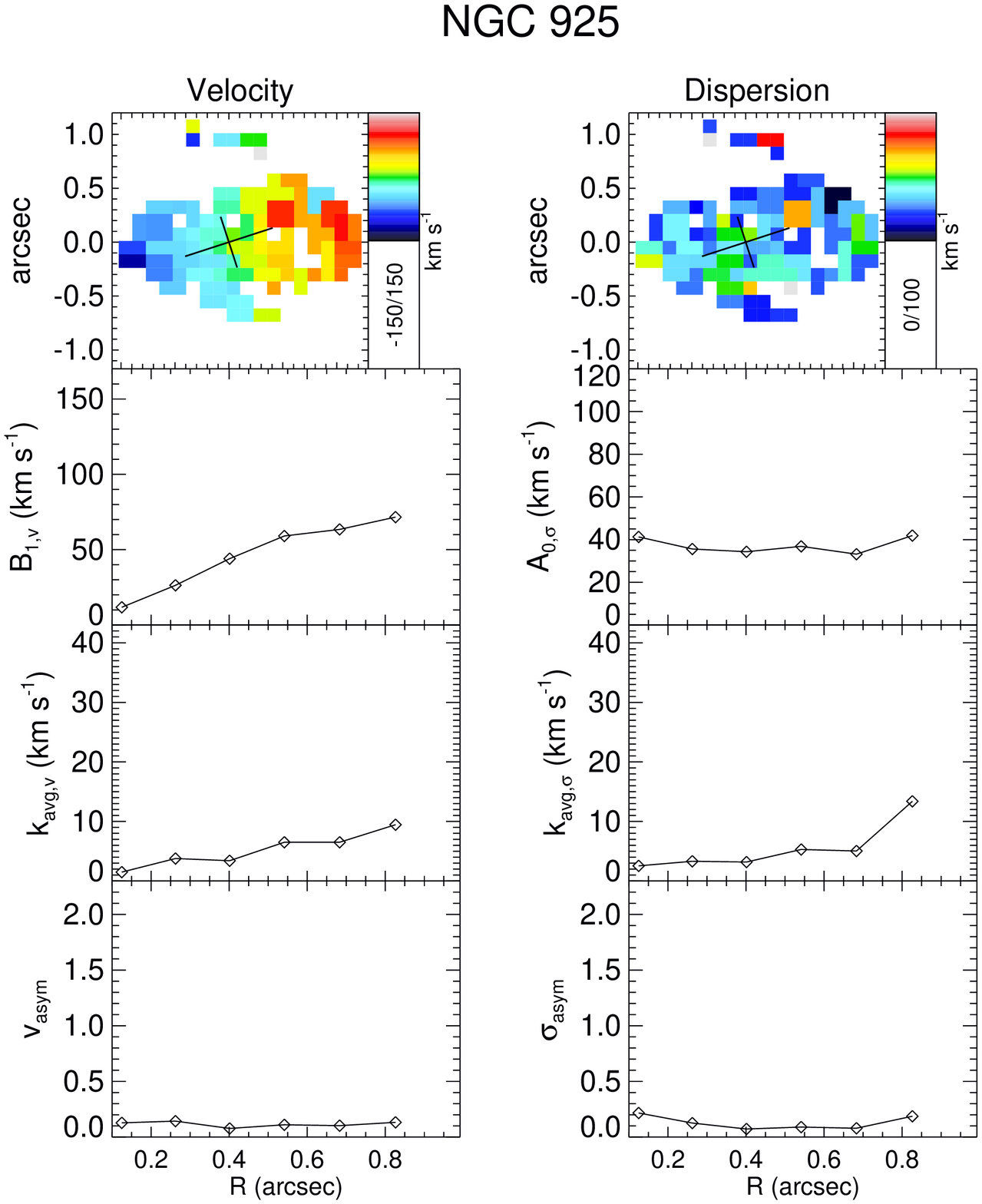}              
  \includegraphics[width=0.3\textwidth]{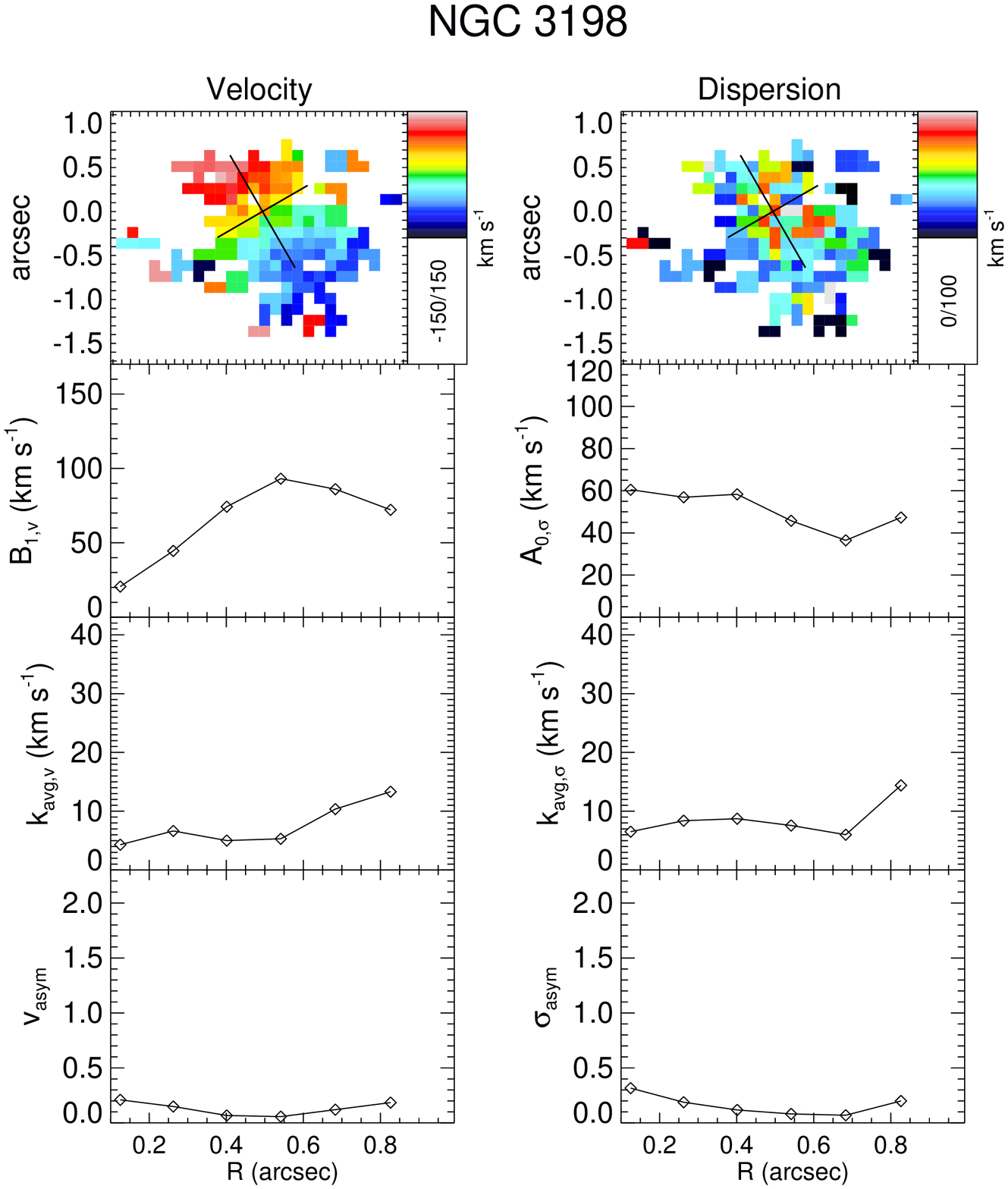}
  \includegraphics[width=0.3\textwidth]{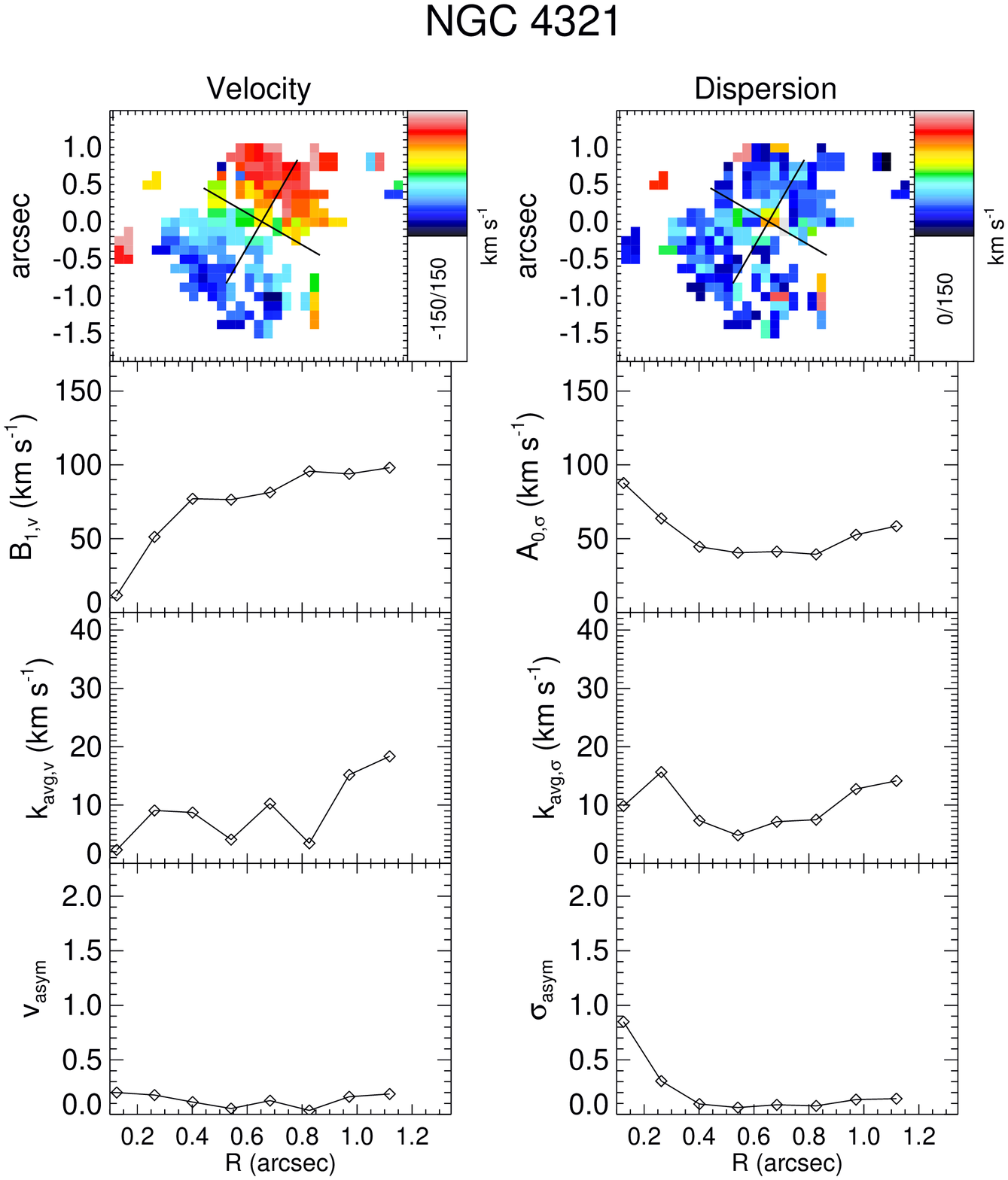}
  \includegraphics[width=0.3\textwidth]{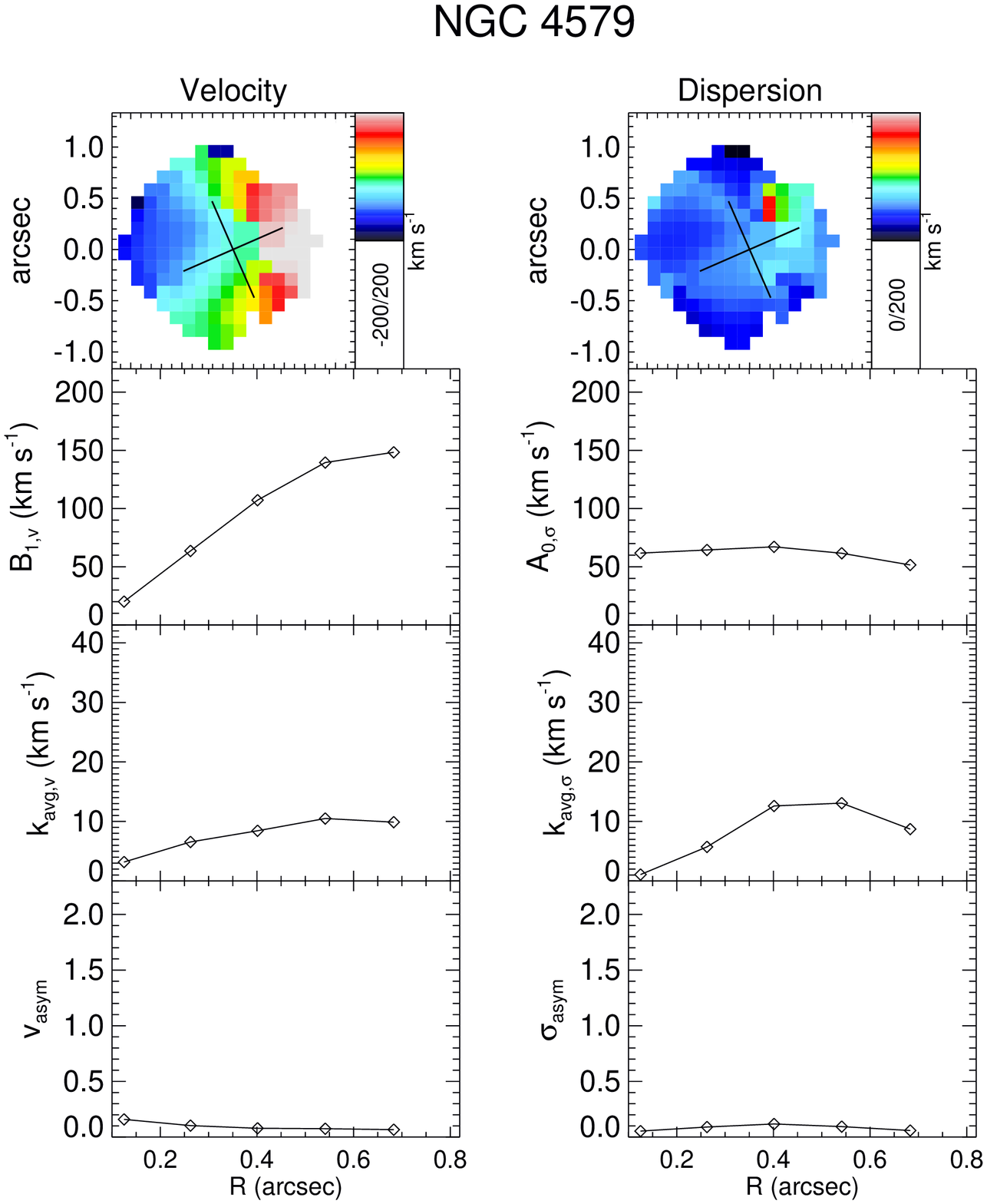}                
  \includegraphics[width=0.3\textwidth]{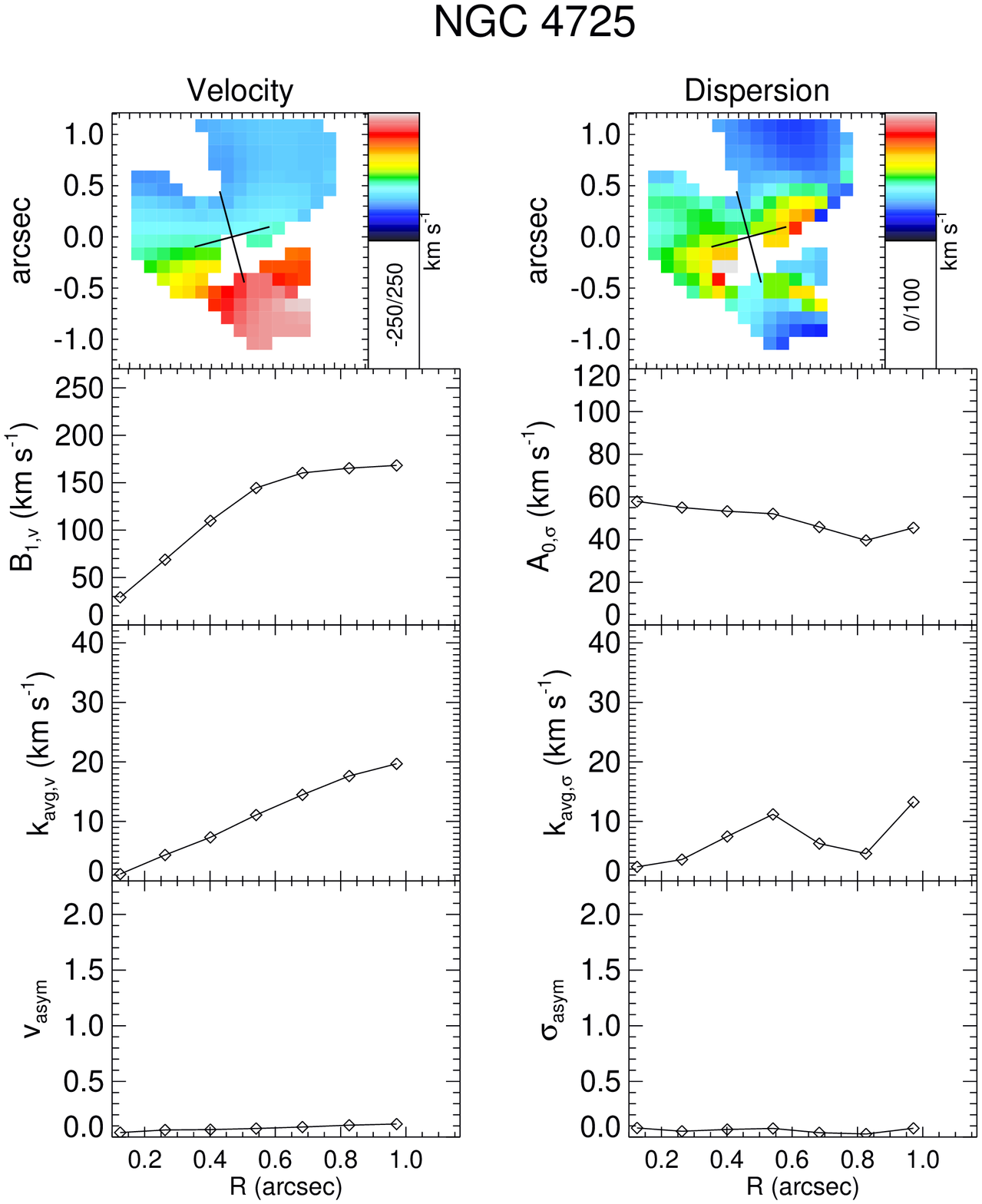}
  \includegraphics[width=0.3\textwidth]{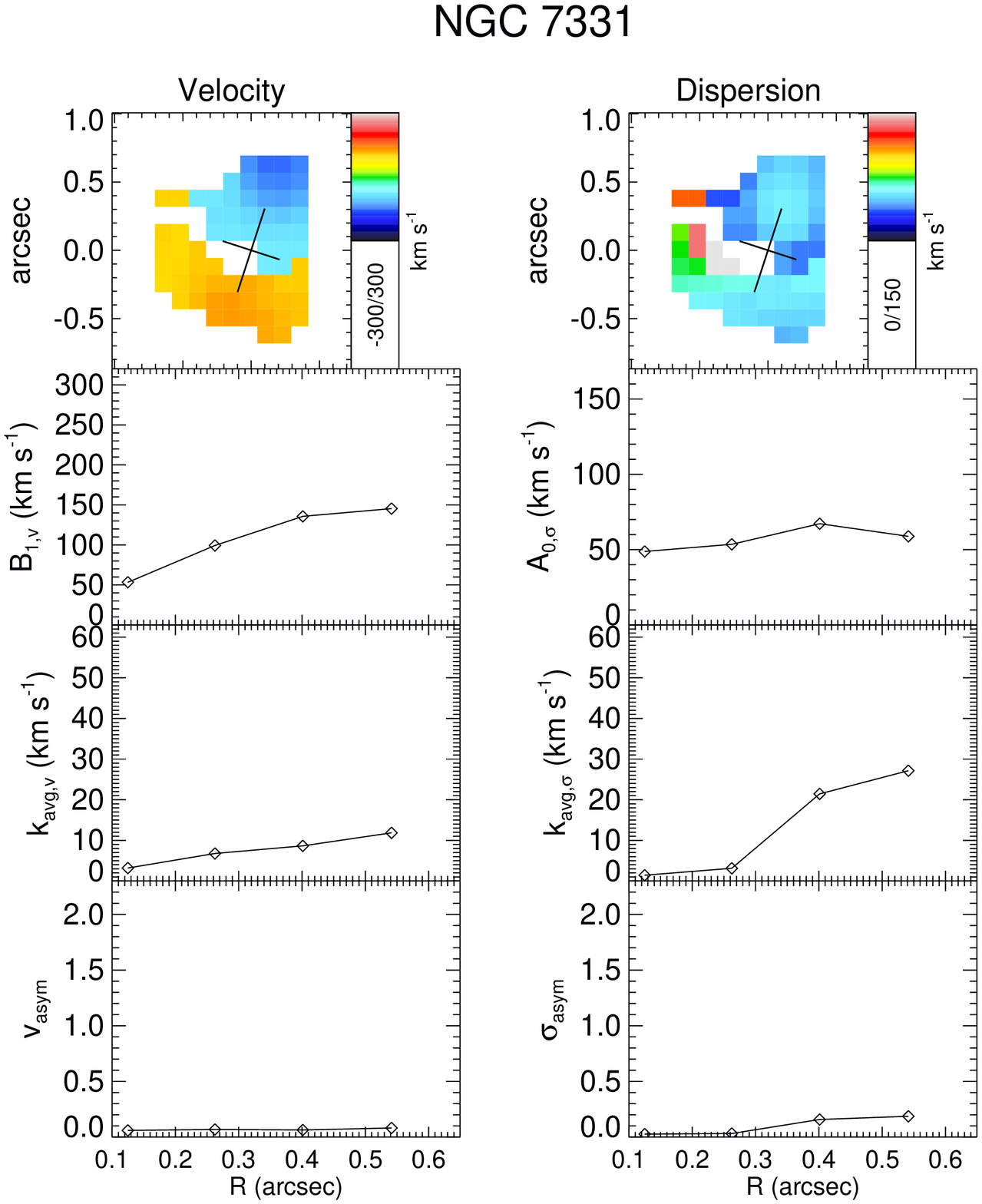}
  \caption{Kinemetry of ``redshifted" observed template disks at $z \sim 2$.  For each system, the velocity and dispersion fields are shown, followed by results from the kinemetric expansion.  Overplotted on the velocity and dispersion fields are the major and minor axes of the kinemetry ellipses used in the expansion, centered on the continuum center of the system (see text for definition).  Because the analysis described in Sections \ref{KinHighZ} and \ref{Criteria} does not lend itself to straightforward error propagation, we do not include error bars on these figures.  The most reliable measurement of the kinemetry errors comes from the Monte Carlo realizations, whose results are summarized in Table \ref{tabkin}.}
  \label{fig:spirz2}
\end{figure*}

\begin{figure*}
  \centering
  \includegraphics[width=0.3\textwidth]{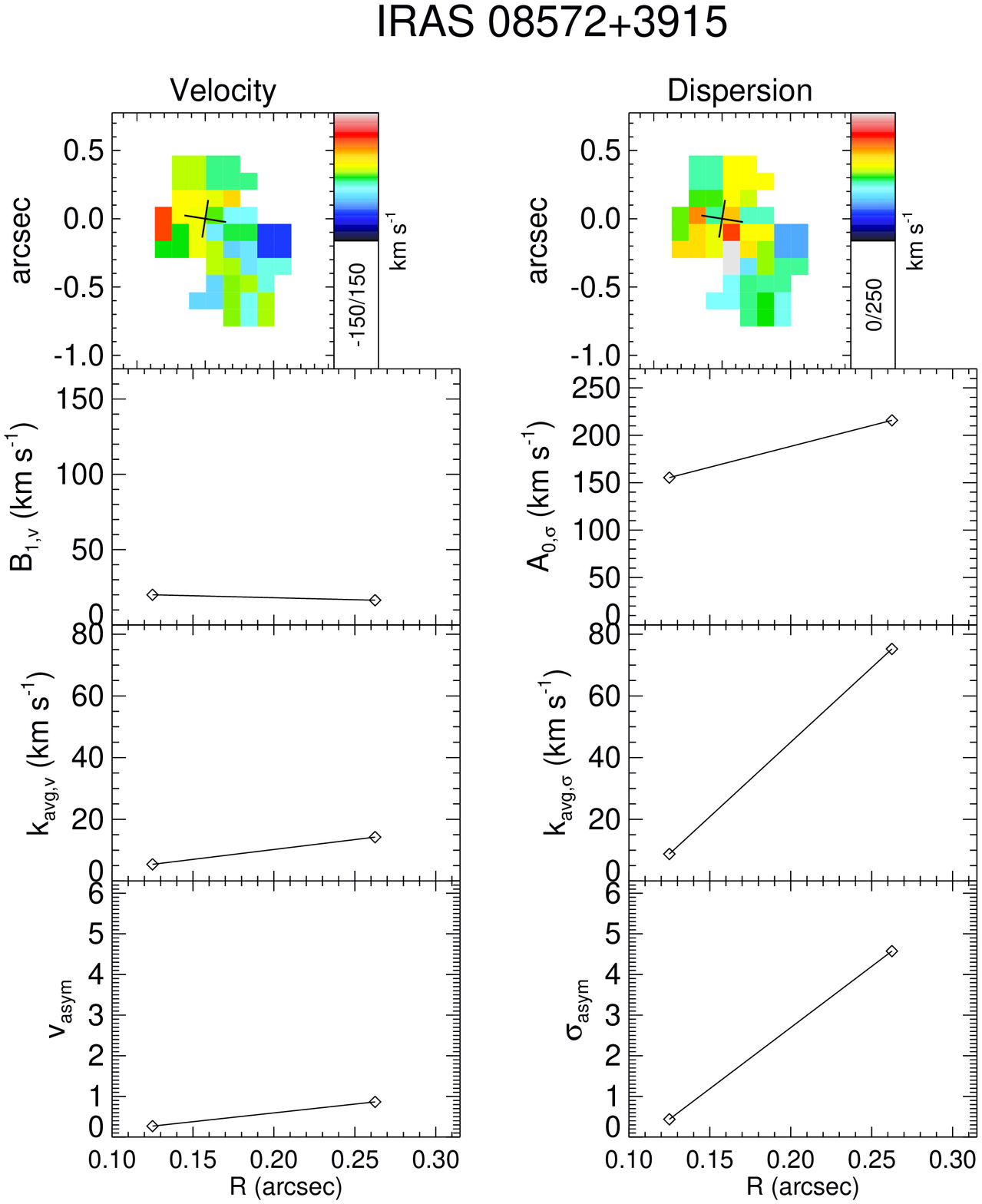}
  \includegraphics[width=0.3\textwidth]{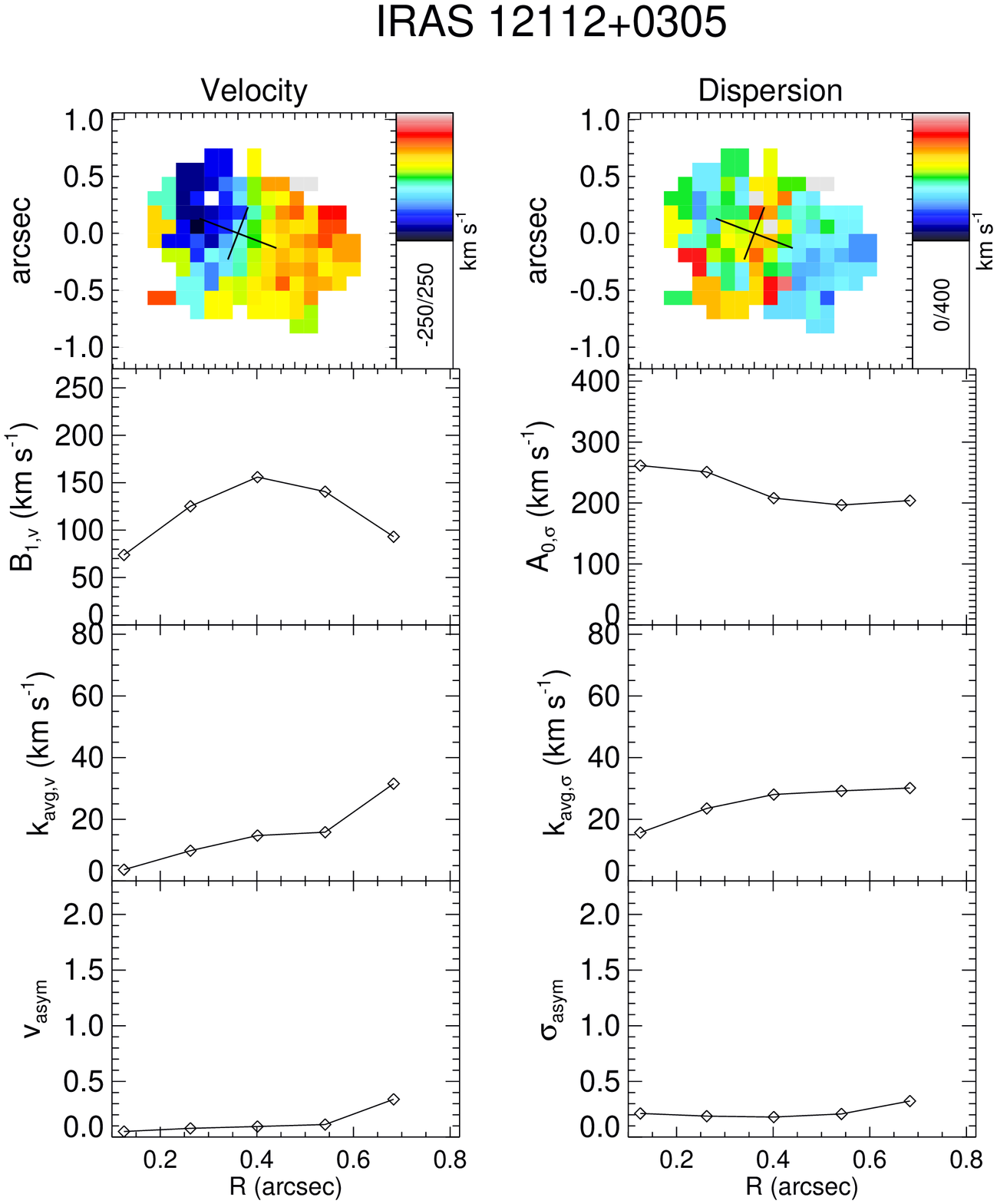}
  \includegraphics[width=0.3\textwidth]{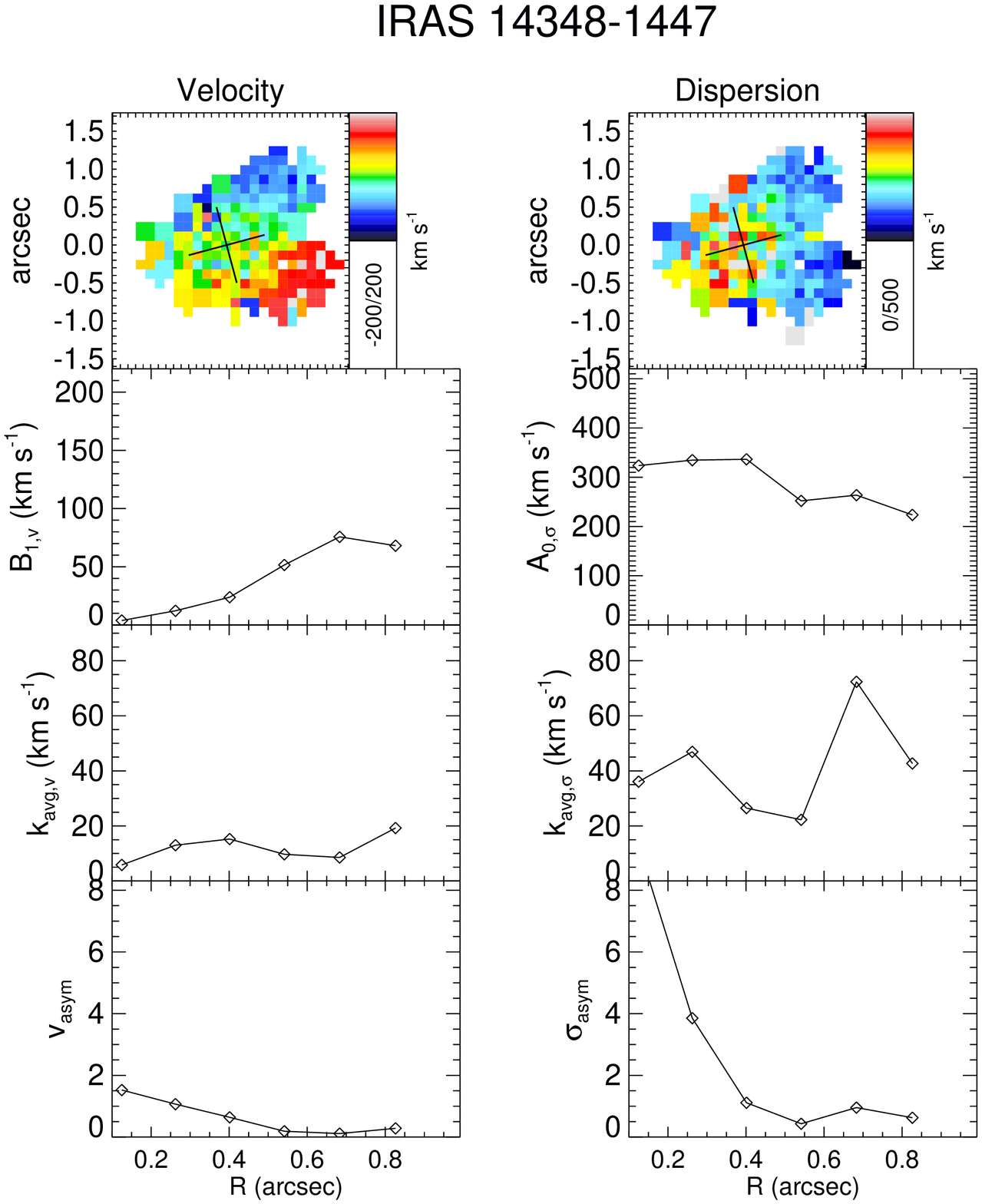}
  \includegraphics[width=0.3\textwidth]{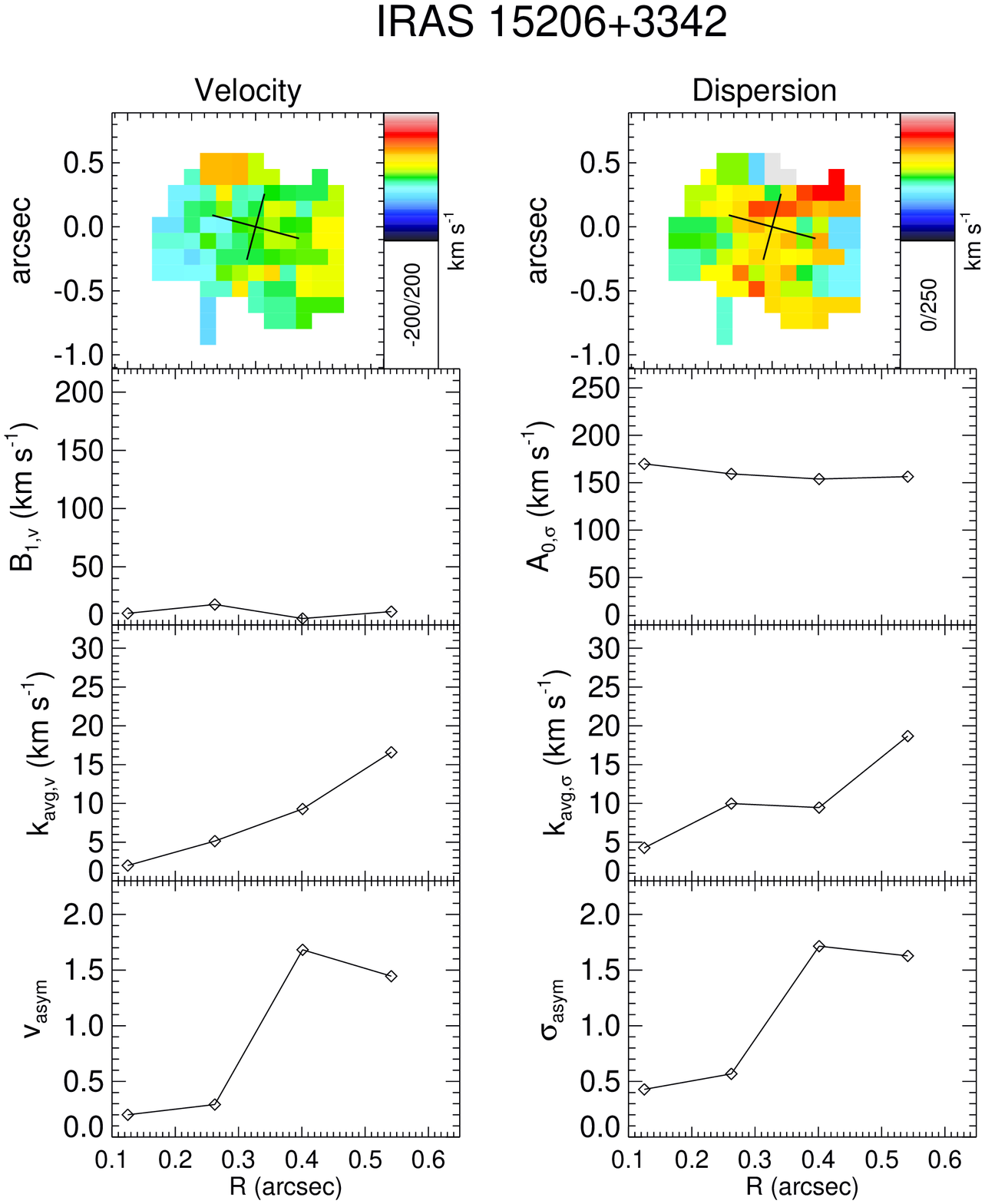}
  \includegraphics[width=0.3\textwidth]{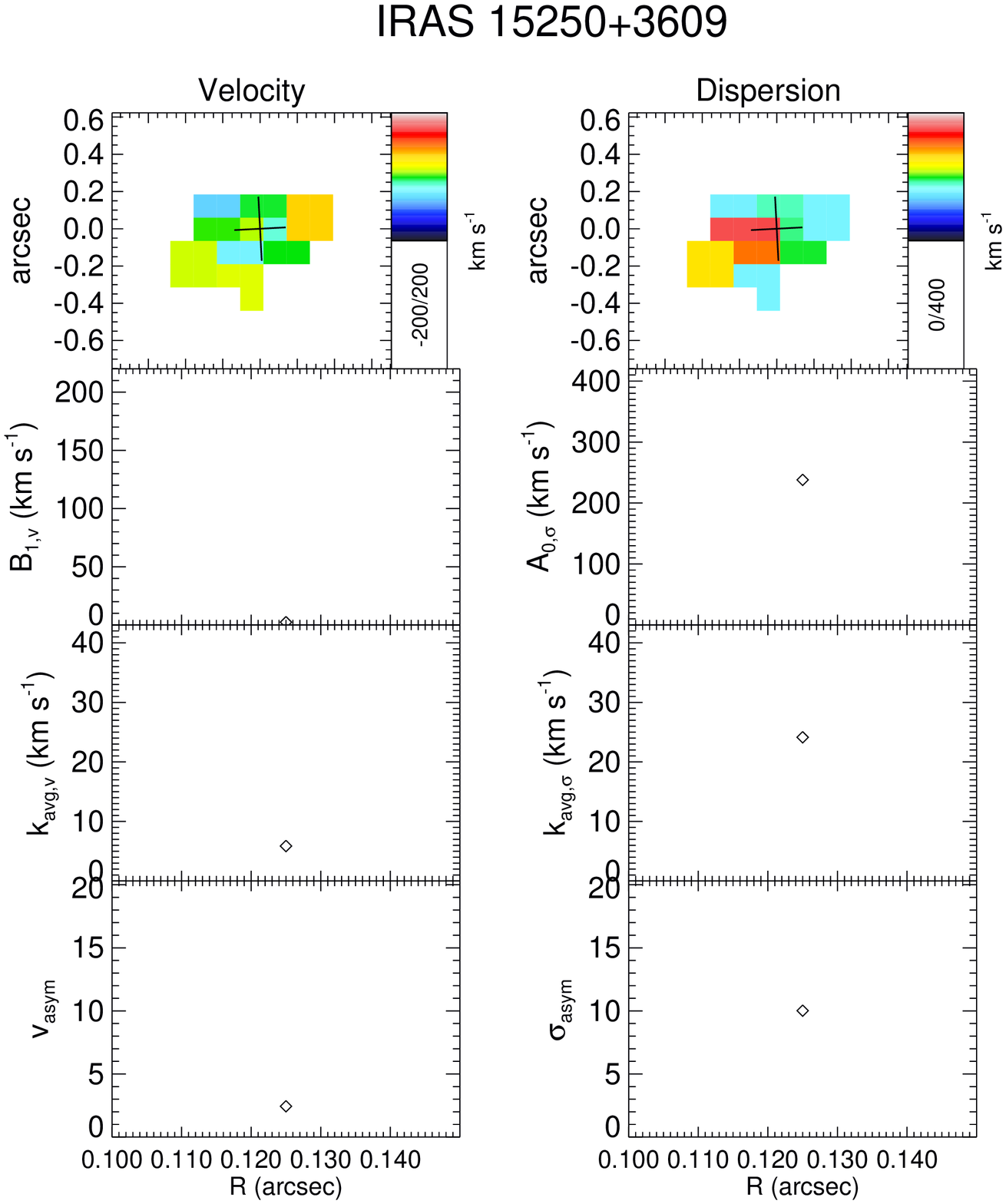}
  \includegraphics[width=0.3\textwidth]{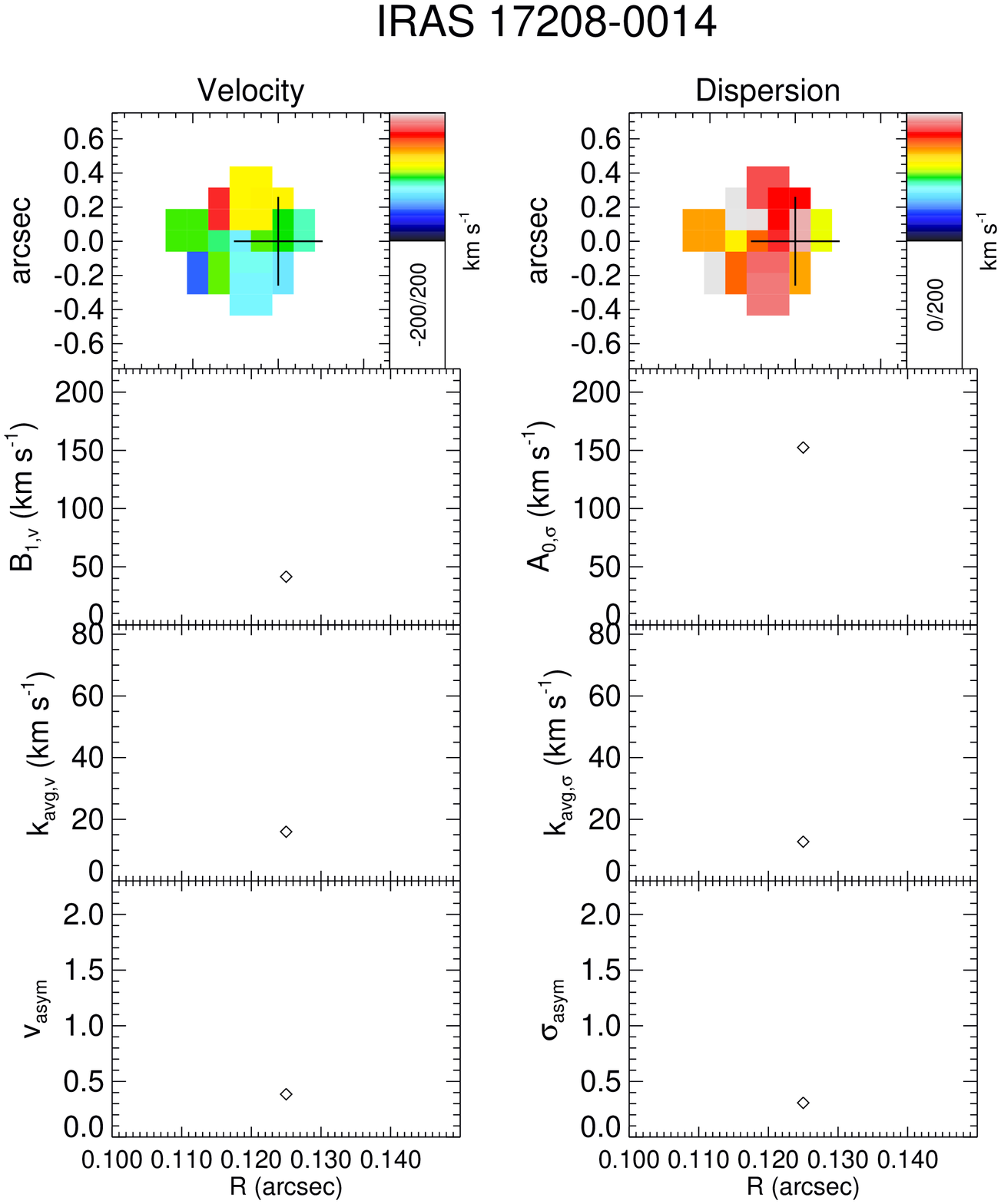}
  \includegraphics[width=0.3\textwidth]{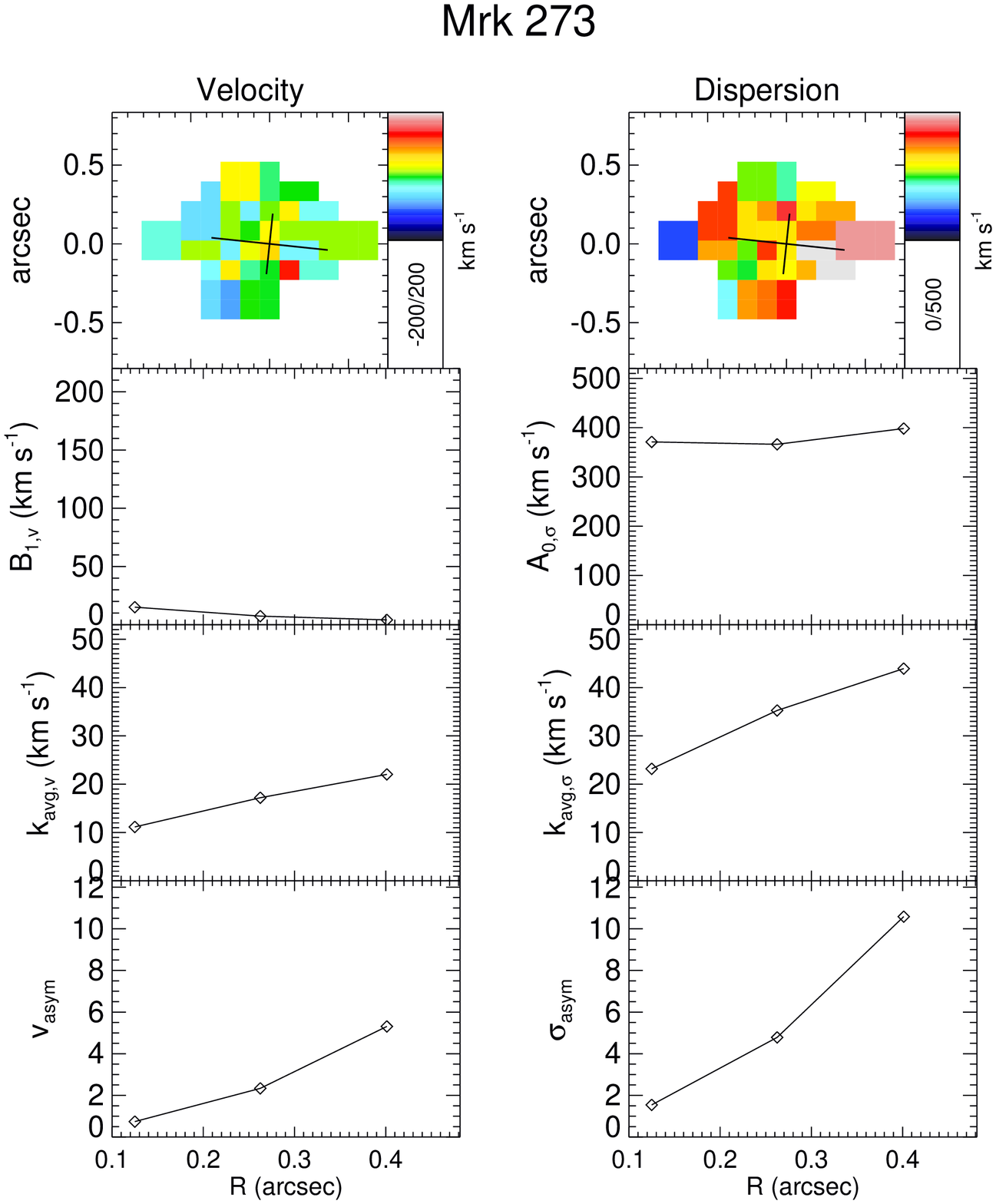}
  \includegraphics[width=0.3\textwidth]{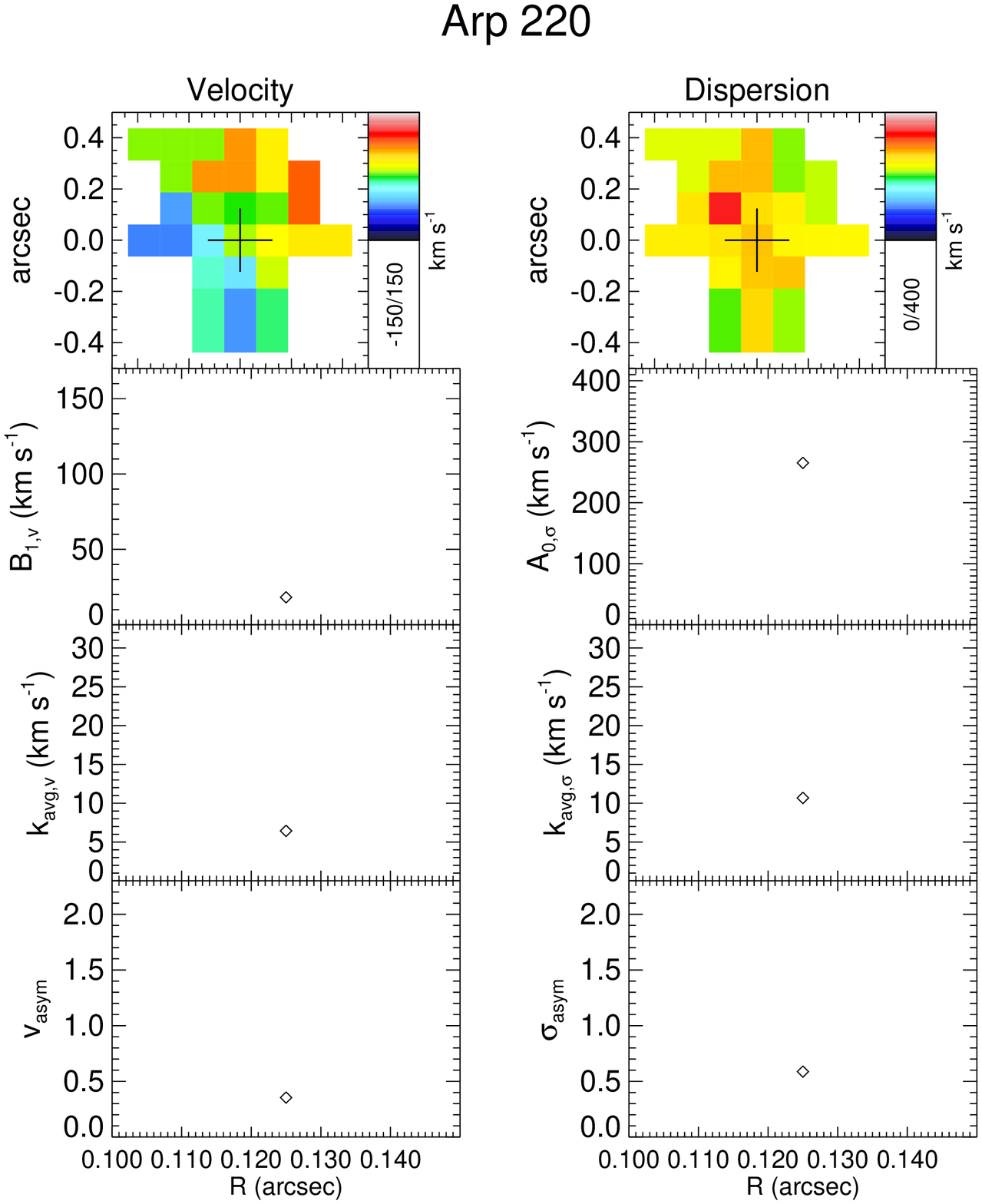}
  \caption{Kinemetry of ``redshifted" template mergers at $z \sim 2$.}
  \label{fig:mergz2}
\end{figure*}

\begin{figure*}
  \centering
  \includegraphics[width=0.3\textwidth]{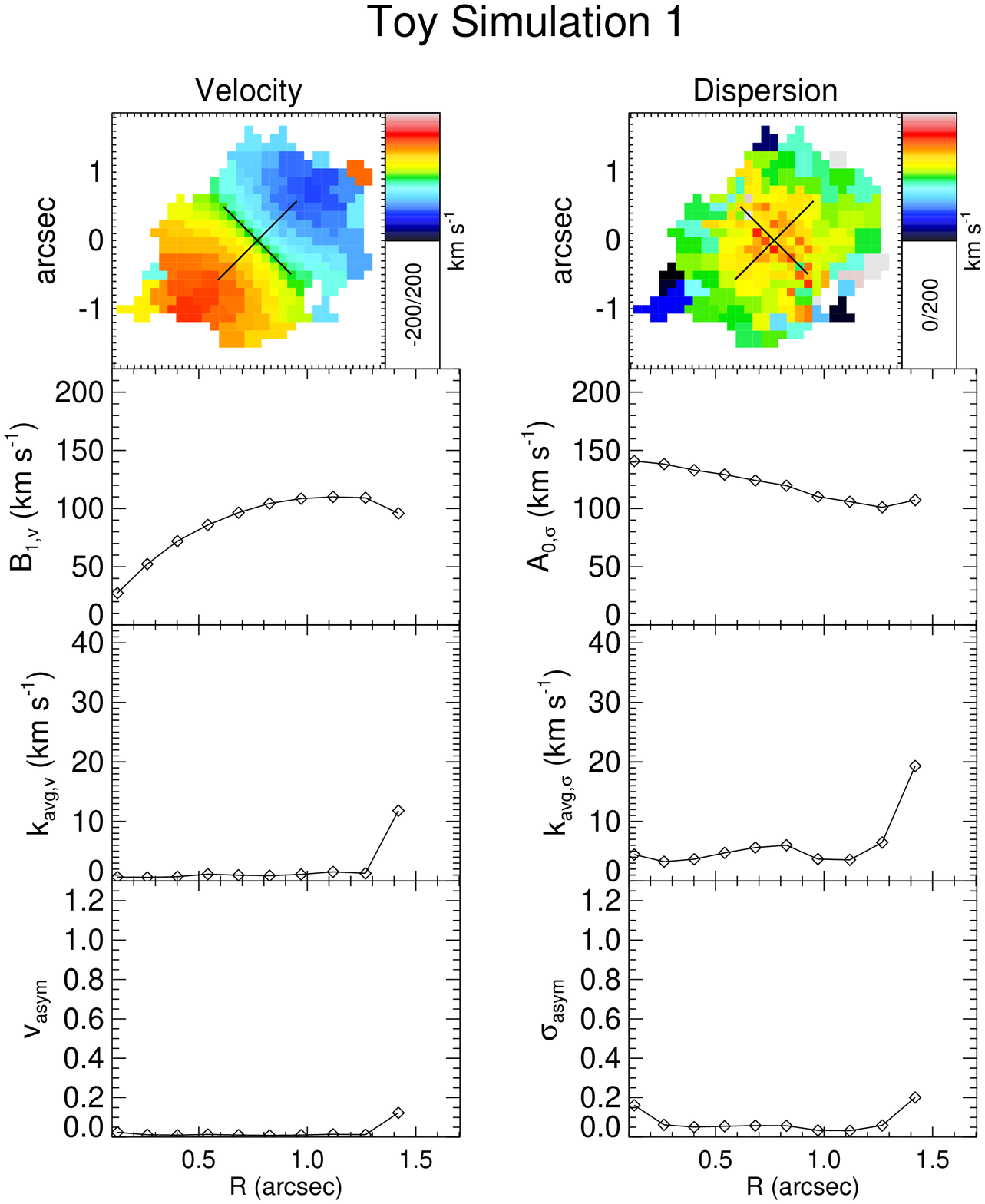}
  \includegraphics[width=0.3\textwidth]{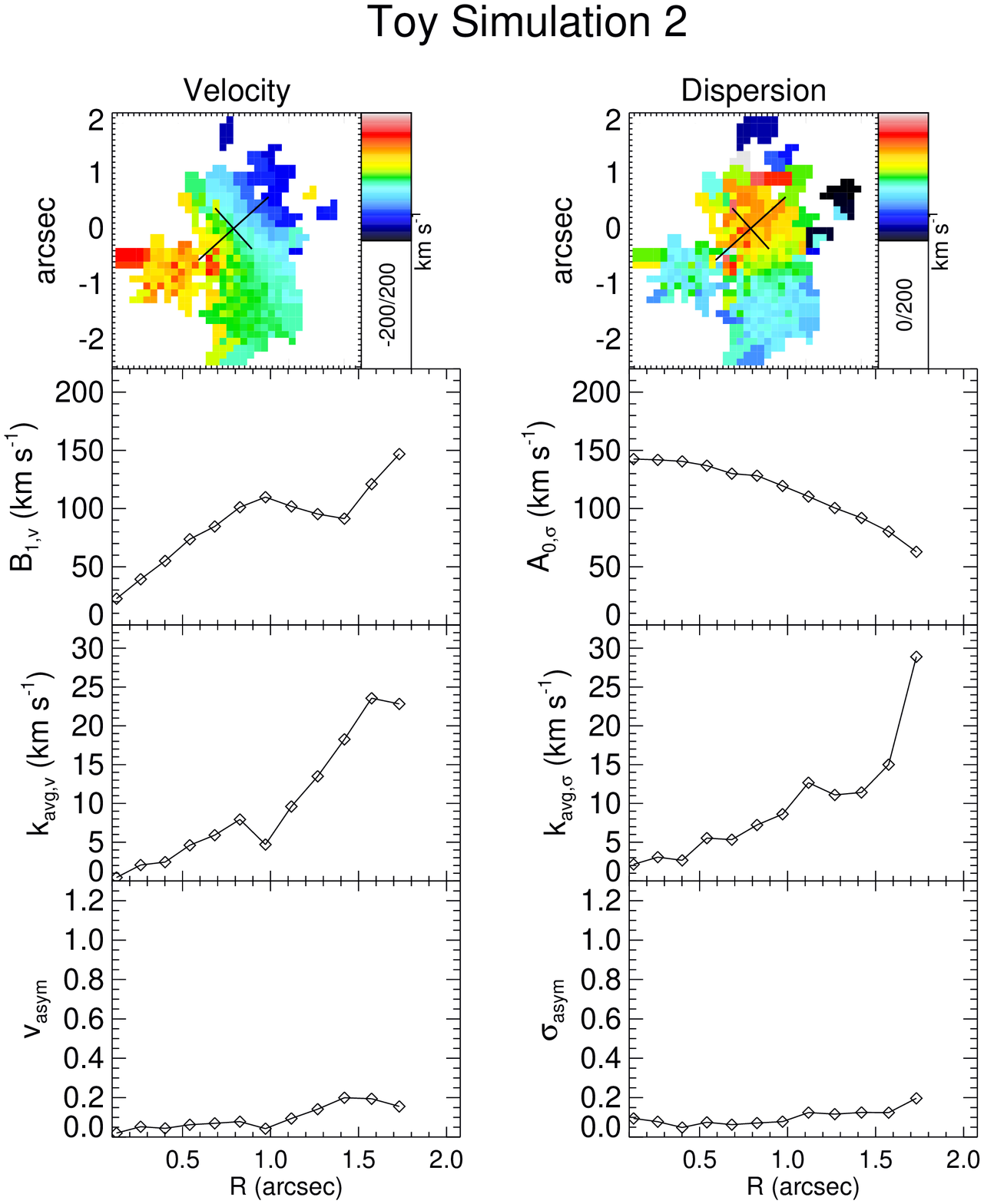}
  \includegraphics[width=0.3\textwidth]{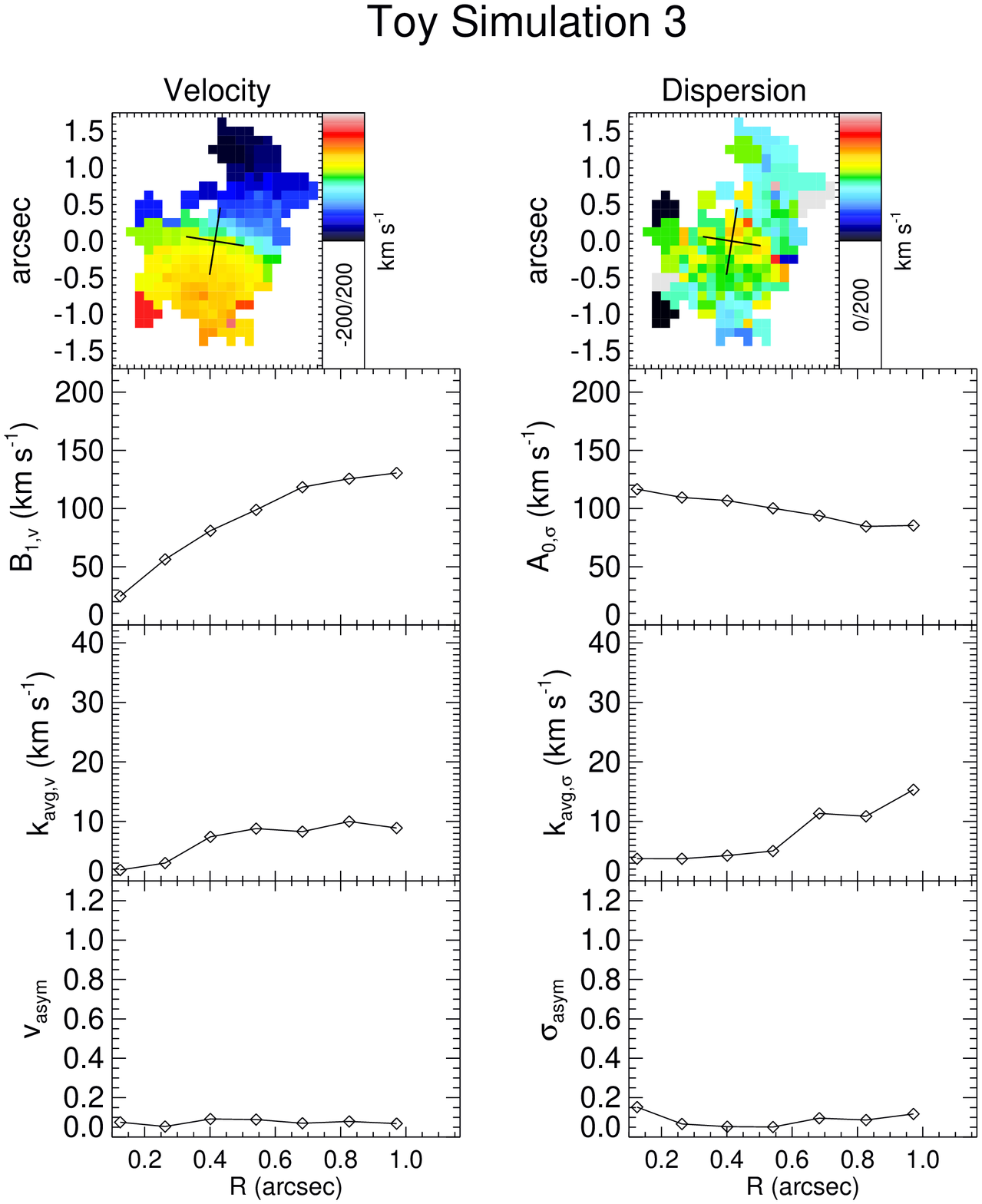}
  \includegraphics[width=0.3\textwidth]{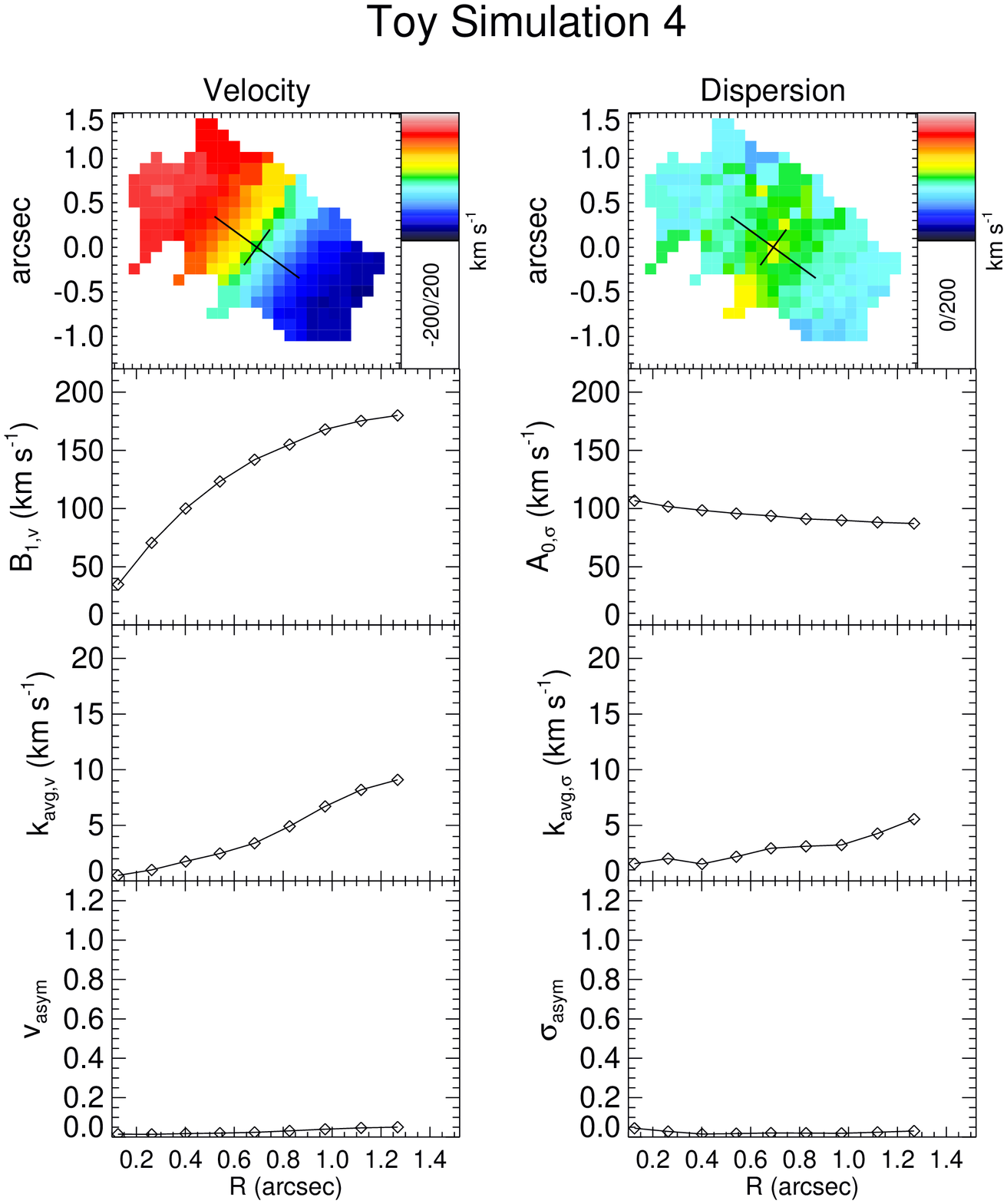}
  \includegraphics[width=0.3\textwidth]{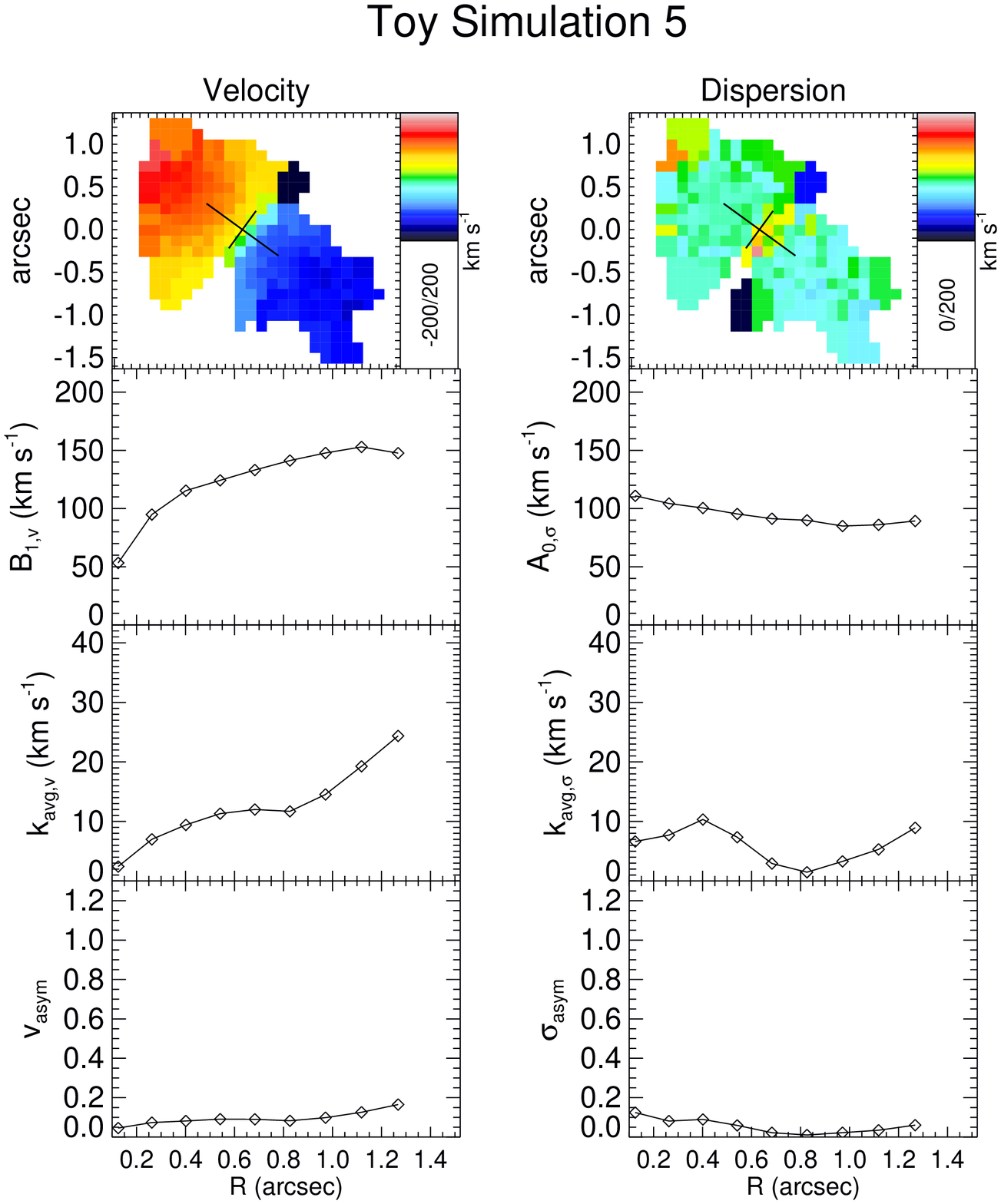}
  \caption{Kinemetry of template toy model disks at $z \sim 2$.  Simulation 1 is of intermediate inclination with a centrally-peaked azimuthally symmetric light distribution.  Simulation 2 is of intermediate inclination and has a light distribution much more extended on one side of the galaxy.  Simulation 3 is of intermediate inclination, with the light distribution illuminating only one side of the galaxy.  Simulation 4 is nearly edge-on with a centrally-peaked, azimuthally symmetric light distribution.  Simulation 5 is nearly edge-on with a varying light distribution. }
  \label{fig:toy}
\end{figure*}

\begin{figure*}
  \centering
  \includegraphics[width=0.3\textwidth]{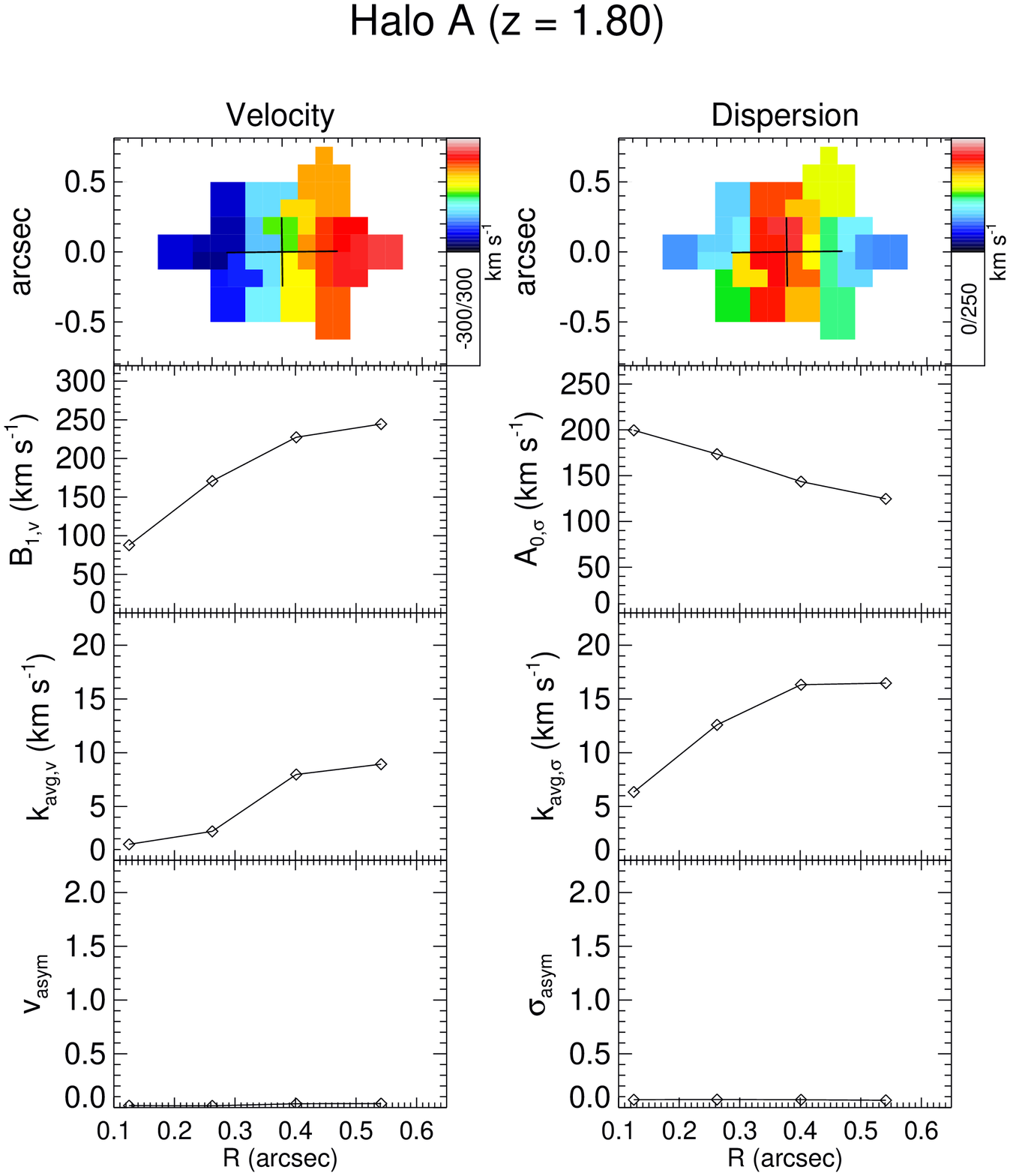}
  \includegraphics[width=0.3\textwidth]{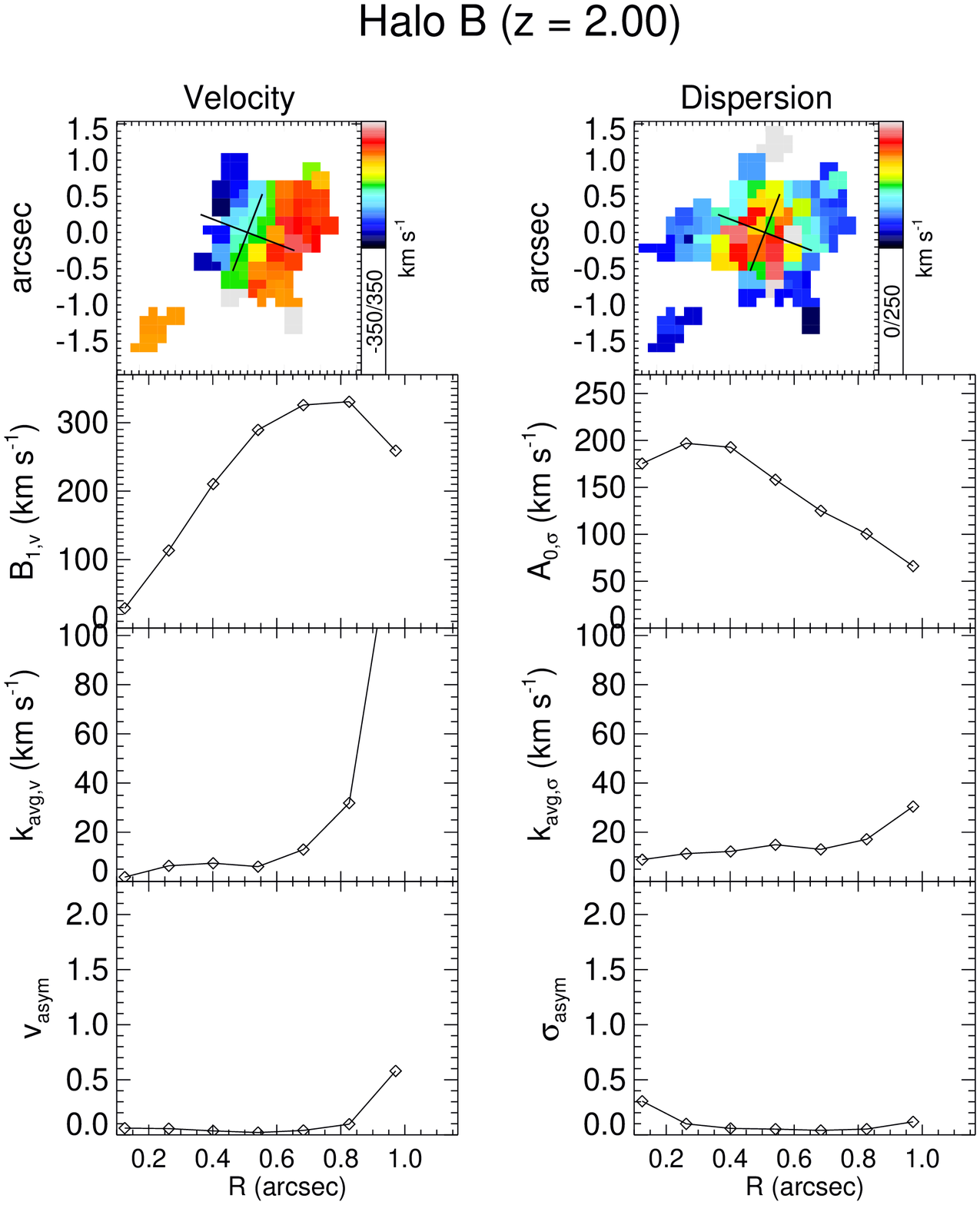}
  \includegraphics[width=0.3\textwidth]{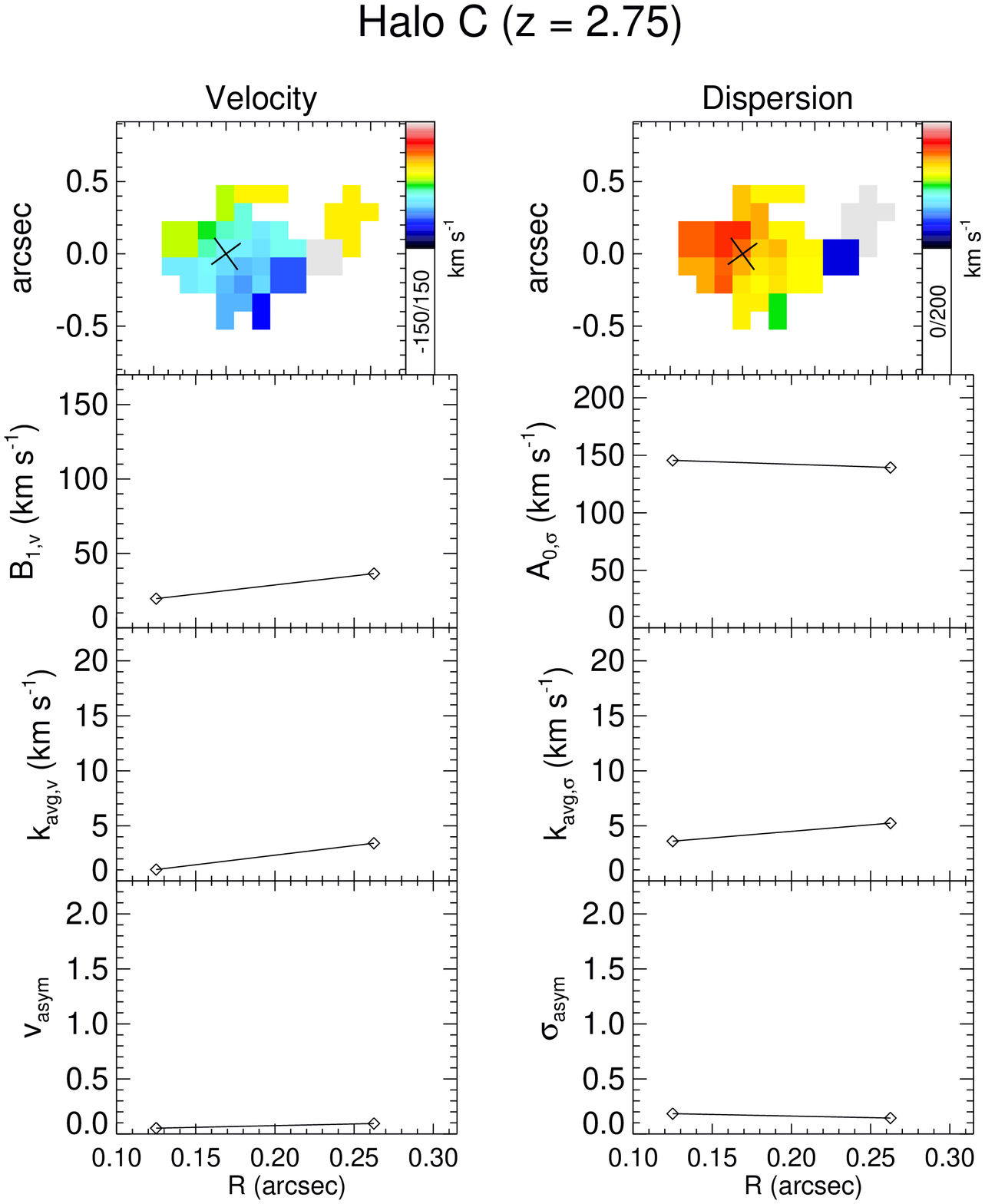}
  \includegraphics[width=0.3\textwidth]{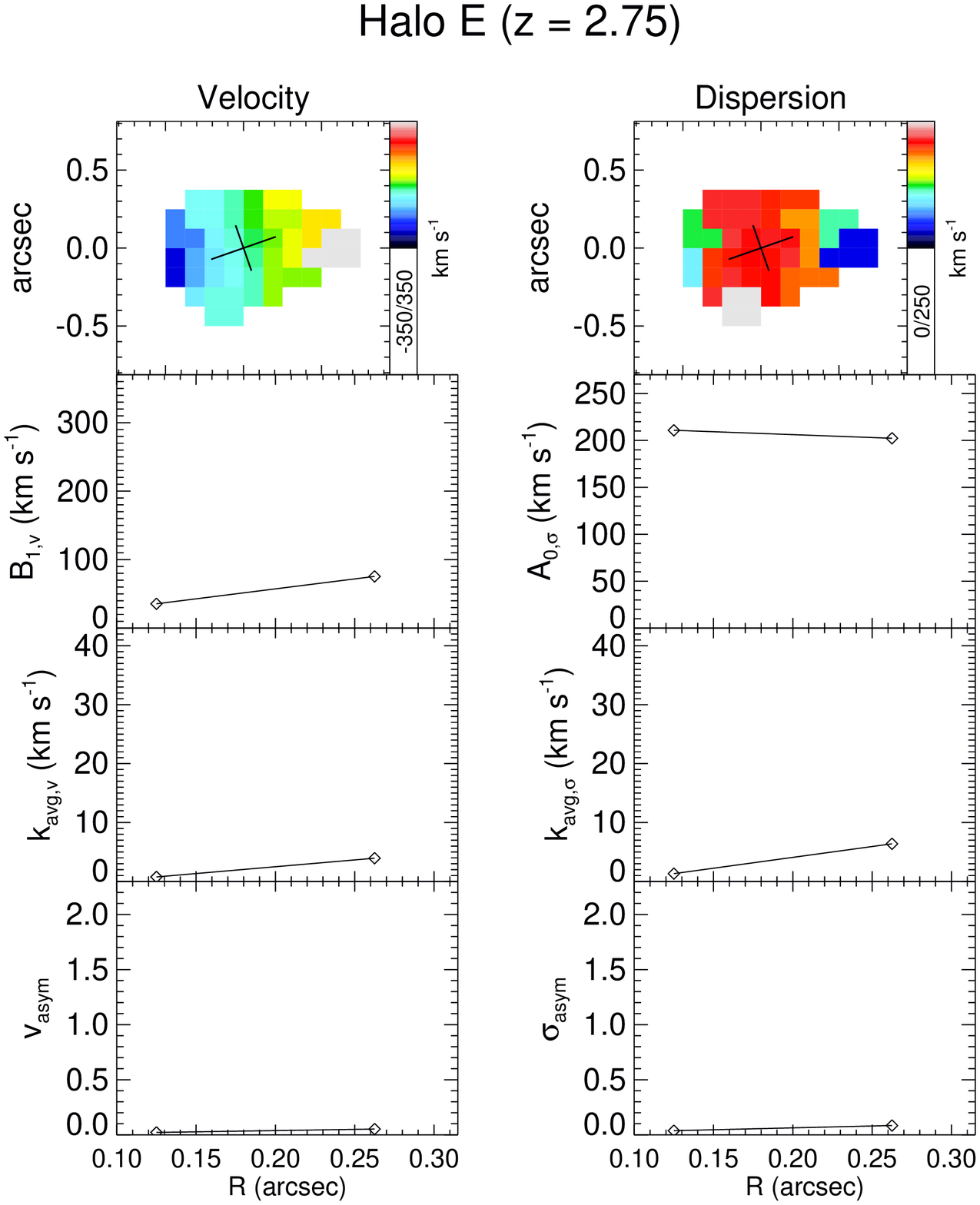}
  \includegraphics[width=0.3\textwidth]{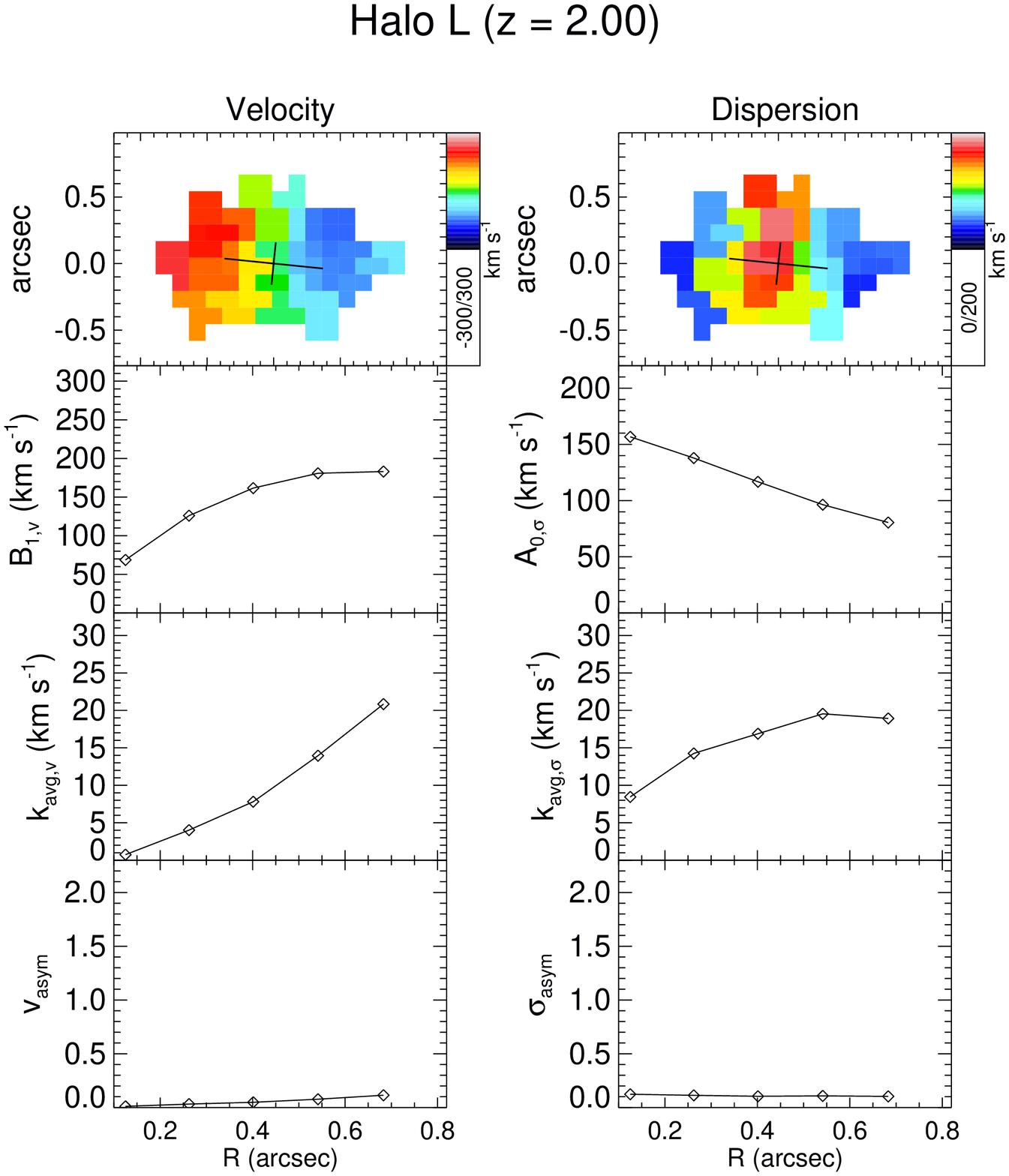}
  \includegraphics[width=0.3\textwidth]{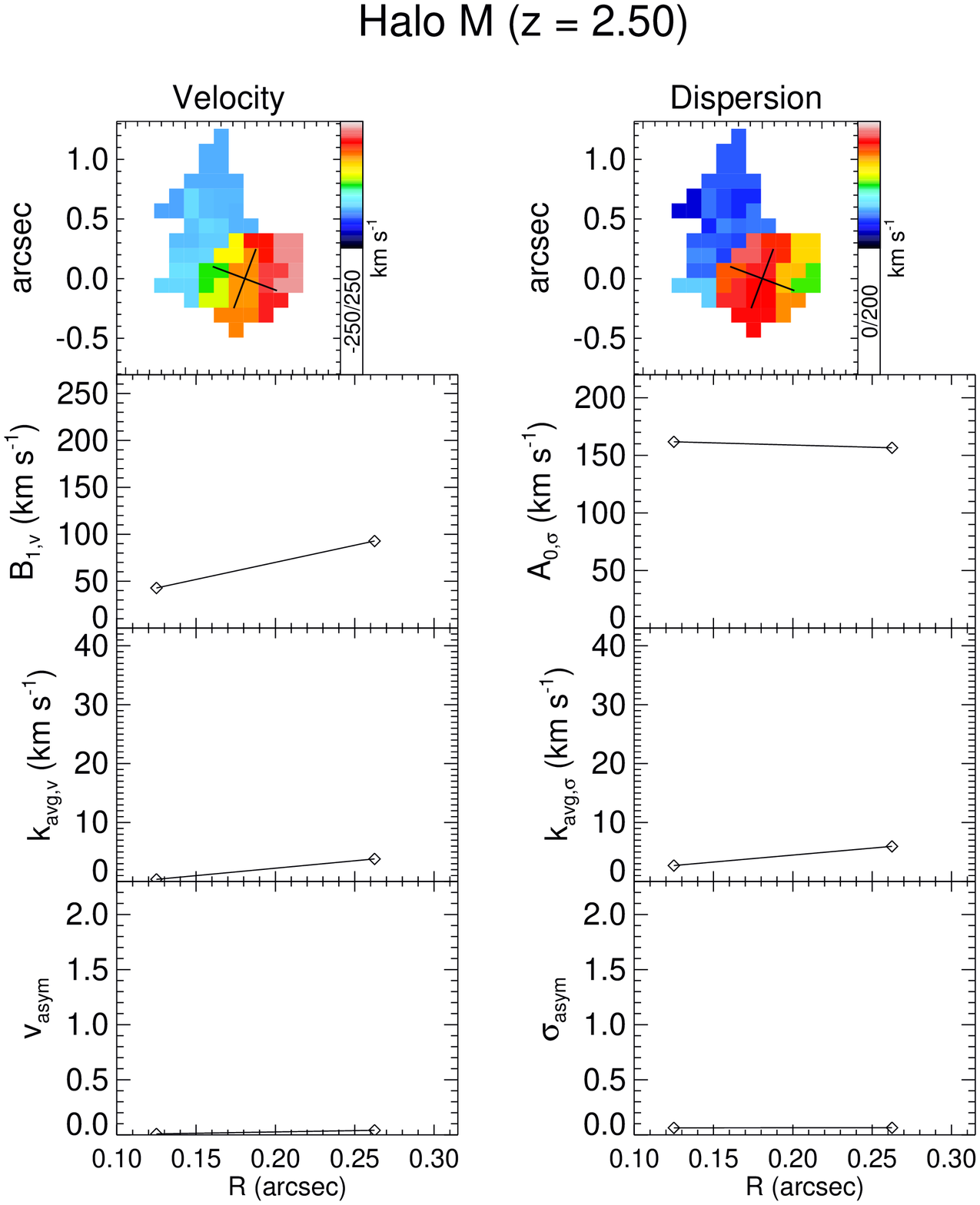}
  \caption{Kinemetry of cosmologically-simulated disks at $z \sim 2$.  The halo lettering is taken from \citet{Naa+07}, who describe the evolution of Halos A, C, and E in detail.  The snapshots at specific redshifts were selected such that the halos are accreting smoothly, with no major mergers, at this point in their histories.}
  \label{fig:cd}
\end{figure*}

\begin{figure*}
  \centering
  \includegraphics[width=0.45\textwidth]{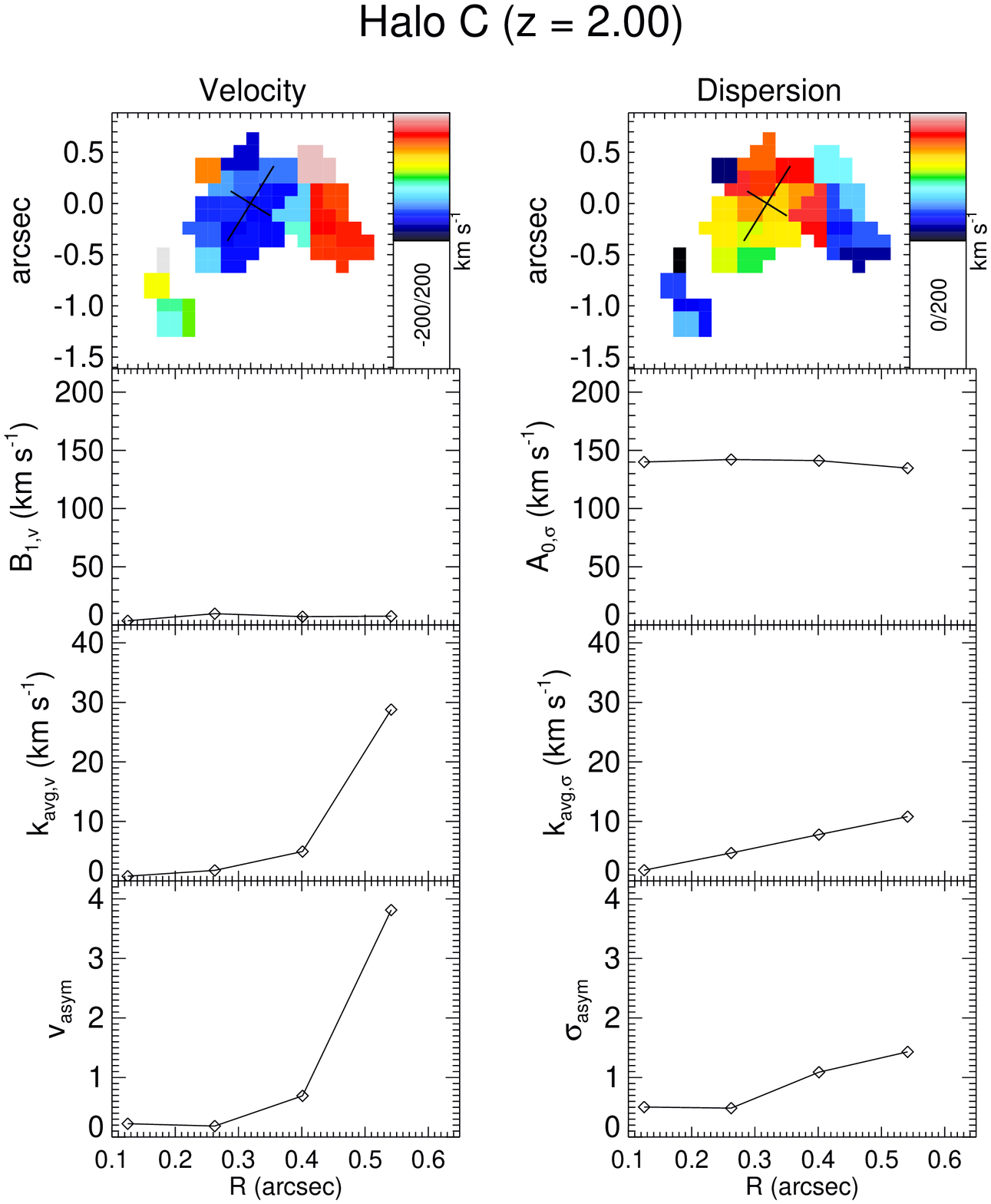}
  \includegraphics[width=0.45\textwidth]{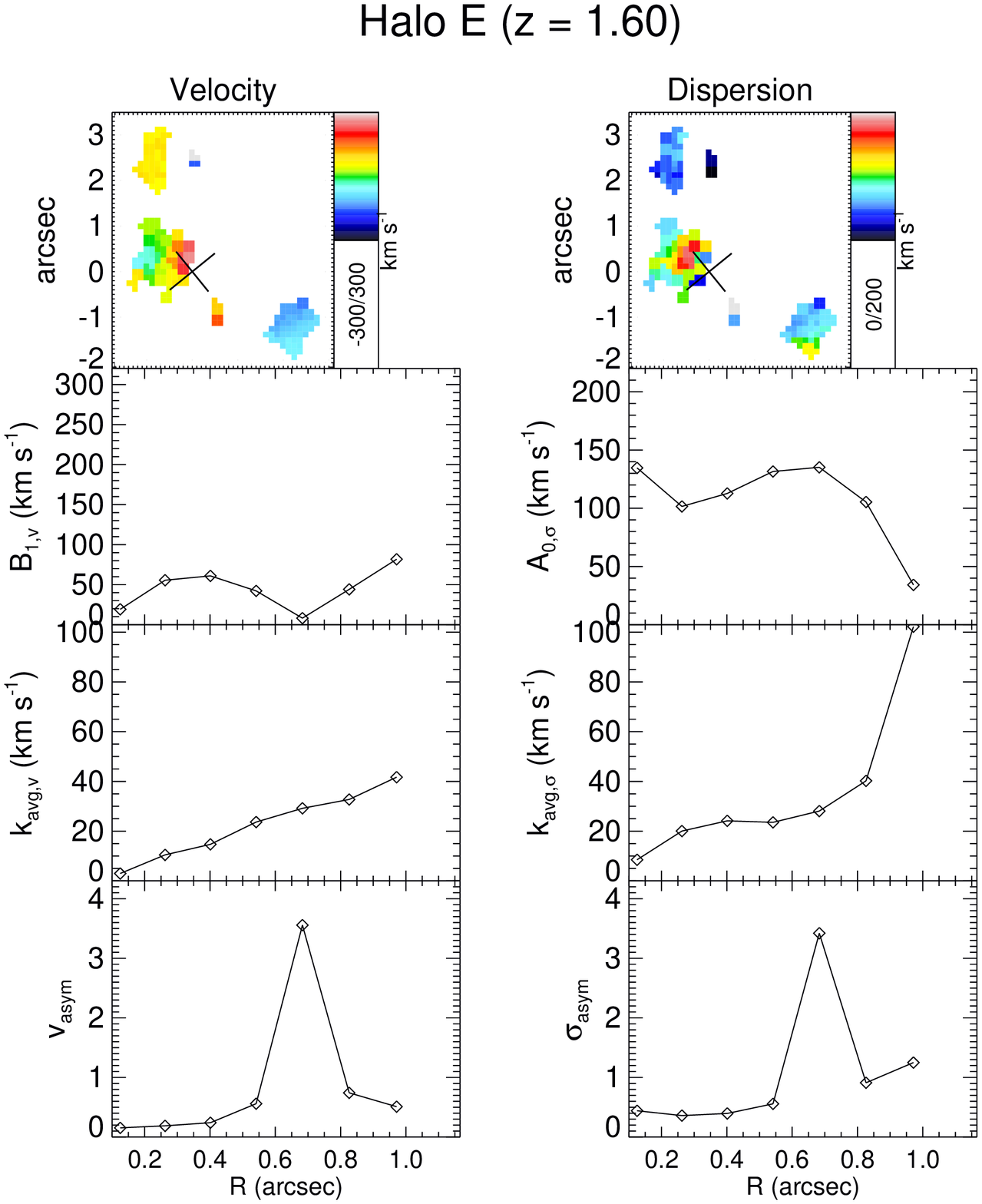}
  \includegraphics[width=0.45\textwidth]{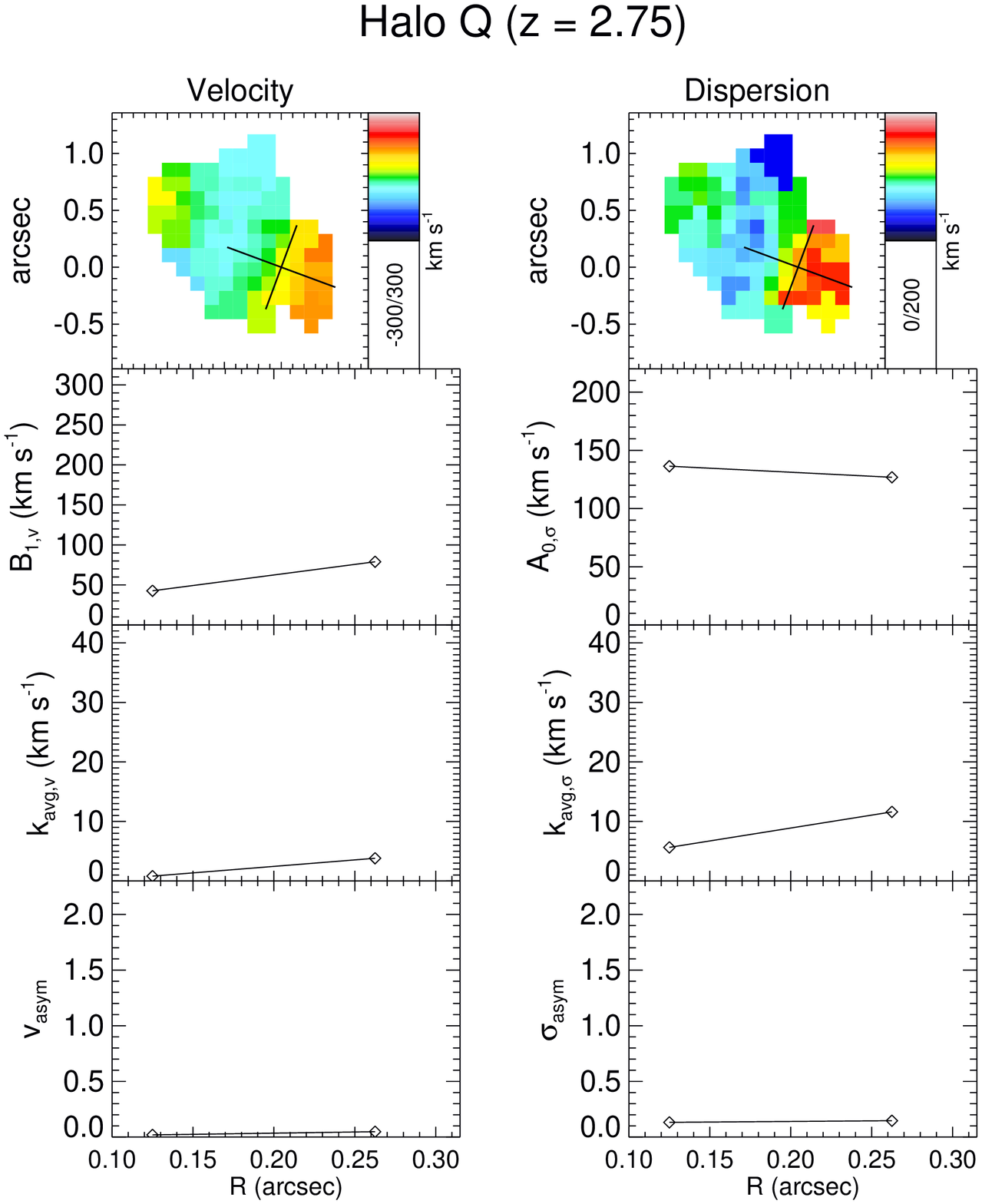}
  \includegraphics[width=0.45\textwidth]{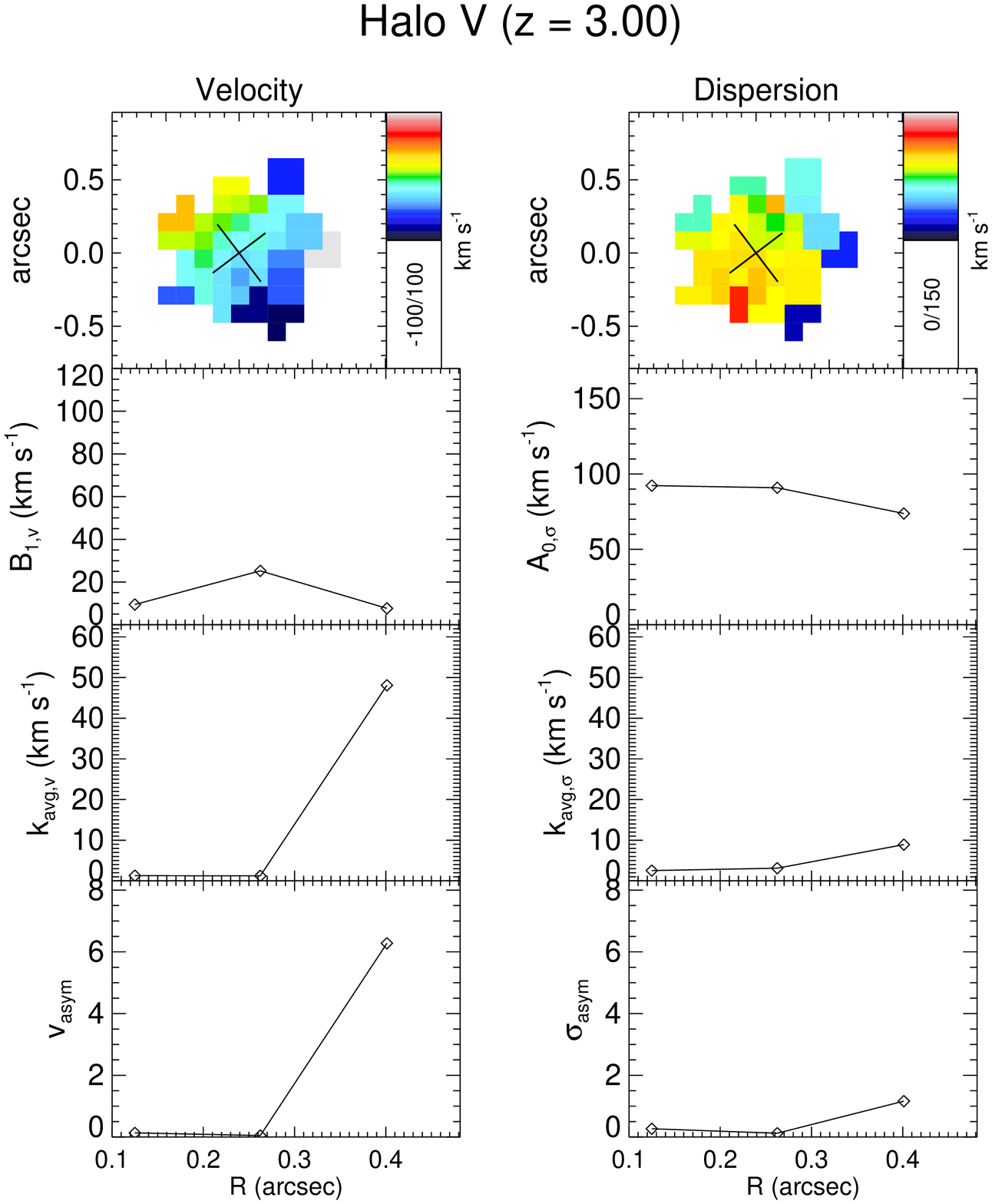}                
  \caption{Kinemetry of cosmologically-simulated mergers at $z \sim 2$.  The halo lettering is taken from \citet{Naa+07}, who describe the evolution of Halos A, C, and E in detail.  The snapshots at specific redshifts were selected such that the halos are undergoing major mergers at this point in their histories.}
  \label{fig:cm}
\end{figure*}

\begin{figure*}
  \centering
  \includegraphics[width=0.3\textwidth]{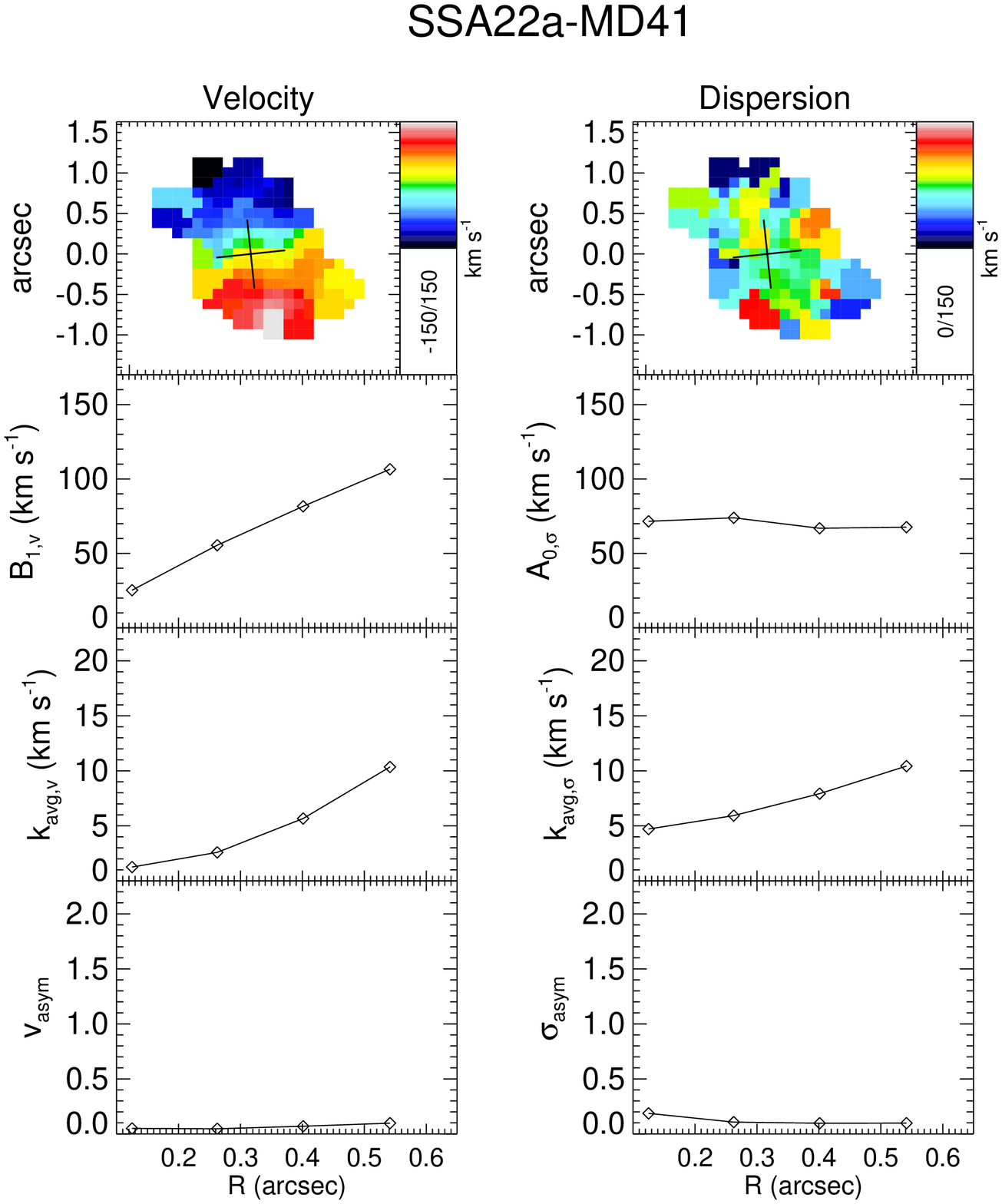}
  \includegraphics[width=0.3\textwidth]{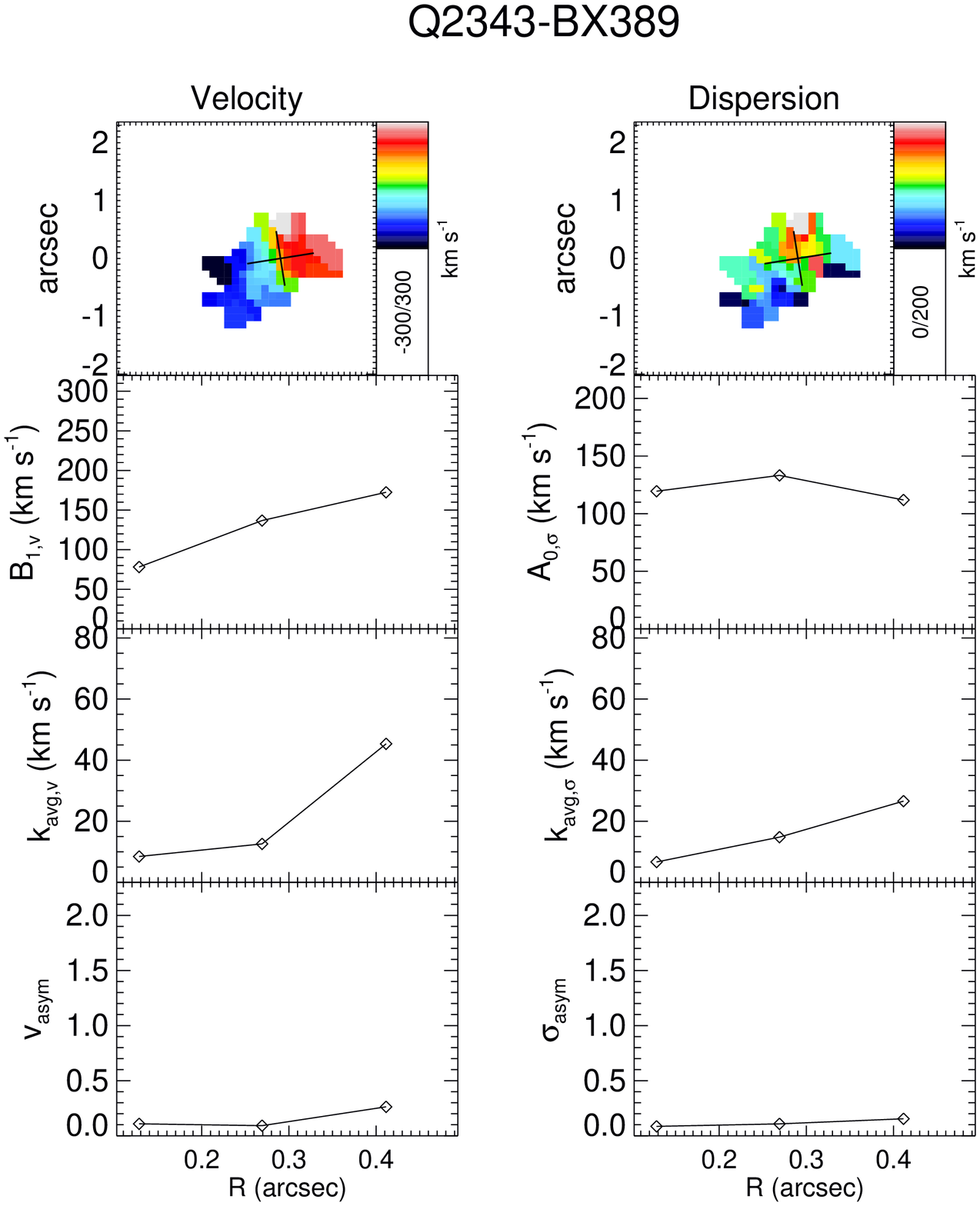}
  \includegraphics[width=0.3\textwidth]{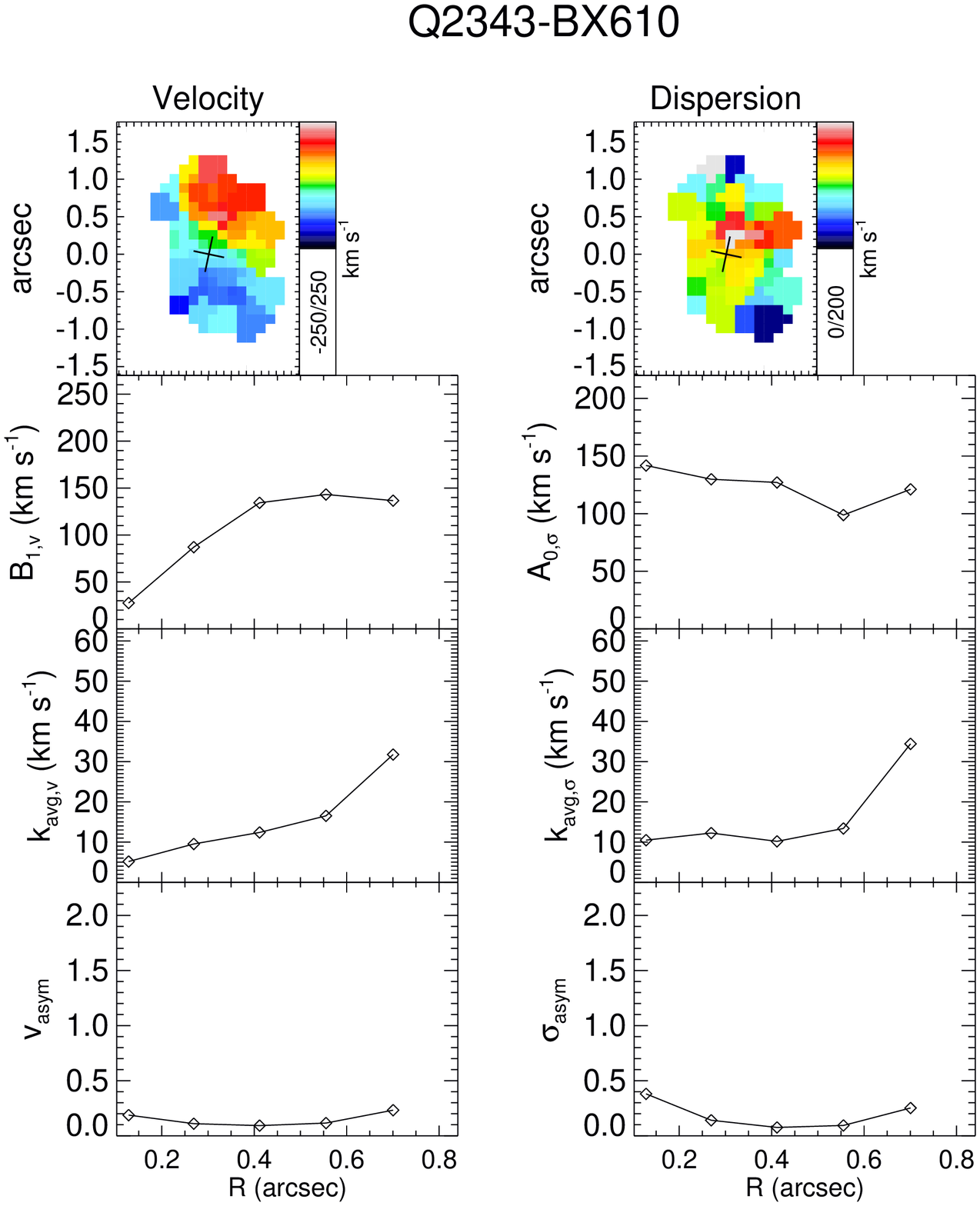}
  \includegraphics[width=0.3\textwidth]{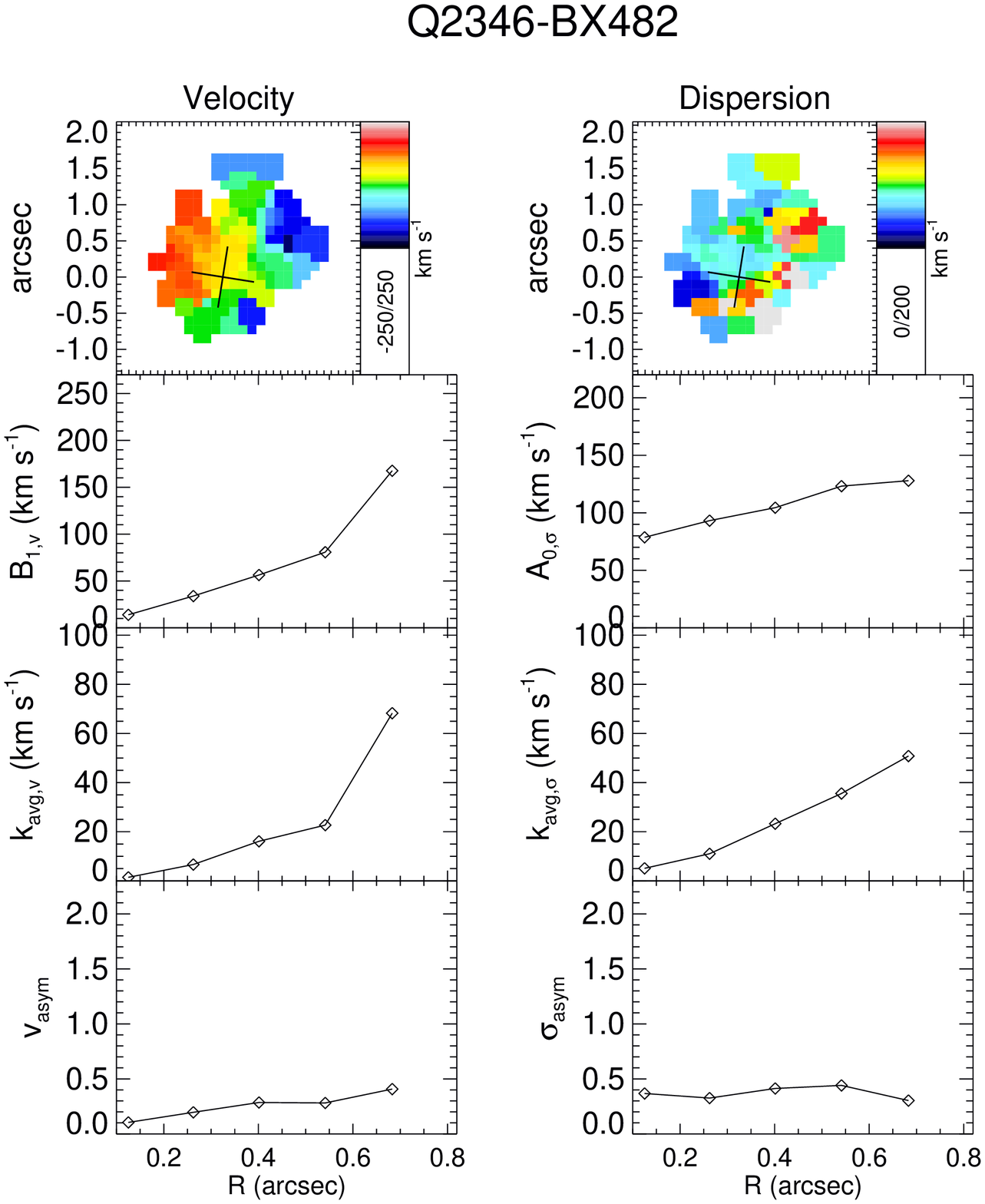}
  \includegraphics[width=0.3\textwidth]{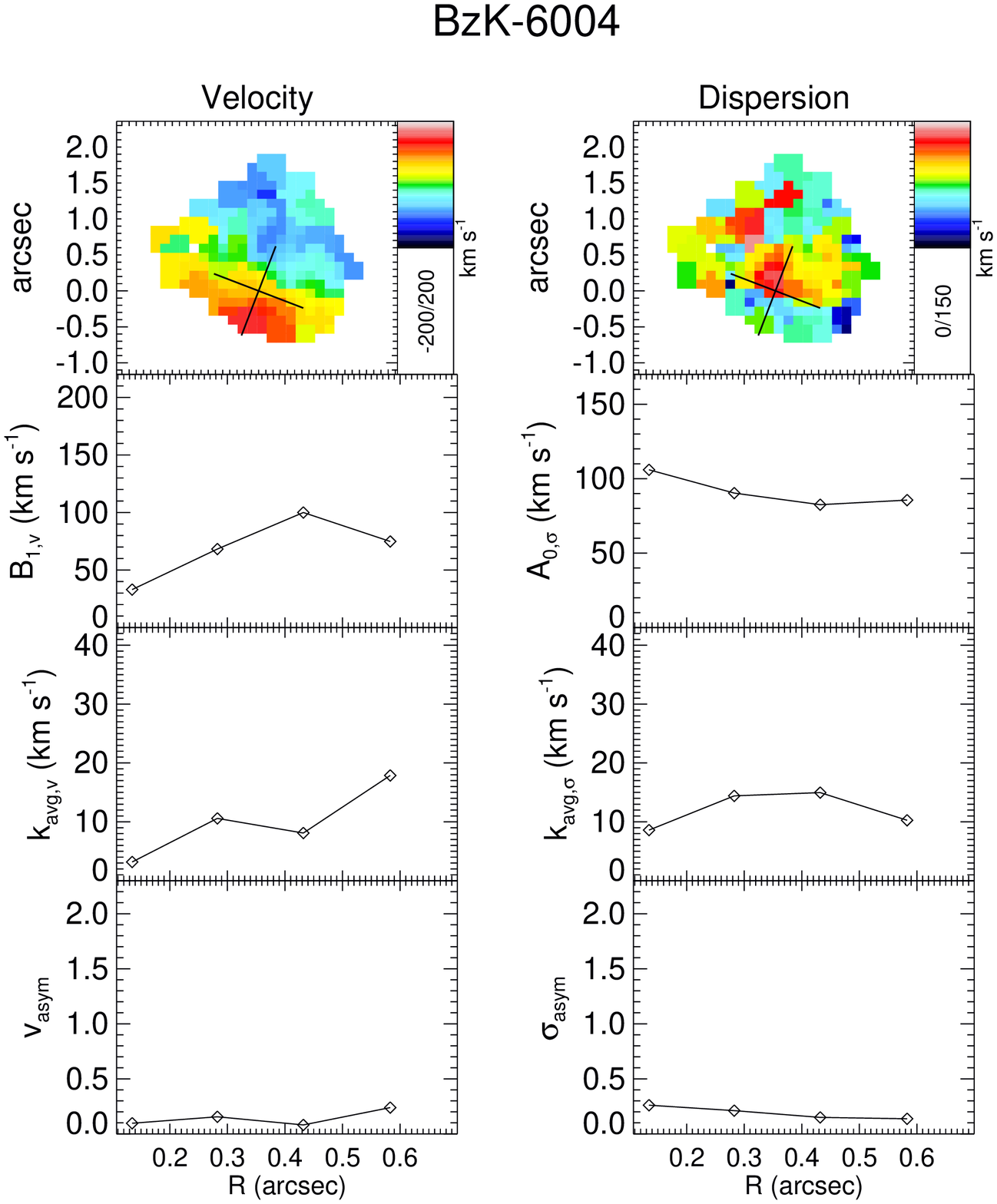}
  \includegraphics[width=0.3\textwidth]{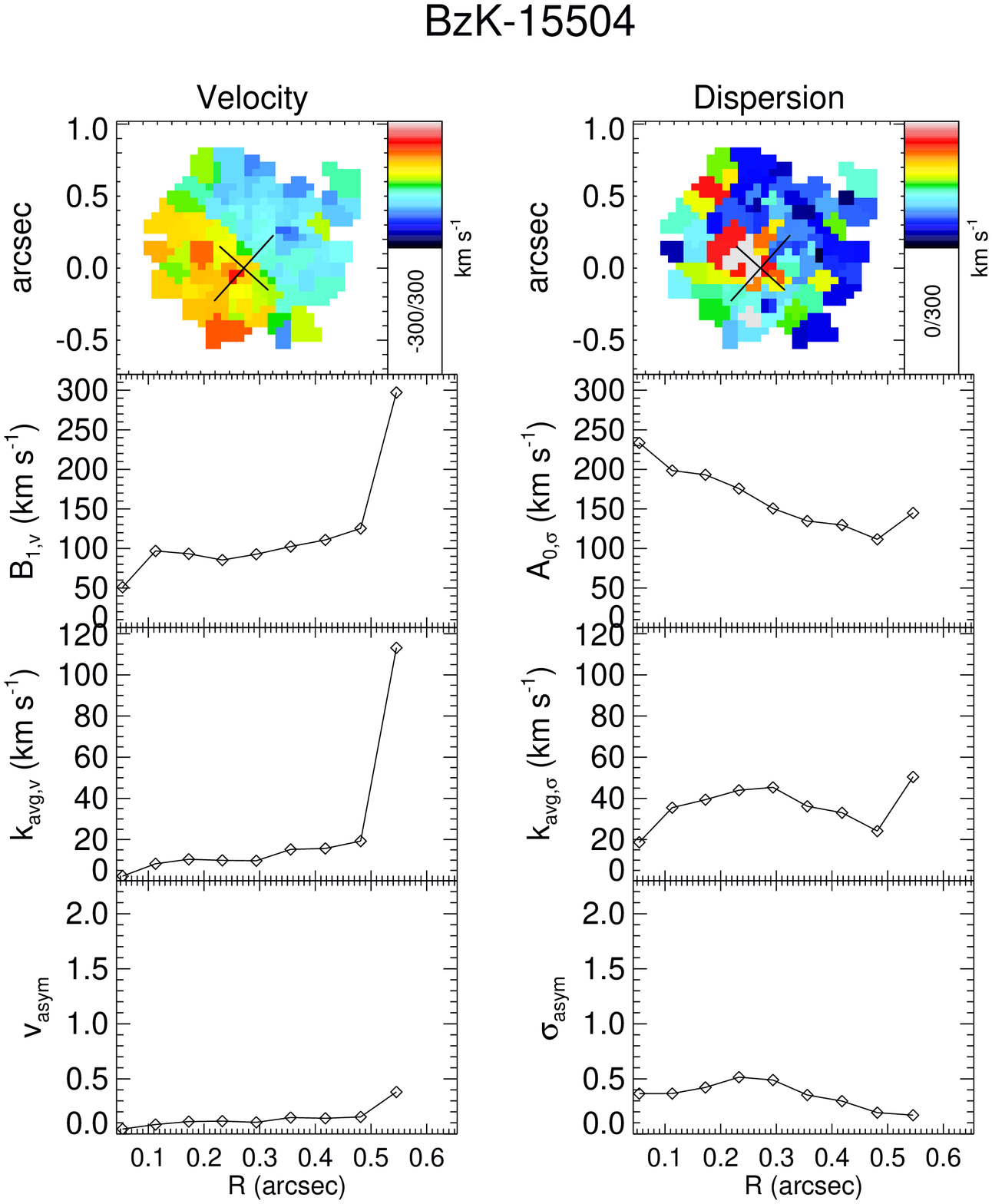}
  \includegraphics[width=0.3\textwidth]{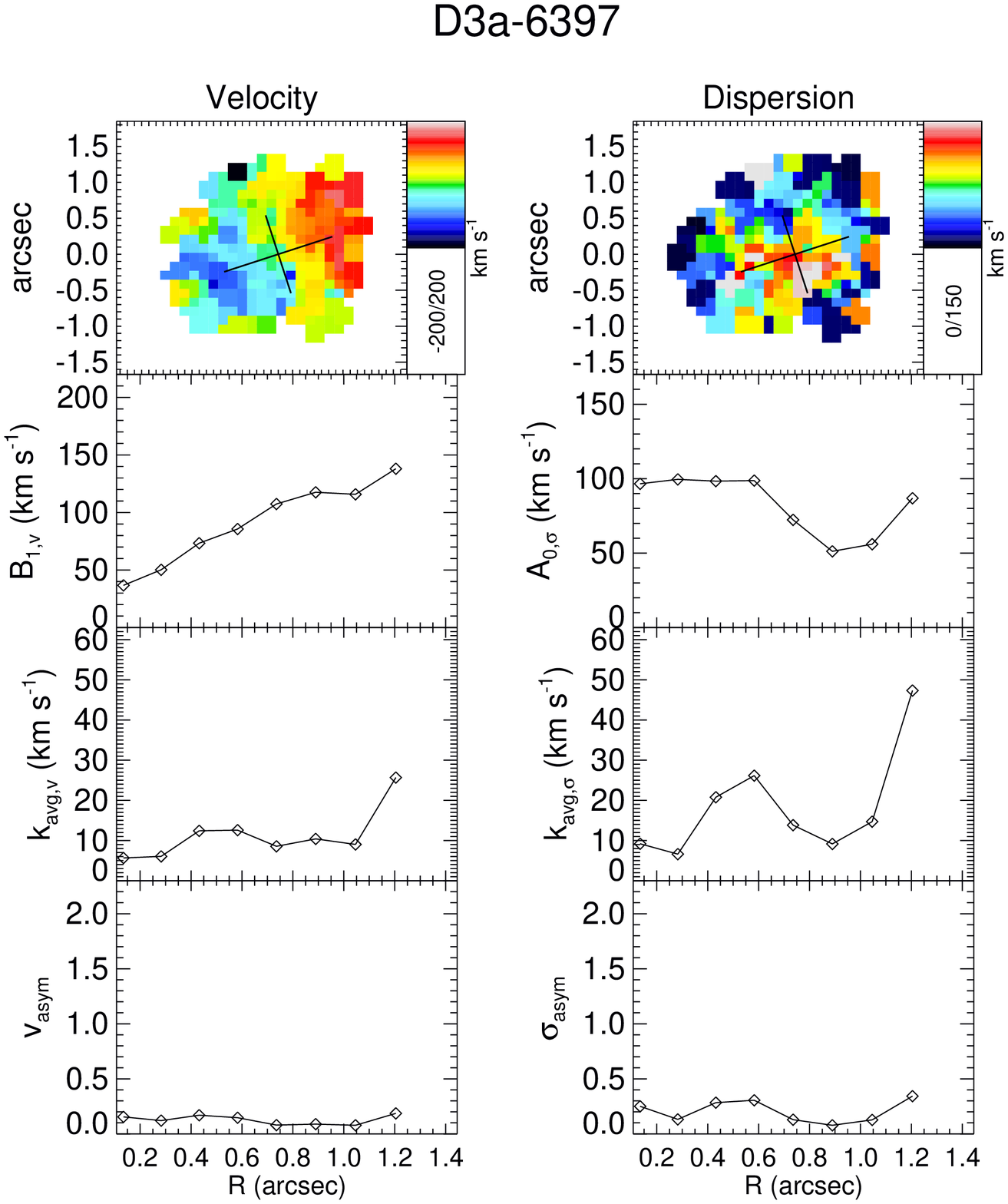}
  \includegraphics[width=0.3\textwidth]{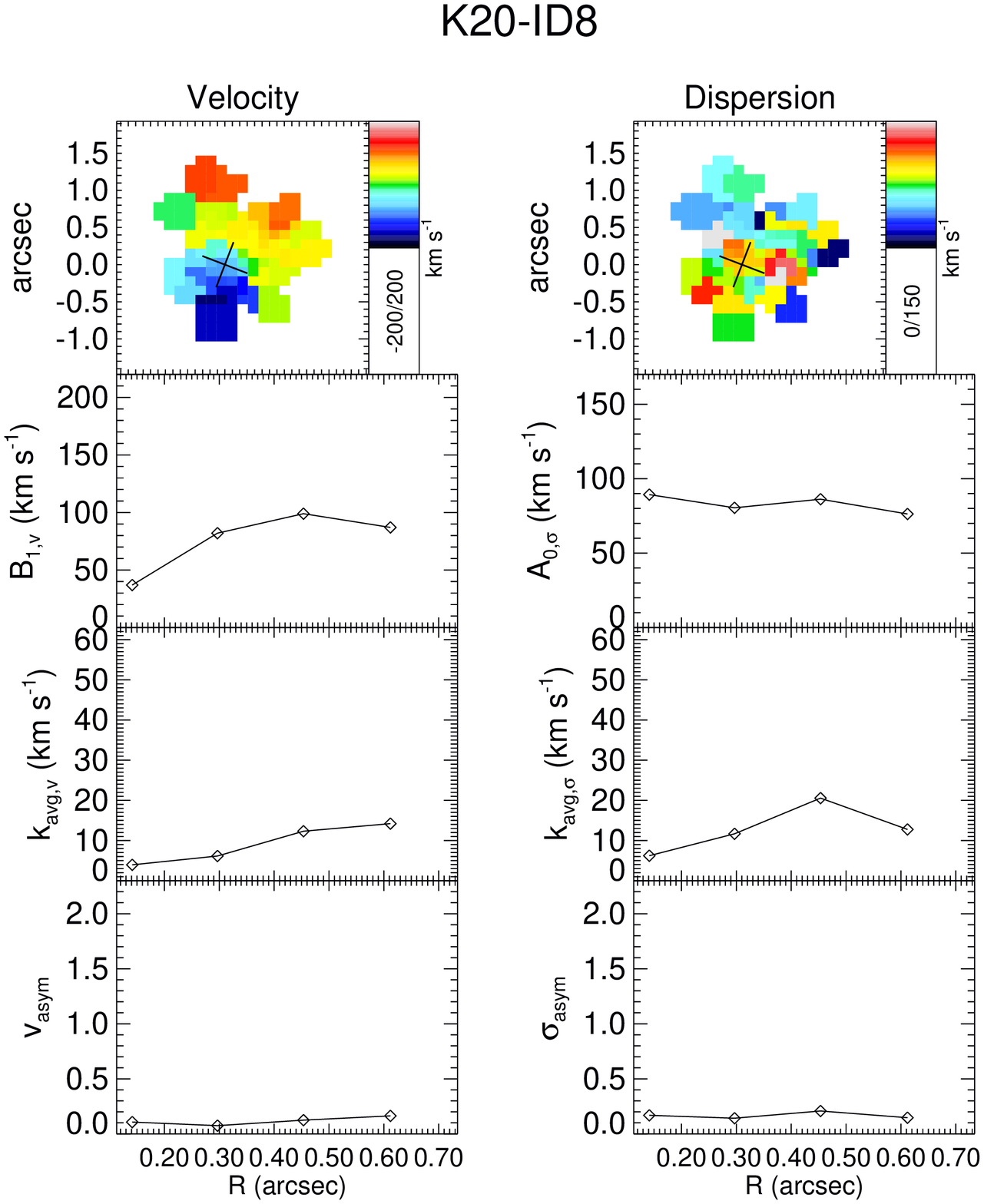}
  \caption{Kinemetry of SINS galaxies found to be disk-like.}
  \label{fig:zd}
\end{figure*}

\begin{figure*}
  \centering
  \includegraphics[width=0.45\textwidth]{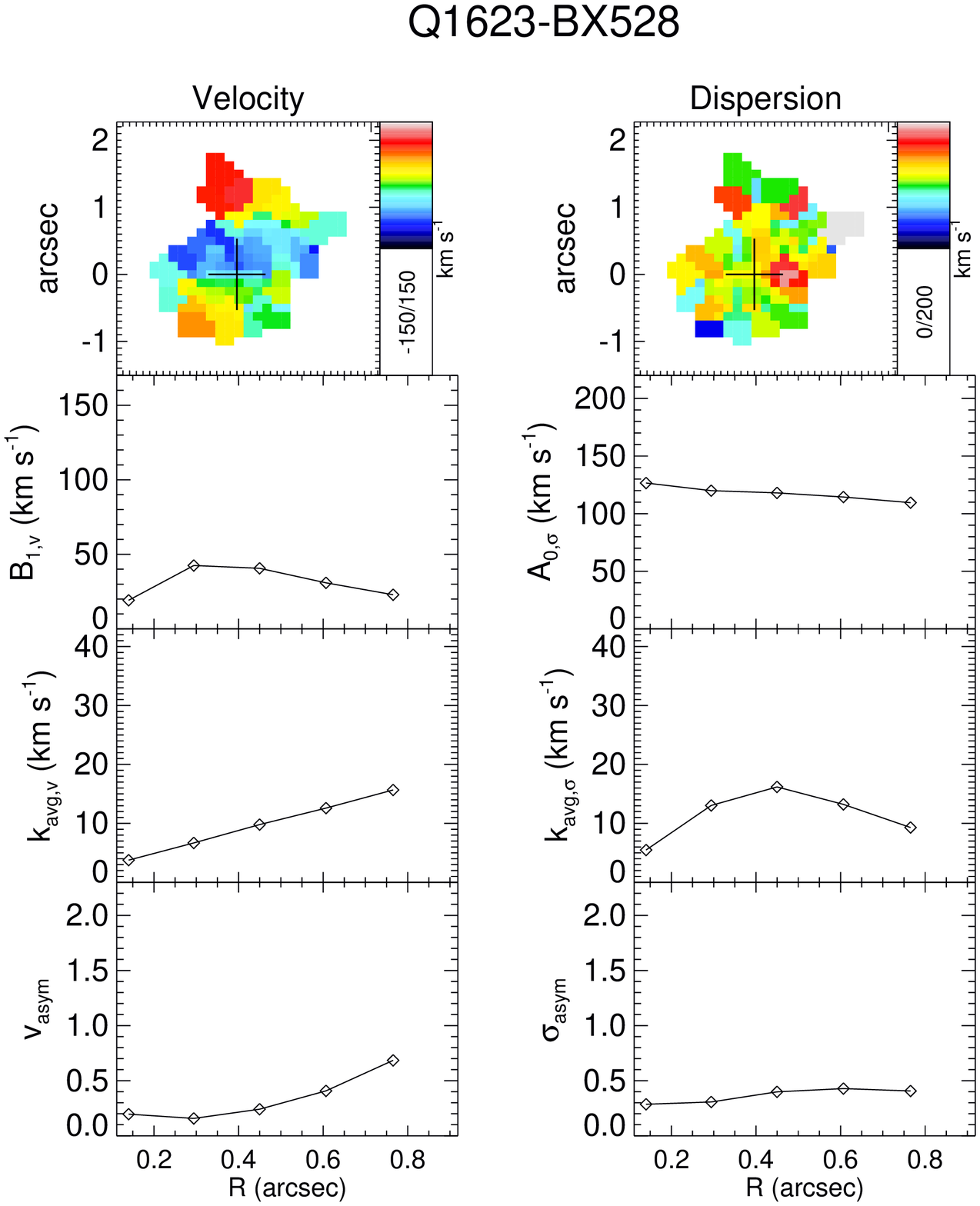}
  \includegraphics[width=0.45\textwidth]{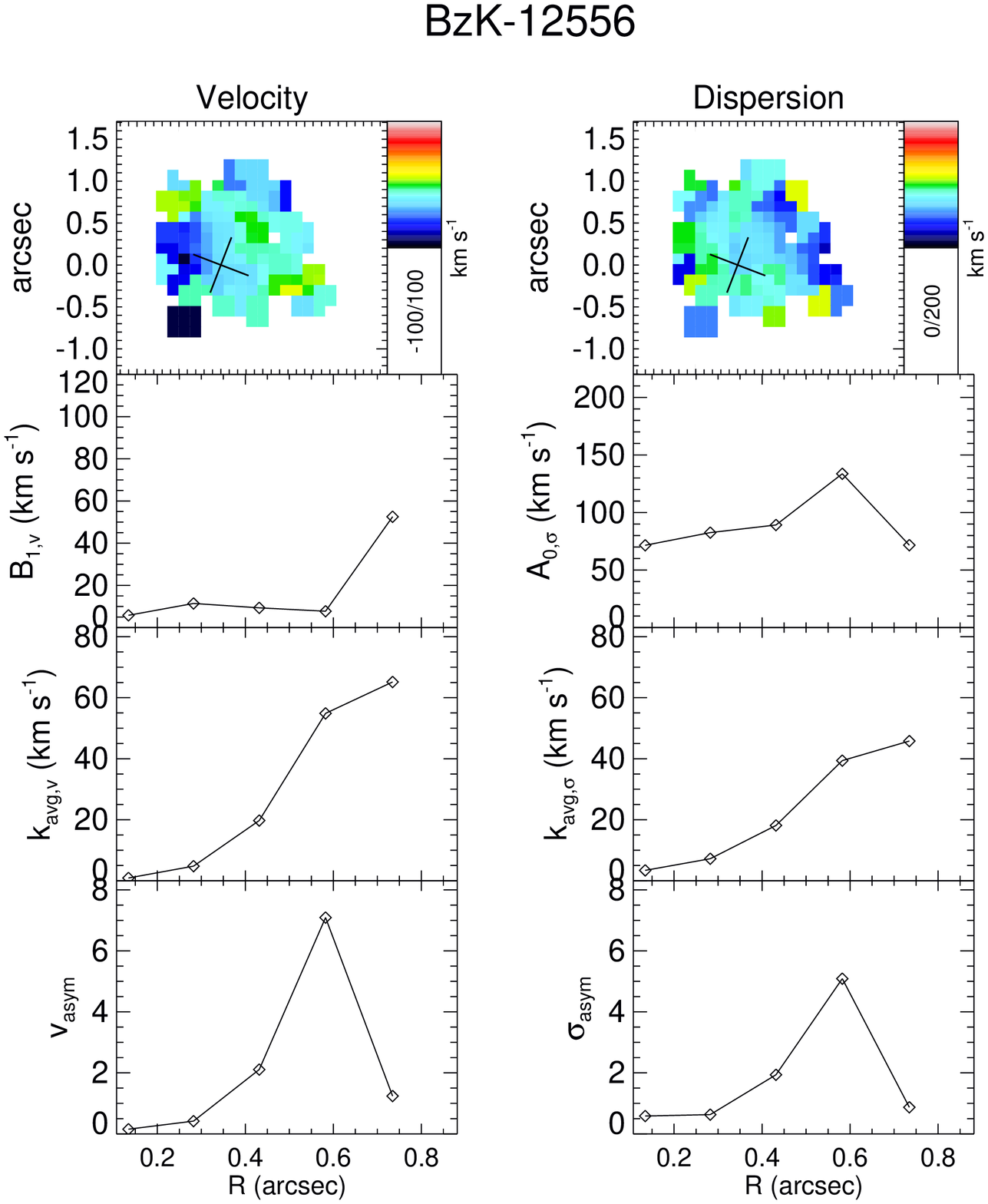}
  \includegraphics[width=0.45\textwidth]{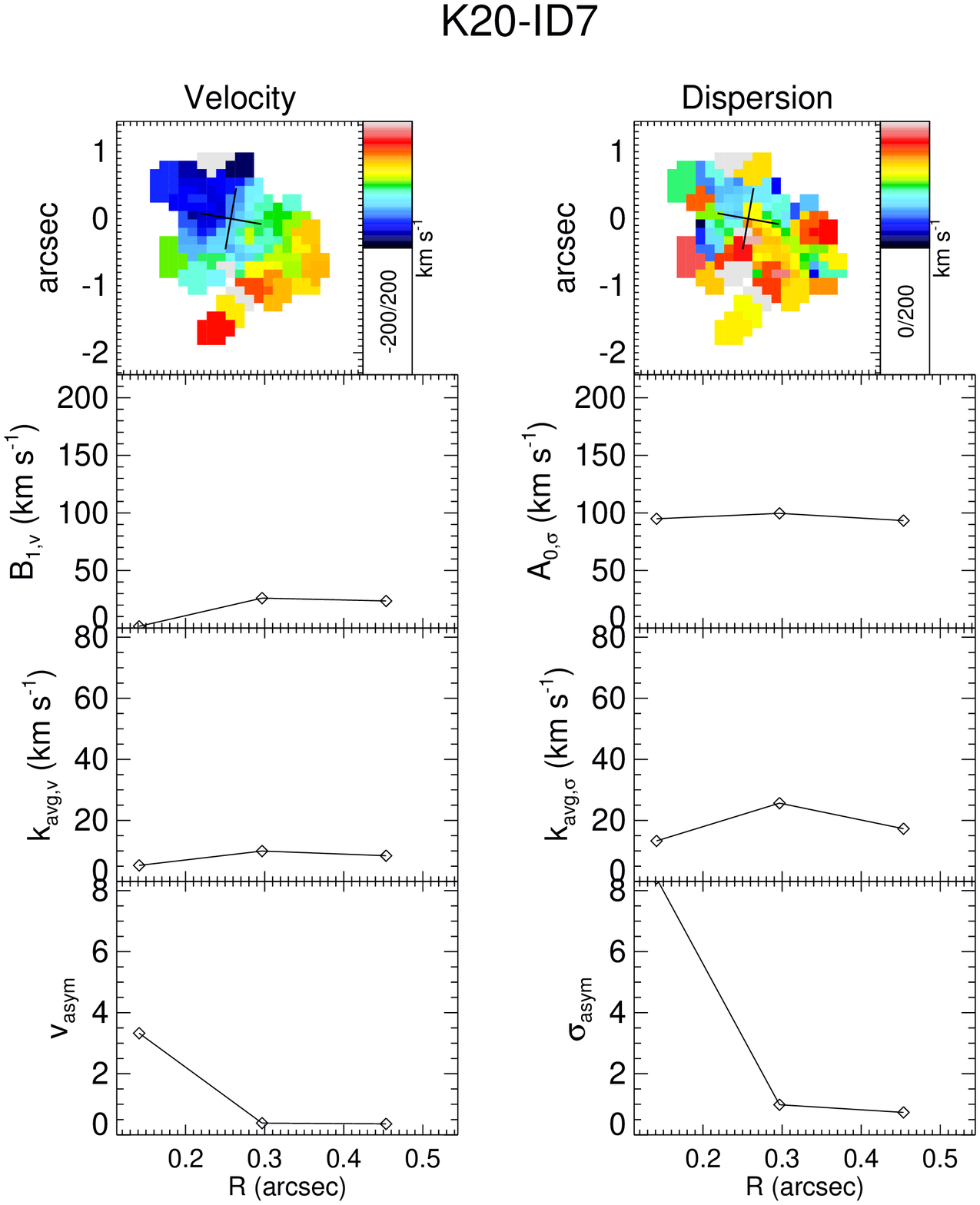}
  \caption{Kinemetry of SINS galaxies found to be merger-like.}
  \label{fig:zm}
\end{figure*}

In this appendix, we present the kinemetric analysis on all template and SINS galaxies.  These results are summarized in Table \ref{tabkin} and are shown for each system in Figures \ref{fig:spirz2} through \ref{fig:zm}.

\clearpage
\begin{table*}
\begin{center}
\caption{Kinemetry Results for the Templates and the High-$z$ Galaxies}
\label{tabkin}
\begin{minipage}{17cm}
\begin{center}
\begin{tabular*}{17cm}{cccccccccccc}
\hline
Galaxy & Type & \hspace{0.3cm} & & \vasym\footnotemark[a] & & \hspace{0.7cm} & & \sasym\footnotemark[a] & & \hspace{0.3cm} & Classification \\
\hline
\hline
NGC 925		& SABd; HII	&	& 0.14 & 0.18 & 0.25	 &	& 0.18 & 0.26 & 0.37 &	& Disk \\
NGC 3198	& SBc; non-active &	& 0.11 & 0.14 & 0.20	 &	& 0.16 & 0.21 & 0.32	 &	& Disk \\
NGC 4321	& SABbc; LINER &	& 0.13 & 0.16 & 0.21	 &	& 0.24 & 0.37 & 0.61	 &	& Disk \\
NGC 4579	& SABb; Seyfert	&	& 0.24 & 0.43 & 1.11	 &	& 0.13 & 0.21 & 0.47	 &	& Disk \\
NGC 4725	& SABab; Seyfert &	& 0.08 & 0.08 & 0.10	 &	& 0.06 & 0.08 & 0.11	 &	& Disk \\
NGC 7331	& SAb; LINER	&	& 0.06 & 0.07 & 0.10	 &	& 0.10 & 0.23 & 0.49	 &	& Disk \\
 & & & & & & & & & & & \\
IRAS 08572+3915	& ULIRG; HII	& & 0.20 & 0.42 & 1.34 &		& 0.75 & 1.58 & 4.40	&	& Merger \\
IRAS 12112+0305	& ULIRG; LINER & & 0.18 & 0.22 & 0.26 &	& 0.22 & 0.26 & 0.31	&	& Disk \\
IRAS 14348-1447	& ULIRG; LINER & & 0.56 & 0.99 & 2.47 &	& 1.35 & 2.25 & 5.02	&	& Merger \\
IRAS 15206+3342	& ULIRG; LINER & & 0.57 & 1.07 & 2.96 &	& 0.78 & 1.57 & 4.43	&	& Merger \\
IRAS 15250+3609	& ULIRG; LINER & & 0.17 & 0.47 & 2.10 &	& 0.56 & 1.52 & 6.42	&	& Merger \\
IRAS 17208-0014	& ULIRG; LINER & & 0.19 & 0.34 & 0.95 &	& 0.64 & 1.12 & 2.92	&	& Merger \\
Mrk 273			& ULIRG; Seyfert & & 0.55 & 1.10 & 3.25 &	& 0.96 & 1.97 & 5.82	&	& Merger \\
Arp 220			& ULIRG; HII	& & 0.18 & 0.27 & 0.51 &		& 0.24 & 0.34 & 0.64	&	& Merger \\
 & & & & & & & & & & & \\
Toy Simulation 1	& Toy Disk Model &	& 0.03 & 0.03 & 0.04	 &	& 0.10 & 0.11 & 0.16	&	& Disk \\
Toy Simulation 2	& Toy Disk Model &	& 0.13 & 0.15 & 0.17	 &	& 0.16 & 0.20 & 0.28	&	& Disk \\
Toy Simulation 3	& Toy Disk Model &	& 0.07 & 0.08 & 0.09	 &	& 0.11 & 0.13 & 0.16	&	& Disk \\
Toy Simulation 4	& Toy Disk Model &	& 0.03 & 0.03 & 0.03	 &	& 0.03 & 0.03 & 0.03	&	& Disk \\
Toy Simulation 5	& Toy Disk Model &	& 0.10 & 0.10 & 0.11	 &	& 0.05 & 0.07 & 0.08	&	& Disk \\
 & & & & & & & & & & & \\
Halo A ($z = 1.80$)	& Simulated Disk &	& 0.03 & 0.04 & 0.04	 &	& 0.07 & 0.07 & 0.08	&	& Disk \\
Halo B ($z = 2.00$)	& Simulated Disk &	& 0.11 & 0.12 & 0.12	 &	& 0.07 & 0.09 & 0.10	&	& Disk \\
Halo C ($z = 2.75$)	& Simulated Disk &	& 0.10 & 0.13 & 0.18	 &	& 0.19 & 0.24 & 0.31	&	& Disk \\
Halo E ($z = 2.75$)	& Simulated Disk &	& 0.08 & 0.10 & 0.13	 &	& 0.11 & 0.13 & 0.16	&	& Disk \\
Halo L ($z = 2.00$)	& Simulated Disk &	& 0.05 & 0.06 & 0.07	 &	& 0.09 & 0.10 & 0.11	&	& Disk \\
Halo M ($z = 2.50$)	& Simulated Disk &	& 0.08 & 0.09 & 0.10 &	& 0.08 & 0.09 & 0.11	&	& Disk \\
 & & & & & & & & & & & \\
Halo C ($z = 2.00$)	& Simulated Merger	& & 1.01 & 1.71 & 2.84 &	& 0.40 & 0.70 & 2.22	 &	& Merger \\
Halo E ($z = 1.60$)	& Simulated Merger	& & 0.30 & 0.51 & 1.28 &	& 0.24 & 0.57 & 1.59	 &	& Merger \\
Halo Q ($z = 2.75$)	& Simulated Merger	&  & 0.55 & 0.82 & 1.97 &	& 0.28 & 0.42 & 0.92	 &	& Merger \\
Halo V ($z = 3.00$)	& Simulated Merger	& & 0.85 & 1.43 & 4.80 &	& 0.32 & 0.46 & 1.19	 &	& Merger \\
 & & & & & & & & & & & \\
SSA22a-MD41 	& SINS Galaxy	&	& 0.08 & 0.10 & 0.12	 &	& 0.12 & 0.14 & 0.17	 &	& Disk \\
Q2343-BX389  		& SINS Galaxy	&	& 0.13 & 0.16 & 0.21	 &	& 0.11 & 0.14 & 0.19	 &	& Disk \\
Q2343-BX610 		& SINS Galaxy	&	& 0.14 & 0.17 & 0.21	 &	& 0.17 & 0.23 & 0.31	 &	& Disk \\
Q2346-BX482 		& SINS Galaxy	&	& 0.20 & 0.23 & 0.29	 &	& 0.26 & 0.31 & 0.38	 &	& Disk \\
BzK-6004 		& SINS Galaxy	&	& 0.15 & 0.17 & 0.23	 &	& 0.23 & 0.29 & 0.39	 &	& Disk \\
BzK-15504 		& SINS Galaxy &	& 0.15 & 0.18 & 0.21	 &	& 0.28 & 0.34 & 0.44	 &	& Disk \\
D3a-6397 		& SINS Galaxy	&	& 0.15 & 0.18 & 0.21	 &	& 0.21 & 0.25 & 0.30	 &	& Disk \\
K20-ID8			& SINS Galaxy	&	& 0.12 & 0.15 & 0.21	 &	& 0.16 & 0.19 & 0.27	 &	& Disk \\
 & & & & & & & & & & & \\
Q1623-BX528 		& SINS Galaxy	&	& 0.35 & 0.47 & 0.97	 &	& 0.34 & 0.48 & 0.94 &	& Merger \\
BzK-12556 		& SINS Galaxy	&	& 1.36 & 2.27 & 6.71	 &	& 1.13 & 1.91 & 5.18	 &	& Merger \\
K20-ID7 			& SINS Galaxy	&	& 0.31 & 0.46 & 0.97	 &	& 0.44 & 0.73 & 1.45	 &	& Merger \\
\hline
\end{tabular*}
\end{center}
\footnotetext[a]{Listed here is the 68\% confidence interval, as derived from 1000 Monte Carlo realizations, with the three columns indicating the lower bound on the confidence interval, the median value, and the upper bound on the confidence interval, respectively.}
\end{minipage}
\end{center}
\end{table*}

 
 \end{document}